\documentclass{aa}

\usepackage{comment}
\usepackage{ulem}
\usepackage{color}
\usepackage{txfonts}
\usepackage{lipsum}
\usepackage[]{hyperref}
\hypersetup{colorlinks=true,linkcolor=black,citecolor=blue}

\def\vec#1{\mbox{\boldmath $#1$}}

\usepackage{xcolor}
\colorlet{Mycolor1}{orange!100!}
\def\revise#1{#1}
\def\sout#1{}

\begin{document}

    \title{Corona and XUV emission modelling of the Sun and Sun-like stars}

   \subtitle{}

   \author{Munehito Shoda
          \inst{1}
          \and
          Shinsuke Takasao
          \inst{2}
          }

   \institute{
    National Astronomical Observatory of Japan, 
    National Institutes of Natural Sciences, 2-21-1 Osawa, Mitaka, Tokyo, 181-8588, Japan
    \email{munehito.shoda@nao.ac.jp}
    \and
    Department of Earth and Space Science, Graduate School of Science, Osaka University, Toyonaka, Osaka 560-0043, Japan
             }

   \date{Received month dd, yyyy; accepted month dd, yyyy}

    \abstract
    {
    The X-ray and extreme-ultraviolet (EUV) emissions from the low-mass stars significantly affect the evolution of the planetary atmosphere.
    However, it is, observationally difficult to constrain the stellar high-energy emission because of the strong interstellar extinction of EUV photons.
    In this study, we simulate the XUV (X-ray+EUV) emission from the Sun-like stars by extending the solar coronal heating model that self-consistently solves, with sufficiently high resolution, the surface-to-coronal energy transport, turbulent coronal heating, and coronal thermal response by conduction and radiation.
    The simulations are performed with a range of loop lengths and magnetic filling factors at the stellar surface.
    With the solar parameters, 
    the model reproduces the observed solar XUV spectrum below the Lyman edge, thus validating its capability of predicting the XUV spectra of other Sun-like stars.
    The model also reproduces the observed nearly-linear relation between the unsigned magnetic flux and the X-ray luminosity.
    From the simulation runs with various loop lengths and filling factors, 
    \sout{the following scaling relations are found.
    where $L_{\rm EUV}$ and $L_{\rm X}$ are the cgs-unit luminosity in the EUV ($100 {\rm \ \AA} < \lambda \le 912 {\rm \ \AA}$) and X-ray ($5 {\rm \ \AA} < \lambda \le 100 {\rm \ \AA}$) range, respectively, and $\Phi_{\rm photon}^{\rm EUV}$ is the total number of EUV photons emitted per second.}
    \revise{we also find a scaling relation, namely $\log L_{\rm EUV} = 9.93 + 0.67 \log L_{\rm X}$, where $L_{\rm EUV}$ and $L_{\rm X}$ are the luminosity in the EUV ($100 {\rm \ \AA} < \lambda \le 912 {\rm \ \AA}$) and X-ray ($5 {\rm \ \AA} < \lambda \le 100 {\rm \ \AA}$) range, respectively, in cgs.}
    By assuming a power--law relation between the Rossby number and the magnetic filling factor, we reproduce the renowned relation between the Rossby number and the X-ray luminosity.
    We also propose an analytical description of the energy injected into the corona, which, in combination with the conventional Rosner--Tucker--Vaiana scaling law, semi-analytically explains the simulation results.
    This study refines the concepts of solar and stellar coronal heating and derives a theoretical relation for estimating the hidden stellar EUV luminosity from X-ray observations.
    }
  

    \maketitle
    
\section{Introduction}

The Sun has an aura of hot plasma called the corona, which has a temperature of a few million Kelvin \citep{Edlen_1943}.
The image of the corona has been captured by the Atmospheric Imaging Assembly \citep{Lemen_2012_SolPhys} of the NASA’s Solar Dynamics Observatory \citep{Pesnell_2012_SolPhys}.
The corona is not unique to the Sun and has been observed to surround low-mass main-sequence stars in general \citep{Gudel_1997_ApJ}.
Due to its high temperature, the corona is the principal source of stellar XUV (X-ray + \revise{EUV:} extreme ultraviolet) emissions.

Stellar XUV emissions drive the expansion and thermal escape of planetary atmospheres \citep{Vidal-Madjar_2003_Nature,Lecavelier_des_Etangs_2012_AA,Owen_2013_ApJ,Ehrenreich_2015_Nature,Airapetian_2017_ApJ}.
Therefore, describing the stellar XUV emission in terms of the stellar fundamental parameters (luminosity, mass, radius, etc.) is essential in \sout{the arguments} \revise{understanding} the evolution of a planet and its habitability.
Since stellar XUV emissions decrease over time \citep{Gudel_1997_ApJ,Ribas_2005_ApJ,Telleschi_2005_ApJ,Claire_2012_ApJ,Guinan_2016_ApJ} \revise{presumably} in response to the stellar spin-down \citep{Kraft_1967_ApJ,Skumanich_1972_ApJ,Barnes_2003_ApJ,Irwin_2009_proceedings,Matt_2015_ApJ} caused by magnetised stellar wind \citep[][]{Weber_1967_ApJ,Kawaler_1988_ApJ,Shoda_2020_ApJ},
the long-term evolution of the XUV activity of the host star needs to be described as well \citep{Tu_2015_AA,Johnstone_2021_AA}.

While the observational characteristics of stellar X-ray emissions have been established, including their correlations with the rotation \citep{Pallavicini_1981_ApJ,Wright_2011_ApJ} and magnetic field of the star \citep{Pevtsov_2003_ApJ,Vidotto_2014_MNRAS,Kochukhov_2020_AA}, stellar EUV emissions are poorly characterised because they are difficult to observe.
EUV photons suffer from strong absorption by the interstellar medium \citep{Rumph_1994_AJ},
for which stellar EUV spectra are observable only for nearby stars in the limited range of wavelength \citep[$\le 360 \ \AA$, ][]{Ribas_2005_ApJ,Johnstone_2021_AA}.
One thus needs to indirectly estimate or reconstruct the EUV spectrum of a star from other observables.
The proposed reconstruction methods include the inversion of the differential emission measure from UV and/or X-ray observations \citep{Sanz-Forcada_2011_AA,Duvvuri_2021_arXiv}, the empirical correlation between other observable lines and XUV emission \citep{Linsky_2014_ApJ,Youngblood_2017_ApJ,France_2018_ApJ,Sreejith_2020_AA}, 
and/or a combination of them \citep{Diamond-Lowe_2021_arXiv}.
However, the empirical estimations of stellar EUV emission are based on observations with large uncertainty and a limited number of samples, and therefore require further validation from different perspectives.

To circumvent the intrinsic difficulty of stellar EUV observation, this study uses the solar-stellar connection to theoretically estimate the stellar XUV emission.
The solar-stellar theoretical connection is a natural strategy to predict the stellar properties.
For example, by extending the solar coronal theory, \citet{Shibata_2002_ApJ} modelled the X-ray characteristics of the stellar coronae,
which was later extended by \citet{Takasao_2020_ApJ} considering the size distribution of active regions.
In this study, by extending the solar atmospheric model, the structure of the upper stellar atmosphere and the XUV emission are predicted.
To this end, the coronal heating problem must be explicitly solved.

To solve the coronal heating problem, the following three issues must be addressed \citep{Klimchuk_2006_SolPhys}.
\begin{enumerate}
    \item Energy generation and transfer: the source of the coronal thermal energy probably comes from the magneto-convection on the surface \citep{Steiner_1998_ApJ}.
    The energy generation and transfer to the corona, possibly in the form of Alfv\'en waves \citep{Alfven_1947_MNRAS,Osterbrock_1961_ApJ,Kudoh_1999_ApJ,De_Pontieu_2007_Science,McIntosh_2011_Nature,Srivastava_2017_ScientificReports}, needs to be solved. \\[-1.0em]
    \item Energy dissipation in the corona: the magnetic (and kinetic) energy in the corona needs to dissipate to sustain the high-temperature corona.
    The field-braiding process must be considered as a promising mechanism of magnetic-energy dissipation \citep{Parker_1972_ApJ,Parker_1988_ApJ}. \\[-1.0em]
    \item Thermal response to the heating: the density and temperature of a coronal loop are determined by the energy balance among heating, conduction and radiation.
    The Rosner-Tucker-Vaiana (RTV) scaling law \citep{Rosner_1978_ApJ} originates from the thermal response to coronal heating. 
    Thus, the RTV scaling law or its generalised form \citep{Serio_1981_ApJ,Zhuleku_2020_AA} should be reproduced by the model \citep{Antolin_2010_ApJ}.
\end{enumerate}
For these issues, a model of the coronal heating should 1. include the photosphere (stellar surface) and chromosphere, 2. appropriately consider the field-braiding process in the corona, and 3. implement the thermal conduction and radiative cooling. 
Classically, numerical models of a corona have often focused on the thermal responses to heating events by one-dimensional (1D) (expanding) flux-tube models \citep{Antiochos_1978_ApJ,Peres_1982_ApJ,Antiochos_1999_ApJ}.
A zero-dimensional model of the coronal thermal evolution is also proposed \citep{Klimchuk_2008_ApJ}.
The advancement of numerical techniques and increase in computational power have made models highly sophisticated.
Several models deal with the realistic energy generation by explicitly solving the magneto-convection \citep{Hansteen_2015_ApJ,Rempel_2017_ApJ}.
Other models have focused more on energy dissipation with simpler numerical settings \citep{Moriyasu_2004_ApJ,Rappazzo_2008_ApJ,van_Ballegooijen_2011_ApJ,Dahlburg_2016_ApJ}.
These models have predicted that the signature of coronal heating could be explained by convection-driven energy injection.
By extending these studies, we aimed to construct a stellar coronal model that would satisfy the three requirements.

In deriving the X-ray and EUV spectra from simulations,
care needs to be taken in the spatial resolution at the transition region (the temperature-jump region between the chromosphere and the corona).
The numerical resolution around the transition region is found to significantly affect the coronal density \citep{Bradshaw_2013_ApJ}, and the coronal emission measure distribution (EMD). 
Because the coronal emission is proportional to the EMD, 
it means that the XUV emission predicted by simulation significantly depends on the numerical resolution.
The required resolution is in the order of ${\rm km}$ or less, which is impractical in realistic three-dimensional (3D) simulations.
A previous study attempted to solve this ``transition-region problem'' by introducing the artificial broadening of the transition region by tuning the magnitude of radiative cooling and thermal conduction \citep{Johnston_2017_AA,Johnston_2019_ApJ,Iijima_2021_arXiv,Johnston_2021_arXiv}.
However, this treatment may yield an unrealistic EMD in the transition-region temperature, and therefore is inappropriate for the calculation of the XUV emission that includes a significant contribution from the transition region.

\begin{table}[t!]
\centering
  \begin{tabular}{p{5.5em} p{16em}}
    \begin{tabular}{l} \hspace{-1em} notation \end{tabular} 
    & \begin{tabular}{l} \hspace{-1em} meaning \end{tabular}
    \rule[-5.5pt]{0pt}{20pt} \\ \hline \hline
    $r$
    & radial distance from the stellar centre
    \rule[-5.5pt]{0pt}{20pt} \\
    $s$
    & coordinate along the flux-tube axis
    \rule[-5.5pt]{0pt}{20pt} \\
    $G$
    & gravitational constant
    \rule[-5.5pt]{0pt}{20pt} \\
    $k_B$
    & Boltzmann constant
    \rule[-5.5pt]{0pt}{20pt} \\ 
    $m_H$
    & hydrogen mass
    \rule[-5.5pt]{0pt}{20pt} \\ 
    $m_e$
    & electron mass
    \rule[-5.5pt]{0pt}{20pt} \\ 
    $h$
    & Planck constant
    \rule[-5.5pt]{0pt}{20pt} \\ 
    $M_\odot$
    & solar mass
    \rule[-5.5pt]{0pt}{20pt} \\
    $R_\odot$
    & solar radius
    \rule[-5.5pt]{0pt}{20pt} \\
    $T_\odot$
    & solar effective temperature
    \rule[-5.5pt]{0pt}{20pt} \\
    $\Omega_\odot$
    & solar angular rotation rate
    \rule[-5.5pt]{0pt}{20pt} \\ \hline
  \end{tabular}
  \vspace{1.0em}
  \caption{Notations of the coordinates ($r$ and $s$) and constant parameters.
  Note that subscript $\odot$ denotes the solar fundamental parameter.}
  \vspace{0em}
  \label{table:notation}
\end{table}

Considering the difficulty of the TR problem, we perform a series of 1D, high-resolution magnetohydrodynamic (MHD) simulations for a wide range of parameters.
This 1D model facilitates a coronal loop simulation with sufficiently high numerical resolution.
The field-braiding process (or turbulence), which is essentially 3D, must be appropriately modelled when solving the coronal heating problem by 1D simulation.
In this study, an approximated formulation of turbulent dissipation, developed in previous studies \citep{Dmitruk_2002_ApJ,Shoda_2018_ApJ_a_self-consistent_model}, is employed as it is likely to reproduce the average heating rate of the coronal loop \citep{van_Ballegooijen_2011_ApJ}.
For simplicity, we focus on the Sun-like stars that exhibit solar mass, luminosity, radius, and metallicity.

The rest of the manuscript is structured as follows.
In Section \ref{sec:model}, the coronal model and numerical methods are detailed.
Section \ref{sec:simulation_result} produces the numerical results, focusing on the dependencies of coronal properties on the loop length and magnetic filling factor that are likely to vary with the star \citep{Reale_1998_AA,Reale_2004_AA,Reiners_2009_ApJ,See_2019_ApJ}.
The XUV spectra obtained from the simulations are also presented. 
In Section \ref{sec:analytical_model}, analytical arguments on the energy flux injected into the corona are presented.
In Section \ref{sec:discussion}, the interpretations and limitations of the proposed model are discussed.
Section \ref{sec:conclusion} summarises the study.
Several details are provided in the Appendix, including the resolution dependence of the model (see Appendix \ref{app:transition_region_problem}).

\section{Model} \label{sec:model}

\subsection{Model overview and notation}

As mentioned earlier, the aim of this study is to model the XUV (X-ray+EUV) emissions from the Sun and Sun-like stars (including young Sun) with a range of magnetic activity level.
To this end, the luminosity, mass, radius, and metallicity of the stars are fixed to the solar value, i.e.,
\begin{align}
    L = L_\odot, \ \ \ M = M_\odot, \ \ \ R = R_\odot, \ \ \ Z = Z_\odot.
\end{align}
\revise{Dependence on metallicity is considered in the radiative loss function (Section \ref{sec:radiation}) and spectrum calculation (Section \ref{sec:result_fiducial}).}
We study the dependence of coronal properties on the coronal loop length and the filling factor of magnetic fields.
Because the filling factor tends to increase with the stellar rotation rate \citep[or the Rossby number, ][]{Saar_2001_proceedings,Reiners_2009_ApJ}, 
the coronal dependence on the stellar rotation rate is implicitly investigated.

We model a single coronal loop rooted in the stellar surface, and self-consistently solve the energetics and dynamics inside the loop.
In other words, we solve the time-dependent, 1D MHD equations for an expanding coronal loop.
The differential emission measure (DEM) of a single coronal loop is directly obtained from the simulation, which is then converted to the XUV spectrum by prescribing the chemical composition and integrating the continuum and line emissions as a function of wavelength using the CHIANTI atomic database \revise{version 10.0} \citep{Dere_1997_AA,Del_Zanna_2021_ApJ}.

Notations of the coordinates and constant parameters used in this study are listed in Table \ref{table:notation}.
$X_\odot$ denotes the solar value and $X_\ast$ the value of $X$ measured at the photosphere.

\begin{figure}[t!]
\centering
\includegraphics[width=85mm]{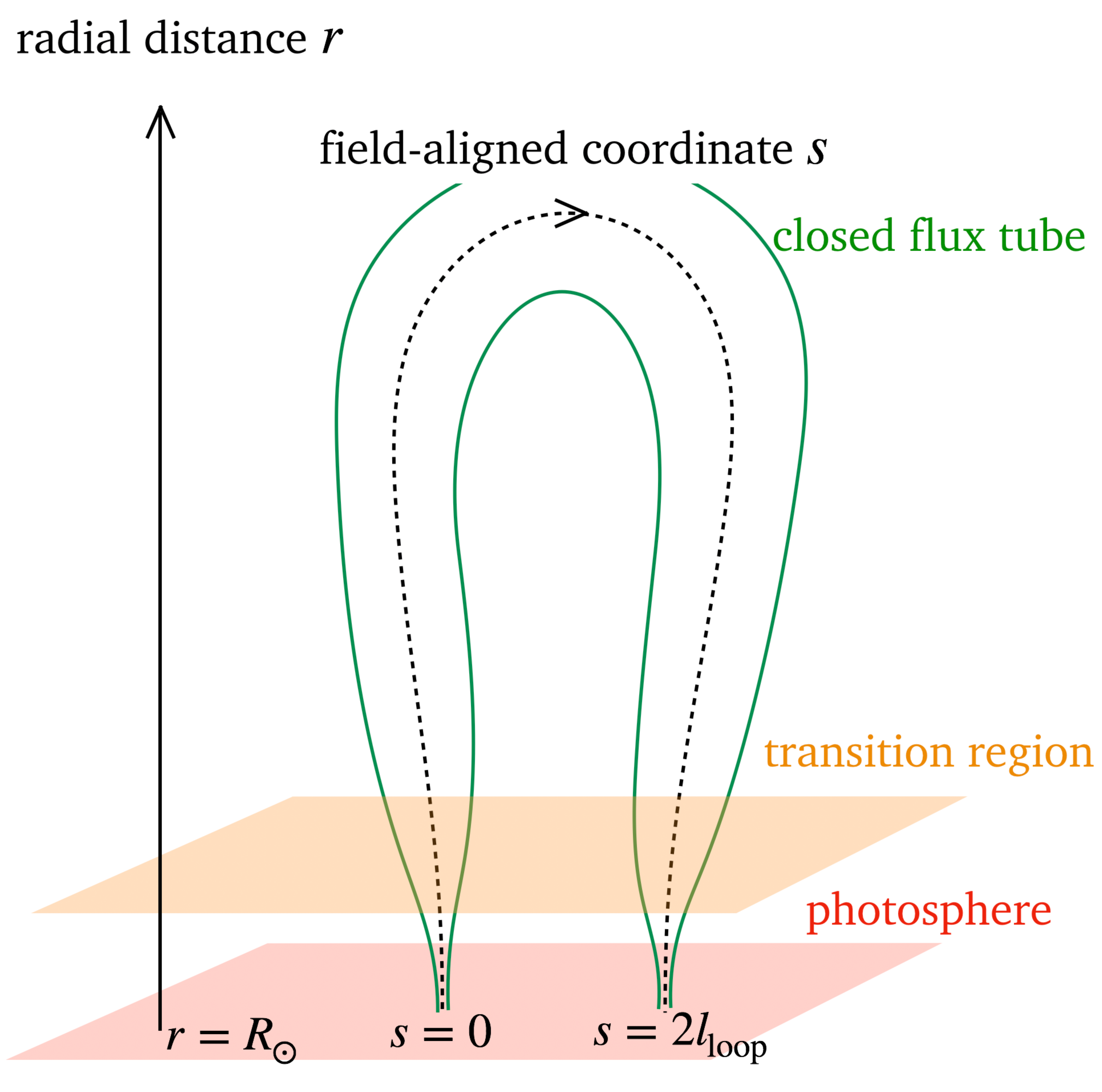}
\vspace{0.5em}
\caption{
A schematic picture of the system.
One-dimensional dynamics along the axis of the closed flux tube is simulated. 
The axis is indicated by the dotted line, while the flux tube surface is denoted by green lines.
The flux tube is intended to be nearly vertical and super-radially expanding.
The geometry of the loop is defined by the filling factor $f$ and radial distance $r$ as functions of field-aligned distance $s$.
}
\label{fig:loop_structure}
\vspace{0em}
\end{figure}

\subsection{Model of the closed flux tube}

A single coronal loop is modelled by a 1D expanding flux tube rooted in the photosphere.
Figure \ref{fig:loop_structure} illustrates a schematic of the model.
As shown in Table~\ref{table:notation}, the coordinate along the axis of the flux tube is denoted by $s$.
The spatial variation only along the loop is assumed to be nonzero, that is, $\partial/\partial s \neq 0$.
Similar 1D flux tube models were used in the models of solar spicules \citep{Hollweg_1982_SolPhys,Kudoh_1999_ApJ,Matsumoto_2010_ApJ}, solar and stellar coronal heating \citep{Moriyasu_2004_ApJ,Washinoue_2019_ApJ}, and solar wind acceleration \citep{Suzuki_2005_ApJ,Shoda_2018_ApJ_a_self-consistent_model}.

Hereinafter, the two perpendicular directions shall be denoted by $x$ and $y$. Thus, the local $xy$ plane is perpendicular to the axis of the loop.
The flux-tube expansion is incorporated through the scale factors in the $x$ and $y$ directions: $h_{x,y}$.
For simplicity, the loop is assumed to expand isotropically in the perpendicular directions.
In terms of scale factors, the isotropic expansion is represented by
\begin{align}
    h_x = h_y \propto \sqrt{A(s)}, \label{eq:scale_factors}
\end{align}
where $A(s)$ is the cross section of the coronal loop.
As $h_s =1$ by definition, Eq. (\ref{eq:scale_factors}) results in
\begin{align}
    \nabla \cdot \vec{X} &= \frac{1}{A(s)} \frac{\partial}{\partial s} \left( X_s A(s) \right), \nonumber \\
    \nabla \times \vec{X} &= \frac{1}{\sqrt{A(s)}} \frac{\partial}{\partial s} \left( X_x \sqrt{A(s)} \right) \vec{e}_y \label{eq:nabla_operations} \\
    &- \frac{1}{\sqrt{A(s)}} \frac{\partial}{\partial s} \left( X_y \sqrt{A(s)} \right) \vec{e}_x \nonumber
\end{align}
for any vector field $\vec{X}$,
where $\vec{e}_{x,y}$ represent the unit vectors in the $x,y$ directions.
The 1D spherical coordinate system is reproduced as a special case of $A(s) = s^2$.

The flux tube expands in the chromosphere as a response to the exponential decrease in the ambient gas pressure \citep{Cranmer_2005_ApJ,Ishikawa_2021_arXiv}.
As a result of this expansion, the filling factor of the magnetic field (flux tube) $f$ should increase nearly exponentially with altitude.
Under this assumption, we model the filling factor $f$ as
\begin{align}
    f = \min \left[1,f_\ast \exp \left( \frac{r-R_\odot}{H_{\rm mag}} \right) \right],
    \ \ \ \ H_{\rm mag} = c_{\rm mag} H_\ast , \label{eq:filling_factor_formulation}
\end{align}
where $f_\ast$ is the magnetic filling factor on the photosphere and
\begin{align}
    H_\ast = \frac{k_B T_\odot R_\odot^2}{GM_\odot m_H} \label{eq:scale_height_photosphere}
\end{align}
is the pressure scale height at the photosphere.
By this formulation, we assume that the loop expands only in the chromosphere and exhibits a uniform cross section in the corona,
which is supported by \sout{the solar observation} \revise{some solar observations} \citep[][\revise{but see a recent discussion by} \citet{Malanushenko_2021_arXiv}]{Klimchuk_1992_PASJ}.
Given that the pressure scale height is uniform from the photosphere up to the chromosphere, $c_{\rm mag} = 2$ yields a flux-tube expansion with a constant plasma beta in altitude.
In this work, we set $c_{\rm mag} = 2.5$ that realizes a slightly low-beta chromosphere.
We have confirmed that the choice of $c_{\rm mag}$ does not have a significant influence over the simulation results.

When the coronal loop extends to a region far above the surface, the flux tube also undergoes the radial expansion $\propto r^2$, where $r$ is the radial distance from the stellar centre.
Considering the chromospheric and radial expansions, 
the cross section $A$ is expressed as
\begin{align}
    A \propto r^2 f. \label{eq:cross_section}
\end{align}

\begin{figure}[t!]
\centering
\includegraphics[width=80mm]{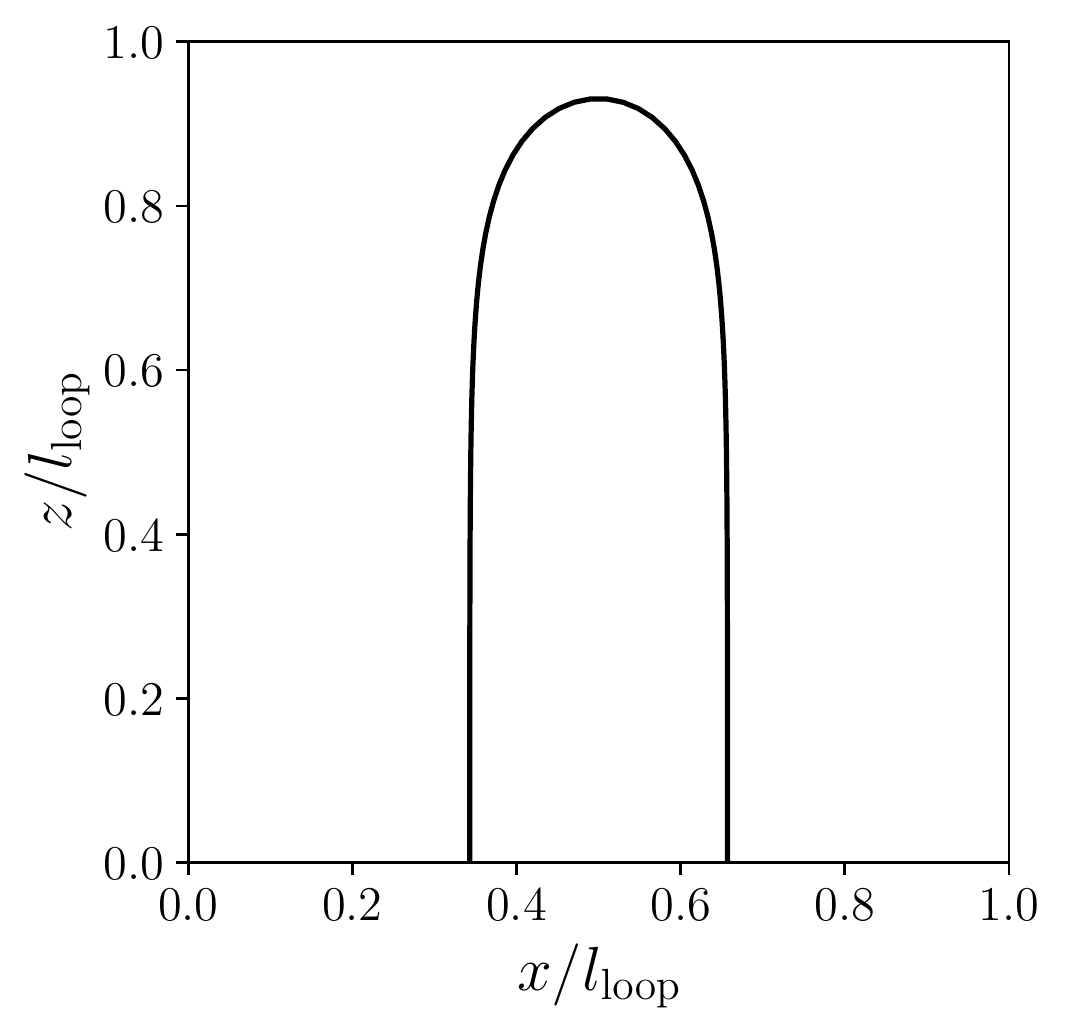}
\vspace{0.5em}
\caption{
\revise{Shape of the flux-tube axis defined by Eq. (\ref{eq:r_s_relation}). $x$ and $z$ denote the horizontal and vertical coordinates, repectively.}
}
\label{fig:flux_tube_shape}
\vspace{0em}
\end{figure}

We define the inclination of the flux tube by prescribing $r$ as a function of $s$.
We consider nearly vertical flux tubes, so that the vertical line of sight is nearly aligned with the axis of the flux tube (Figure \ref{fig:loop_structure}).
It is easier to calculate the DEM along the vertical line of sight for this structure.
In particular, we set
\begin{align}
    \frac{dr}{ds}  = \tanh \left[ \frac{10 \left( l_{\rm loop} -s \right)}{l_{\rm loop}} \right], \ \ \ \ \left. r \right|_{s=0} = R_\odot \label{eq:r_s_relation}
\end{align}
where $l_{\rm loop}$ is the half-loop length.
\revise{The actual shape of the flux-tube axis is displayed in Figure \ref{fig:flux_tube_shape}.}
Combining Eq.s (\ref{eq:filling_factor_formulation})--(\ref{eq:r_s_relation}),
the cross section $A(s)$ is well defined as a function of $s$.

\subsection{Basic equations}

The MHD equations with the equation of state of partially ionised hydrogen, gravity, thermal conduction, radiative cooling, and phenomenology of turbulent heating are selected as the basic equations of the model, which are expressed in the form of conservation law (for derivation, see Appendix \ref{app:basic_equation})
\begin{align}
    \frac{\partial}{\partial t} \vec{U} + \frac{1}{r^2 f} \frac{\partial}{\partial s} \left( \vec{F} r^2 f \right) = \vec{S}.
    \label{eq:basic_equation_conservation_form}
\end{align}
The conserved variables $\vec{U}$ and the corresponding fluxes $\vec{F}$ are given by
\begin{align}
    \vec{U} =
    \left(
    \begin{array}{c}
    \rho \\
    \rho v_s \\
    \rho v_x \\
    \rho v_y \\
    B_x \\
    B_y \\
    e
    \end{array}
    \right), \hspace{1em}
    \vec{F} =
    \left(
    \begin{array}{c}
    \rho v_s \\
    \rho v_s^2 + p_T \\
    \rho v_s v_x - B_s B_x / (4\pi) \\
    \rho v_s v_y - B_s B_y / (4\pi) \\
    v_s B_x - v_x B_s \\
    v_s B_y - v_y B_s \\
    \left( e + p_T \right) v_s - B_s \left(\vec{v}_\perp \cdot \vec{B}_\perp \right)/(4\pi)
    \end{array}
    \right), \label{eq:conserved_variable_flux}
\end{align}
where
\begin{align}
    &\vec{v}_\perp = v_x \vec{e}_x + v_y \vec{e}_y, \hspace{2em}
    \vec{B}_\perp = B_x \vec{e}_x + B_y \vec{e}_y, \\
    &p_T = p + \frac{\vec{B}_\perp^2}{8\pi}, \hspace{2em} e = e_{\rm int} + \frac{1}{2} \rho \vec{v}^2 + \frac{\vec{B}_\perp^2}{8\pi}.
\end{align}
$e_{\rm int}$ denotes the internal energy per unit volume and is defined in Section \ref{sec:eos}.
The source term $\vec{S}$ is given by
\begin{align}
    \vec{S}
    =
    \left(
    \begin{array}{c}
    0 \\
    \left( p +\dfrac{1}{2} \rho \vec{v}_\perp^2 \right)/L - \rho \dfrac{GM_\odot}{r^2} \dfrac{dr}{ds}  \\
    \dfrac{1}{2L} \left( - \rho v_s v_x + \dfrac{B_s B_x}{4\pi} \right)  + \rho D^v_x \\
    \dfrac{1}{2L} \left( - \rho v_s v_y + \dfrac{B_s B_y}{4\pi} \right)  + \rho D^v_y \\
    \dfrac{1}{2L} \left( v_s B_x - v_x B_s \right)  + \sqrt{4 \pi \rho} D^b_x\\
    \dfrac{1}{2L} \left( v_s B_y - v_y B_s \right)  + \sqrt{4 \pi \rho} D^b_y\\
    - \rho v_s \dfrac{GM_\ast}{r^2} \dfrac{dr}{ds} + Q_{\rm cnd} + Q_{\rm rad}
    \end{array} 
    \right), \label{eq:source_non_rotating_model}
\end{align}
where $L^{-1} = d/ds \ln \left( r^2 f \right)$ denotes the length scale of the flux-tube expansion.
The conduction term $Q_{\rm cnd}$ is defined in terms of conductive flux $q_{\rm cnd}$ as
\begin{align}
    Q_{\rm cnd} = - \frac{1}{r^2 f} \frac{\partial}{\partial s} \left( q_{\rm cnd} r^2 f \right).
\end{align}
Because the mean free path of an electron is generally smaller than the system size, the Spitzer--H\"arm flux \citep{Spitzer_1953_PhysRev} is applied to $q_{\rm cnd}$:
\begin{align}
    q_{\rm cnd} = - \frac{\left| B_s \right|}{\left| \vec{B} \right|} \kappa_{\rm SH} T^{5/2} \frac{\partial T}{\partial s},
\end{align}
where $\kappa_{\rm SH} = 10^{-6} {\rm \ erg \ cm^{-1} \ s^{-1} \ K^{-7/2}}$.
The radiative cooling $Q_{\rm rad}$ and turbulent dissipation $D_{x,y}^{v,b}$ are described in Section \ref{sec:radiation} and \ref{sec:alfven_wave_turbulence}, respectively.
\revise{Because turbulence is considered not as an external force but as a dissipative process, the losses of the kinetic and magnetic energies by turbulence are locally balanced by the gain in the internal energy. 
Thus, the presence of the turbulence terms does not affect the conservation of the total energy.}
\revise{For the same reason, we do not explicitly consider the numerical dissipation of velocity and magnetic field in the energy equation.}
\revise{We note that the numerical dissipation is unlikely to be the dominant heating mechanism because the coronal Alfv\'en wave, which has a typical wavelength of $\sim 100 {\rm \ Mm}$, is resolved by a sufficiently fine grid in the corona ($100 {\rm \ km}$).}

\subsection{Equation of state
\label{sec:eos}}
We assume that the plasma consists of neutral hydrogen atoms, protons, and electrons.
The internal energy per unit volume $e_{\rm int}$ is composed of the conventional thermal energy $p/(\gamma-1)$ and the latent heat of the ionised gas.
\begin{align}
    e_{\rm int} = \frac{p}{\gamma-1} +  n_H \chi I_H, \ \ \ \ n_H = \rho/m_H,
\end{align}
where $\chi$ is the ionisation degree and $n_H$ is the number density of hydrogen atoms (proton + neutral hydrogen).
$I_H$ is the ionisation energy of the hydrogen atom ($I_H = 13.6 {\rm \ eV}$).
We assume that the ionisation degree could be determined from the \revise{approximated version of the} Saha-Boltzmann equation, \sout{that holds in the thermal equilibrium} \revise{in which only the ground state is considered as the bound state (low-temperature limit)}.
\begin{align}
    \frac{\chi^2}{1-\chi} = \frac{2}{n_H \lambda_e^3} \exp \left( - \frac{I_H}{k_B T} \right),
    \label{eq:saha_boltzmann_ionization}
\end{align}
where $\lambda_e$ is the thermal de Broglie wavelength of electron.
\begin{align}
    \lambda_e = \sqrt{\frac{h^2}{2 \pi m_e k_B T}}.
\end{align}
\revise{When chromospheric hydrogen is no longer in thermal equilibrium, the ionisation degree will deviate from the Saha--Boltzmann value \citep{Goodman_2012_ApJ}, which is beyond the scope of this study.}
Once the ionisation degree $\chi$ is obtained, the pressure and temperature are related by
\begin{align}
    p = \left( n_e + n_H \right) k_B T = \left( 1+\chi \right) n_H k_B T.
\end{align}

\subsection{Radiation \label{sec:radiation}}

\begin{figure}[t]
\centering
\includegraphics[width=80mm]{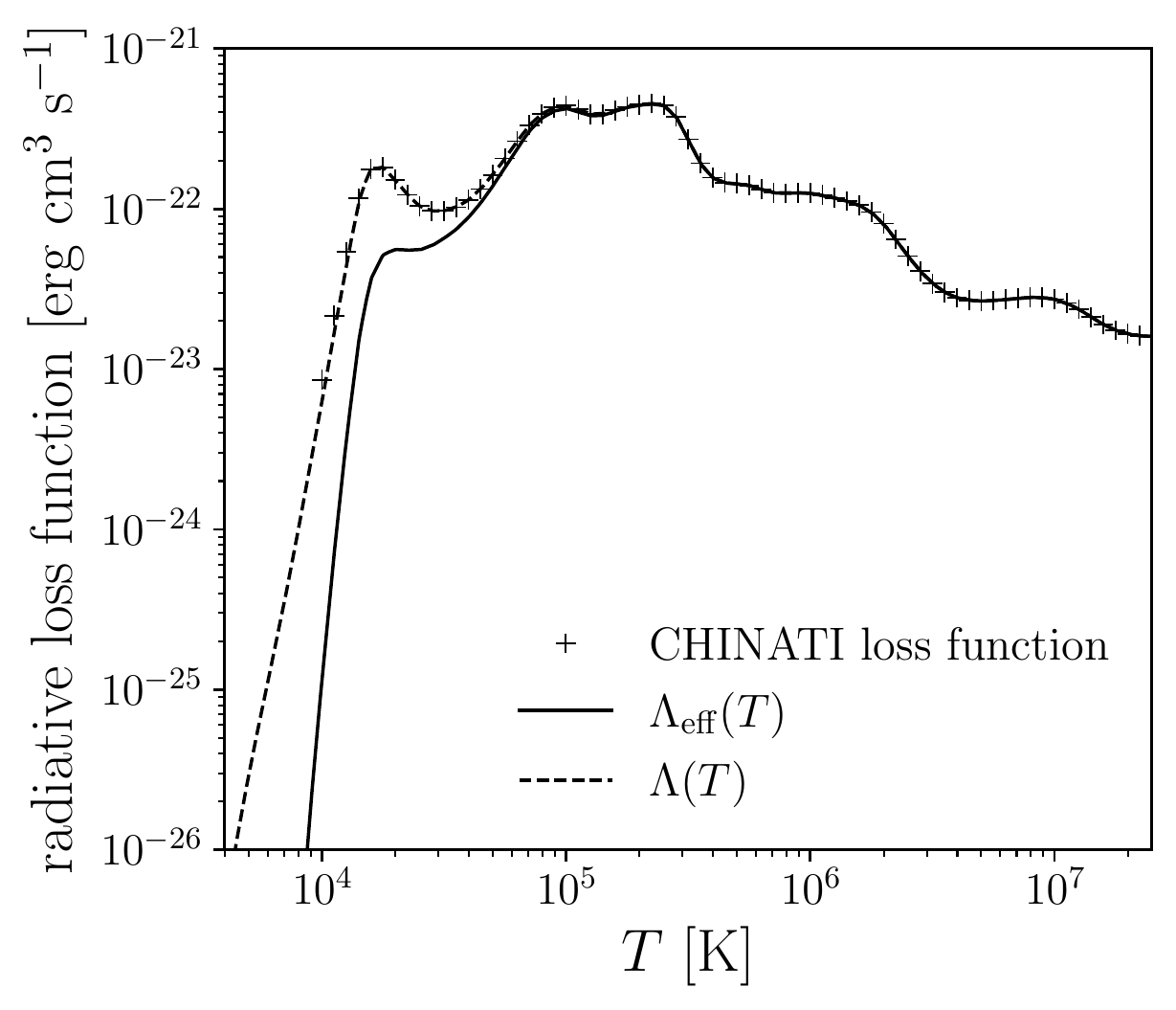}
\caption{
\sout{Black solid line shows the effective optically-thin radiative loss function $\Lambda_{\rm eff} (T)$ implemented in this study.}
\sout{Also shown by diamonds are the loss function by Goodman $\&$ Judge (2012).}
\revise{Solid and dashed lines show the effective and original optically-thin radiative loss functions $\Lambda_{\rm eff} (T)$ and $\Lambda (T)$, respectively.
Also shown by crosses are the loss function from the CHIANTI atomic database with photospheric abundance.}
}
\label{fig:effective_radiative_loss}
\vspace{0em}
\end{figure}

The radiative cooling rate per unit volume $Q_{\rm rad}$ is given by
\begin{align}
    Q_{\rm rad} &= \xi_{\rm rad} Q_{\rm rad}^{\rm thck} + \left( 1 - \xi_{\rm rad} \right) Q_{\rm rad}^{\rm thin}, \\
    \xi_{\rm rad} &= 1-  \exp \left( - \frac{p}{p_{\rm rad}}\right), \ \ \ \ p_{\rm rad} / p_\ast = 0.1,
\end{align}
where $p_\ast$ is the pressure at the surface (photosphere) and $Q_{\rm rad}^{\rm thck}$ and $Q_{\rm rad}^{\rm thin}$ approximate the optically thick and thin cooling rates, respectively.
The optically thick and thin functions are seamlessly connected via $\xi_{\rm rad}$.
Instead of solving the radiative transfer, 
we model $Q_{\rm rad}^{\rm thck}$ and $Q_{\rm rad}^{\rm thin}$ as follows. 

The radiative heating and cooling are approximately balanced near the photosphere to maintain a nearly constant surface temperature.
The optically thick radiative loss is approximated by an exponential cooling function that forces the local temperature to approach the reference value \citep{Gudiksen_2005_ApJ}:
\begin{align}
    Q_{\rm rad}^{\rm thck} = \frac{1}{\tau} \left( e_{\rm int}^{\rm ref} - e_{\rm int} \right),
    \ \ \ \ \tau = 0.1 {\rm \ s} \left( \frac{\rho}{\rho_\ast} \right)^{-1/2},
\end{align}
where $\rho_\ast$ is the photospheric mass density and $e_{\rm int}^{\rm ref}$ is the reference internal energy density corresponding to the reference temperature $T^{\rm ref}$.
We simply assume $T^{\rm ref} = T_\odot$, i.e., 
the stellar atmosphere exhibits isothermal behaviour in the absence of other cooling/heating mechanisms.

The optically thin cooling is expressed in terms of the radiative loss function $\Lambda (T)$ by $n_e n_H \Lambda (T)$.
In this work, we define the loss function over a wide range of temperature ($10^3 {\rm \ K} \le T \le 10^7 {\rm \ K}$) as follows.
\begin{enumerate}
    \item For simplicity, in the high-temperature range ($T \ge 1.5 \times 10^4 {\rm \ K}$), the cooling rate is referred from the CHIANTI atomic database with the photospheric abundance (no first ionisation potential (FIP) effect).
    \item The cooling rate in the low-temperature range ($T \le 1.0 \times 10^4 {\rm \ K}$) is deduced by \citet{Goodman_2012_ApJ}, which partially consider the non-LTE effect.
    \item In the intermediate-temperature range ($1.0 \times 10^4 {\rm \ K} < T < 1.5 \times 10^4 {\rm \ K}$), a bridging law between two loss functions asre employed following the method of \citet{Iijima_2016_PhD}.
\end{enumerate}
The chromospheric heating effect by backward coronal radiation is still missing in $\Lambda (T)$ defined above.
To introduce this effect, we quench $\Lambda (T)$ in the chromospheric temperature range, which gives $Q_{\rm rad}^{\rm thin}$
\begin{align}
    Q_{\rm rad}^{\rm thin} = n_e n_H \Lambda_{\rm eff} (T), \ \ \ \  \Lambda_{\rm eff} (T) = \Lambda (T) \exp \left( -\frac{T_{\rm chr}^2}{T^2} \right),
\end{align}
where $T_{\rm chr} = 2.0 \times 10^4 {\rm \ K}$.
The effective radiative loss function $\Lambda_{\rm eff} (T) $ is displayed in Figure \ref{fig:effective_radiative_loss}, along with the 
\sout{chromospheric loss function by Goodman $\&$ Judge (2012)}
\revise{original radiative loss function $\Lambda (T)$ and the CHIANTI loss function defined in $T \ge 10^4 {\rm \ K}$.}

\subsection{Phenomenological model of coronal turbulence
\label{sec:alfven_wave_turbulence}}

Although the mechanism of the coronal heating is debated,
it is of no doubt that magnetic field feeds heat to the corona that maintains the million-Kelvin temperature by dissipation.
The formation of tangential discontinuities (electric current sheets) in response to the continuous shuffling of the foot points of coronal magnetic fields is a plausible mechanism of magnetic-field dissipation \citep{Parker_1972_ApJ,Sturrock_1981_ApJ,Parker_1983_ApJ,van_Ballegooijen_1986_ApJ,Galsgaard_1996_JGR}.
The ubiquitous current-sheet formation should lead to small-scale impulsive energy release, which is likely to feed a sufficient amount of energy to the corona, possibly in the form of micro- and nano-flares \citep{Parker_1988_ApJ,Shimizu_1995_PASJ,Aschwanden_2002_ApJ}.

The formation of current sheets can be interpreted as turbulent cascading \citep{Rappazzo_2007_ApJ,Rappazzo_2008_ApJ,Verdini_2012_AA}.
Because the coronal loop is threaded by a strong mean magnetic field, the MHD turbulence evolving in the coronal loop can be accurately approximated by the reduced-MHD turbulence, in which the energy cascades preferentially in the perpendicular direction \citep{Shebalin_1983_JPP,Cho_2000_ApJ,Cho_2003_MNRAS}.
The energy-cascading (or heating) rate of the reduced-MHD turbulence, $Q_{\rm heat}$, is precisely approximated by the mean-field quantities as follows \citep{Hossain_1995_PhysFluids,Matthaeus_1999_ApJ,Dmitruk_2002_ApJ,Verdini_2007_ApJ}
\begin{align}
    Q_{\rm heat} \approx c_d \rho \frac{z_\perp^+ {z_\perp^-}^2 + z_\perp^- {z_\perp^+}^2}{4 \lambda_\perp}, \label{eq:phenomenology_heating}
\end{align}
where $z_\perp^\pm$ denotes the (root-mean-squared (RMS)) amplitude of the perpendicular Els\"asser variables and
$\lambda_\perp$ is the correlation length of the Els\"asser variable (Alfv\'en wave) perpendicular to the mean field.
$c_d$ is a dimensionless parameter.
By this approximation, we estimate the averaged heating rate using the coronal turbulence (field braiding).

The approximated heating rate in Eq. (\ref{eq:phenomenology_heating}) is implemented by adding the source terms $D_{x,y}^v$, $D_{x,y}^b$ given by \citet{Shoda_2018_ApJ_a_self-consistent_model}.
\begin{align}
    D^v_{x,y} &= - \frac{c_d}{4\lambda_\perp} \left( \left| z_{x,y}^+ \right| z_{x,y}^- + \left| z_{x,y}^- \right| z_{x,y}^+  \right), \label{eq:phenomenological_awt_vsource} \\ 
    D^b_{x,y} &= - \frac{c_d}{4\lambda_\perp} \left( \left| z_{x,y}^+ \right| z_{x,y}^- - \left| z_{x,y}^- \right| z_{x,y}^+  \right), \label{eq:phenomenological_awt_bsource}
\end{align}
where $z_{x,y}^\pm = v_{x,y} \mp B_{x,y}/\sqrt{4 \pi \rho}$.
The role of these terms is explained below.
Without the conservation part $\propto \partial \left( r^2 f \vec{F} \right) / \partial s$, 
the perpendicular components of the equation of motion and induction equation are expressed as
\begin{align}
    \frac{\partial}{\partial t} \left( \rho v_{x,y} \right) = \rho D^v_{x,y}, \ \ \ \ \frac{\partial}{\partial t} B_{x,y} = \sqrt{4 \pi \rho} D^b_{x,y}, \label{eq:phenomenological_awt_formulation}
\end{align}
In the limit of the reduced-MHD approximation (time-independent density, $\partial \rho / \partial t =0$), Eqs. (\ref{eq:phenomenological_awt_vsource}), (\ref{eq:phenomenological_awt_bsource}), and (\ref{eq:phenomenological_awt_formulation}) are reduced to
\begin{align}
    \frac{\partial}{\partial t} z^\pm_{x,y} = - \frac{c_d}{2 \lambda_\perp} \left| z^\mp_{x,y} \right| z^\pm_{x,y},
\end{align}
which yields the energy conservation law of
\begin{align}
    \frac{\partial}{\partial t} e_\perp = - c_d \rho
    \sum_{i=x,y} \frac{\left| z_i^- \right| {z_i^+}^2  + \left| z_i^+ \right| {z_i^-}^2 }{4 \lambda_\perp}, \label{eq:heating_rate_our_formulation}
\end{align}
where $e_\perp$ is the sum of the kinetic and magnetic energies emerging from the fluctuations of the perpendicular components:
\begin{align}
    e_\perp = \frac{1}{4} \rho \left( {z_\perp^+}^2 + {z_\perp^-}^2  \right) = \frac{1}{2} \rho v_\perp^2 + \frac{B_\perp^2}{8 \pi}.
\end{align}
Comparing Eq.s (\ref{eq:phenomenology_heating}) and (\ref{eq:heating_rate_our_formulation}), 
one obtain
\begin{align}
     \frac{\partial}{\partial t} e_\perp \approx - Q_{\rm heat},
\end{align}
indicating that the energy dissipation by the reduced-MHD turbulence is considered appropriately.

The perpendicular correlation length is assumed to increase with the flux-tube radius, i.e.,
\begin{align}
    \lambda_\perp = \lambda_{\perp,\ast} \sqrt{\frac{A}{A_\ast}} = \lambda_{\perp,\ast} \frac{r}{R_\odot} \sqrt{\frac{f}{f_\ast}},
\end{align}
where the perpendicular correlation length at the photosphere is set equal to the typical width of the inter-granular lane: $\lambda_{\perp,\ast} = 150 {\rm \ km}$.
However, the best possible free parameter $c_d$ is still debated.
In this study, we infer $c_d=0.1$ from the previous studies of the solar-wind turbulence \citep{van_Ballegooijen_2017_ApJ,Chandran_2019_JPP,Verdini_2019_SolPhys}.

\subsection{Boundary condition and simulation setting}

Both boundaries of the simulation domain are located at the photosphere, and the photospheric temperature is fixed to the effective temperature, i.e.,
\begin{align}
    T_\ast = T_\odot \revise{ = 5.77 \times 10^3 {\rm \ K}}. 
\end{align}
The photospheric magnetic field is known to form localised kilo-Gauss patches \citep{Spruit_1979_SolPhys,Tsuneta_2008_ApJ}. 
These patches are likely to be in thermal equipartition, which equates the gas and magnetic pressures.
For the non-magnetised photosphere, the thermal equipartition field is given by
\begin{align}
    B_{\rm eq} = 1.34 \times 10^3 {\rm \ G}.
\end{align}
In the magnetised photosphere, as the deeper region tends to be observed \citep[e.g.][]{Keller_2004_ApJ}, the ambient gas is likely to exhibit larger pressure than the non-magnetised photosphere.
Thus, the equipartition magnetic field should be larger than this equipartition value.
Therefore, we set the photospheric axial magnetic field equal to
\begin{align}
    B_{s,\ast} = 1.5 B_{\rm eq} = 2.01 \times 10^3 {\rm \ G}.
\end{align}

We model the energy injection from the photosphere by imposing the velocity and magnetic-field fluctuations.
\revise{Fluctuations are imposed at both ends of the simulation domain.}
The vertical and horizontal velocity fluctuations are modelled separately.
The upward acoustic waves are excited on the photosphere by employing the time-dependent boundary conditions on the density and axial velocity.
\begin{align}
    &\rho_\ast = \overline{\rho}_\ast \left( 1 + \frac{v_{s,\ast}}{a_\ast} \right), \\
    &v_{s,\ast} =  \int_{\omega^l_{\rm min}}^{\omega^l_{\rm max}} d\omega \ \tilde{v}_s \left( \omega \right) \sin \left[ \omega t + \psi^l (\omega) \right], \label{eq:vs_ast}
\end{align}
where $\overline{\rho}_\ast$ is the time-averaged photospheric density, $a_\ast = \sqrt{k_B T_\ast / m_H}$ is the isothermal speed of sound on the photosphere, and $\psi^l \left( \omega \right)$ is a random phase function that ranges between $0$ and $2\pi$.
\revise{In the numerical implementation of the integral in Eq. (\ref{eq:vs_ast}), the frequency range is evenly divided into 21 bins and the corresponding 21 components are summed.}
The time-averaged photospheric density is given by equipartition on the photosphere:
\begin{align}
    \overline{\rho}_\ast k_B T_\ast / m_H= \frac{B_{s,\ast}^2}{8 \pi},
\end{align}
which yields
\begin{align}
    \overline{\rho}_\ast = 4.22 \times 10^{-7} {\rm \ g \ cm^{-3}}.
\end{align}
$\overline{\rho}_\ast$ is larger than the typical mass density on the solar surface because the magnetised photosphere should be deeper and denser.
The (time-averaged) Alfv\'en speed $v_A$ on the photosphere is then expressed as
\begin{align}
    \overline{v_{A,\ast}} \approx \frac{B_{s,\ast}}{\sqrt{4 \pi \overline{\rho_\ast}}} = 8.73 \ {\rm km \ s^{-1}}.
\end{align}
We \revise{arbitrarily} set \revise{$\tilde{v}_s \left( \omega \right) \propto \omega^{-1/2}$} with
\begin{align}
    2 \pi / \omega^l_{\rm min} = 300 {\rm \ s}, \ \ \ \ 
    2 \pi / \omega^l_{\rm max} = 100 {\rm \ s}.
\end{align}
The minimum frequency corresponds to the cut-off frequency of the acoustic wave at the photosphere \citep[e.g.][]{Felipe_2018_AA}.
The magnitude of \revise{$\tilde{v}_s \left( \omega \right)$} is tuned such that the RMS amplitude of $v_{s,\ast}$ at the photosphere is $0.6 {\rm \ km \ s^{-1}}$.
\begin{align}
    \sqrt{ \overline{v_{s,\ast}^2} } = 0.6 {\rm \ km \ s^{-1}}, 
\end{align}
where the overline denotes the time average.
\revise{Although the longitudinal-wave excitation on the photosphere is explicitly considered, the effect of the longitudinal-wave input is insignificant; the coronal temperature decreases by only 2$\%$ when the longitudinal wave injection is terminated.}

The horizontal velocity and magnetic field at the bottom boundary are expressed in terms of the Els\"asser variables, which are defined as
\begin{align}
    z^\pm_{x,y} = v_{x,y} \mp \frac{B_{x,y}}{\sqrt{4 \pi \rho}}.
\end{align}
The free boundary condition is imposed on the downward Els\"asser variables.
\begin{align}
    \left. \frac{\partial}{\partial s} z^-_{x,y} \right|_\ast = 0.
\end{align}
The upward Els\"asser variable is assumed to be non-monochromatic with respect to the frequency.
\begin{align}
    z_{x,y,\ast}^+  = \int_{\omega^t_{\rm min}}^{\omega^t_{\rm max}} d \omega \  \tilde{z}^\pm_{x,y} \left( \omega \right) \sin \left[ \omega t + \psi^t(\omega) \right], \label{eq:zplus_ast}
\end{align}
where $\psi^t(\omega)$ is a random phase function that ranges between $0$ and $2\pi$.
\revise{As with Eq. (\ref{eq:vs_ast}), the frequency range is discretised into 21 bins when numerically implementing the integral in Eq. (\ref{eq:zplus_ast}).}
We set \revise{$\tilde{z}^\pm_{x,y} \left( \omega \right) \propto \omega^{-1/2}$} with
\begin{align}
    2 \pi / \omega^t_{\rm min} = 1000 {\rm \ s}, \ \ \ \ 
    2 \pi / \omega^t_{\rm max} = 100 {\rm \ s}.
\end{align}
\revise{$\tilde{z}^\pm_{x,y} \left( \omega \right) \propto \omega^{-1/2}$ corresponds to the $1/\omega$ energy spectrum discovered in previous solar simulations and observations \citep{Van_Kooten_2017_ApJ}}.
Given that the typical size of a granule is $1,000 {\rm \ km}$ and the typical speed of surface convection is $1 {\rm \ km \ s^{-1}}$ \citep{Chitta_2012_ApJ}, the maximum wave period corresponds to the turn-over time of granular motion.
Similarly, given that the typical size of an inter-granular lane is $100 {\rm \ km}$, the minimum wave period corresponds to the turn-over time of inter-granular motion.
The magnitude of \revise{$\tilde{z}^\pm_{x,y} \left( \omega \right)$} is tuned such that the root-mean-squared amplitude of $z_{x,y,\ast}^+$ at the photosphere is $1.2 {\rm \ km \ s^{-1}}$:
\begin{align}
    \sqrt{ \overline{{z_{x,\ast}^+}^2} } = \sqrt{ \overline{{z_{y,\ast}^+}^2} } = 1.2 {\rm \ km \ s^{-1}},
\end{align}
where the overline denotes the time average.
\revise{Although some solar observations have observed the suppression of convective velocity in large-filling-factor regions \citep[e. g.][]{Katsukawa_2005_ApJ}, we dismiss this effect for simplicity.}

By this formulation, the energy flux of the upward Alfv\'en wave at the footpoint of the flux tube is given by
\begin{align}
    F_{A,\ast} &= \frac{1}{4} \overline{\rho \left( z_{x,\ast}^2 + z_{y,\ast}^2 \right) v_{A,\ast}} \approx \frac{1}{4} \overline{\rho_\ast} \left( \overline{z_{x,\ast}^2} + \overline{z_{y,\ast}^2} \right) \overline{v_{A,\ast}} \nonumber \\
    & = 2.65 \times 10^9 \ {\rm \ erg \ cm^{-2} \ s^{-1}},
\end{align}
which is sufficiently larger than the energy flux required to sustain the solar corona \citep{Withbroe_1977_ARAA}.

\begin{table}[t!]
\centering
  \begin{tabular}{l l}
    \begin{tabular}{c} magnetic filling factor \\ 
    (photosphere) \end{tabular} 
    & \begin{tabular}{c} \hspace{2em} half-loop length \\
    \hspace{2em} $[10^3 {\rm \ km}]$ \end{tabular}
    \rule[-5.5pt]{0pt}{20pt} \\ \hline \hline
    $f_\ast$ = 1
    & $l_{\rm loop}$ = [20, 30, 40]
    \rule[-5.5pt]{0pt}{20pt} \\
    $f_\ast$ = 0.5
    & $l_{\rm loop}$ = [20, 30, 40]
    \rule[-5.5pt]{0pt}{20pt} \\
    $f_\ast$ = 0.333
    & $l_{\rm loop}$ = [20, 30, 40, 60]
    \rule[-5.5pt]{0pt}{20pt} \\
    $f_\ast$ = 0.2
    & $l_{\rm loop}$ = [20, 30, 40]
    \rule[-5.5pt]{0pt}{20pt} \\
    $f_\ast$ = 0.1
    & $l_{\rm loop}$ = [20, 30, 40, 60, 80]
    \rule[-5.5pt]{0pt}{20pt} \\
    $f_\ast$ = 0.05
    & $l_{\rm loop}$ = [20, 30, 40]
    \rule[-5.5pt]{0pt}{20pt} \\
    $f_\ast$ = 0.0333
    & \begin{tabular}{l} \hspace{-0.8em} $l_{\rm loop}$ = [20, 30, 40, 60, \\
    \hspace{2.0em} 80, 120, 160, 240] \end{tabular}
    \rule[-10.0pt]{0pt}{30pt} \\
    $f_\ast = 0.01 = f_\odot$ 
    & \begin{tabular}{l} \hspace{-0.8em} $l_{\rm loop}$ = [20, 30, 40, 60, 80, 120, \\
    \hspace{2.0em} 160, 240, 320, 480, 640] \end{tabular}
    \rule[-10.0pt]{0pt}{30pt} \\ 
    $f_\ast$ = 0.005
    & $l_{\rm loop}$ = [20]
    \rule[-5.5pt]{0pt}{20pt} \\ \hline
  \end{tabular}
  \vspace{1.0em}
  \caption{
  List of the simulation runs conducted in this study.
  The magnetic filling factor on the solar photosphere is set to $f_\odot = 0.01$ \citep{Cranmer_2017_ApJ}.
  }
  \vspace{0em}
  \label{table:run}
\end{table}

\subsection{Numerical method}

A non-uniform grid system is used to resolve the computational domain $0 \le s \le 2 l_{\rm loop}$.
A uniform cell size of $\Delta s_{\rm min}$ is used below the critical height $s<s_{\rm ge}$, above which the cell size expands until it reaches the maximum $\Delta s_{\rm max}$.
In particular, in $0 \le s \le l_{\rm loop}$, the size of the $i$-th cell, $\Delta s_i$, is iteratively defined as
\begin{align}
    &\Delta s_i = \max \left[ \Delta s_{\rm min}, \min \left[ \Delta s_{\rm max}, \Delta s_{\rm min} + \frac{2\varepsilon_{\rm ge}}{2 + \varepsilon_{\rm ge}} \left( s_{i-1} - s_{\rm ge} \right) \right] \right], \nonumber \\
    &s_i = s_{i-1} + \frac{1}{2} \left( \Delta s_{i-1} + \Delta s_i \right),
\end{align}
Letting $N$ be the total number of cells, we express the cell size in the latter half of the domain $l_{\rm loop} \le s \le 2 l_{\rm loop}$ by
\begin{align}
    \Delta s_i = \Delta s_{N+1-i}, \ \ \ \ s_i = s_{i-1} + \frac{1}{2} \left( \Delta s_{i-1} + \Delta s_i \right),
\end{align}
The maximum cell size is fixed to $\Delta s_{\rm max} = 100 {\rm \ km}$.
The relation between the minimum cell size and $f_\ast$ is
\begin{align}
    \Delta s_{\rm min} = 5 {\rm \ km} \ \  
    (f_\ast < 0.05), \ \ \ \ \   
    \Delta s_{\rm min} = 2 {\rm \ km} \ \  
    (f_\ast \ge 0.05).
\end{align}
This relation is used because a higher resolution is required at the transition region in the large-$f_\ast$ runs (for details, see Appendix \ref{app:transition_region_problem}).
The grid expansion rate also depends on $f_\ast$ as $\varepsilon_{\rm ge} = 2.13 \ \ (f_\ast < 0.05)$ and $\varepsilon_{\rm ge} = 1.89 \ \ (f_\ast \ge 0.05)$.
The grid expansion height is fixed to $s_{\rm ge} = 10,000 {\rm \ km}$, which is greater than the typical height of the transition region that needs to be resolved with the minimum cell size.

In numerically solving Eq. (\ref{eq:basic_equation_conservation_form}),
we rewrite the basic equations in terms of the cross-section-weighted conserved variables $\tilde{\vec{U}}$ and the corresponding fluxes $\tilde{\vec{F}}$ defined by
\begin{align}
    \tilde{\vec{U}} =
    \left(
    \begin{array}{c}
    \tilde{\rho} \\
    \tilde{\rho} \tilde{v}_r \\
    \tilde{\rho} \tilde{v}_x \\
    \tilde{\rho} \tilde{v}_y \\
    \tilde{B}_x \\
    \tilde{B}_y \\
    \tilde{e} 
    \end{array}
    \right)
    =
    \left(
    \begin{array}{c}
    \rho r^2 f \\
    \rho v_r r^2 f \\
    \rho v_x r^2 f \\
    \rho v_y r^2 f \\
    B_x r \sqrt{f} \\
    B_y r \sqrt{f} \\
    e r^2 f 
    \end{array}
    \right),
\end{align}
\begin{align}
    \tilde{\vec{F}} =
    \left(
    \begin{array}{c}
    \tilde{\rho} \tilde{v}_r \\
    \tilde{\rho} \tilde{v}_r^2 + \tilde{p}_T \\
    \tilde{\rho} \tilde{v}_r \tilde{v}_x - \tilde{B}_r \tilde{B}_x / (4\pi) \\
    \tilde{\rho} \tilde{v}_r \tilde{v}_y - \tilde{B}_r \tilde{B}_y / (4\pi) \\
    \tilde{v}_r \tilde{B}_x - \tilde{v}_x \tilde{B}_r \\
    \tilde{v}_r \tilde{B}_y - \tilde{v}_y \tilde{B}_r \\
    \left( \tilde{e} + \tilde{p}_T \right) \tilde{v}_r - \tilde{B}_r \left(\tilde{\vec{v}}_\perp \cdot \tilde{\vec{B}}_\perp \right)/(4\pi)
    \end{array}
    \right), 
\end{align}
where
\begin{align}
    \tilde{p}_T = \tilde{p} + \frac{\tilde{\vec{B}}_\perp^2}{8 \pi} = p_T r^2 f.
\end{align}
Using $\tilde{\vec{U}}$ and $\tilde{\vec{F}}$, the basic equation is given by
\begin{align}
    \frac{\partial}{\partial t} \tilde{\vec{U}} + \frac{\partial}{\partial s} \tilde{\vec{F}} = \tilde{\vec{S}},
    \label{eq:basic_equation_csw_conservation_form}
\end{align}
where
\begin{align}
    \tilde{\vec{S}} =
    \left(
    \begin{array}{c}
    0 \\
    \left( \tilde{p} +\dfrac{1}{2} \tilde{\rho} \tilde{\vec{v}}_\perp^2 \right)/L - \tilde{\rho} \dfrac{GM_\ast}{r^2} \dfrac{dr}{ds} \\
    \dfrac{1}{2L} \left( - \tilde{\rho} \tilde{v}_r \tilde{v}_x + \dfrac{\tilde{B}_r \tilde{B}_x}{4\pi} \right)  + \tilde{\rho} D^v_x \\
    \dfrac{1}{2L} \left( - \tilde{\rho} \tilde{v}_r \tilde{v}_y + \dfrac{\tilde{B}_r \tilde{B}_y}{4\pi} \right)  + \tilde{\rho} D^v_y \\
    \sqrt{4 \pi \tilde{\rho}} D^b_x \\
    \sqrt{4 \pi \tilde{\rho}} D^b_y \\
    - \tilde{\rho} \tilde{v}_r \dfrac{GM_\ast}{r^2} \dfrac{dr}{ds} + Q_{\rm cnd} r^2 f + Q_{\rm rad} r^2 f
    \end{array} 
    \right). \label{eq:csw_source}
\end{align}
With this variable conversion, any MHD solver designed for the Cartesian coordinate system can be directly applied to Eq. (\ref{eq:basic_equation_csw_conservation_form}).
In this study, the Harten--Lax--van Leer discontinuities (HLLD) approximated Riemann solver \citep{Miyoshi_2005_JCP} is used to calculate $\tilde{\vec{F}}$ at the cell boundary.
For spatial reconstruction, the fifth-order accurate monotonicity-preserving method \citep{Suresh_1997_JCP} is used to reconstruct the cross-section-weighted conserved variables $\tilde{\vec{U}}$ in $s \le s_{\rm ge}$ and $2l_{\rm loop}-s \le s_{\rm ge}$, whereas the monotonic upstream-centred scheme for the law of conservation \citep{van_Leer_1979_JCP} with a minmod flux limiter is used in $s_{\rm ge} < s < 2 l_{\rm loop} - s_{\rm ge}$.

Thermal conduction and the other parts are solved independently by the second-order operator-splitting procedure as follows.
\begin{enumerate}
    \item thermal conduction is solved for a half step $\Delta t/2$: \\
    $ \tilde{\vec{U}}^n \ \ \xrightarrow{\Delta t/2} \ \ \tilde{\vec{U}}^\ast $ \ \ (thermal conduction only)
    \vspace{1.0em}
    \item the rest of the basic equations are solved for a full step $\Delta t$: \\
    $ \tilde{\vec{U}}^\ast_i \ \ \xrightarrow{\Delta t} \ \ \tilde{\vec{U}}^{\ast \ast}_i $ \ \ (without thermal conduction)
    \vspace{1.0em}
    \item thermal conduction is solved again for a half step $\Delta t/2$: \\
    $ \tilde{\vec{U}}^{\ast \ast}_i \ \ \xrightarrow{\Delta t/2} \ \ \tilde{\vec{U}}^{n+1}_i $ \ \ (thermal conduction only)
\end{enumerate}
where $\tilde{\vec{U}}^n$ is the $n$-th step value of $\tilde{\vec{U}}$.
With this procedure, we avoid the severe constraints on the $\Delta t$ from thermal conduction when updating the MHD equations.

The third-order strong-stability-preserving (SSP) Runge--Kutta method is used in the time integration of the MHD equations \citep{Shu_1988_JCP,Gottlieb_2001_SIAMR}.
The super-time-stepping method \citep{Meyer_2012_MNRAS,Meyer_2014_JCP} is used to solve the thermal conduction, which reduce the numerical cost and time with minimum loss of accuracy.

\section{Simulation result \label{sec:simulation_result}}

\subsection{Fiducial (solar) case: atmosphere and spectrum \label{sec:result_fiducial}}

First, we discuss the simulation run with $l_{\rm loop} = 20 {\rm \ Mm}$ and $f_\ast = 1.0 \times 10^{-2}$ as the fiducial case.
In this case, the coronal field strength is $\approx 20 {\rm \ G}$, 
which is within the range of the solar coronal magnetic field strength measured by the coronal seismology technique \citep{Nakariakov_2001_AA,Verwichte_2004_SolPhys,Jess_2016_NatPhys}, and thus the fiducial case is regarded as the solar case.

\begin{figure}[t!]
\centering
\includegraphics[width=80mm]{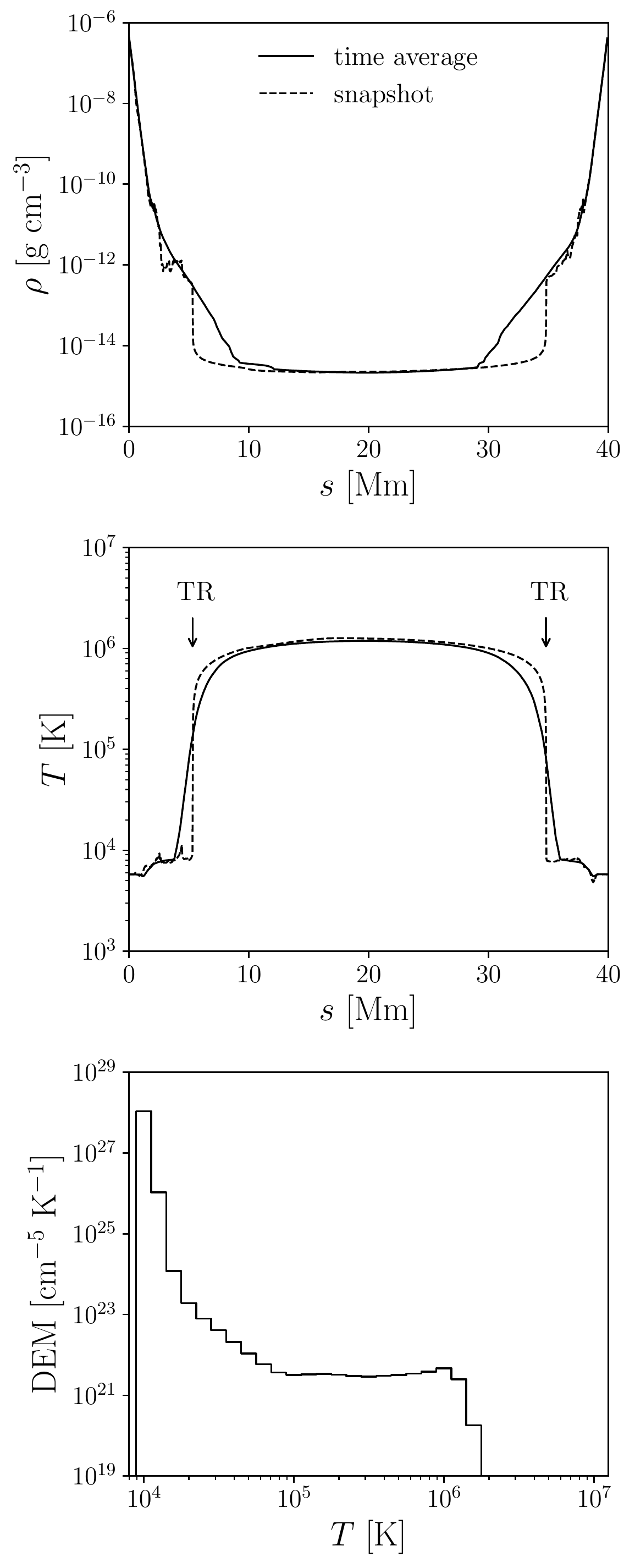}
\vspace{0.5em}
\caption{
Simulated loop properties of the fiducial case.
Top: time-averaged (solid line) and snapshot (dashed line) profiles of mass density along the loop axis.
Middle: time-averaged (solid line) and snapshot (dashed line) profiles of temperature along the loop axis.
Bottom: time-averaged differential emission measure (solid line).
Annotations ``TR'' in the middle panel indicate the locations of the transition region in the snapshot profile.
}
\label{fig:average_fiducial}
\vspace{0em}
\end{figure}

Figure \ref{fig:average_fiducial} illustrates the time-averaged properties of a quasi-steady coronal loop.
Panels show the mass density (top) and temperature (middle) along the loop axis and the differential emission measure (DEM, bottom), defined by
\begin{align}
    {\rm DEM} (T) = n_e^2 \left( T \right) \frac{dl_{\rm los}}{dT},
\end{align}
where $n_e (T)$ is the electron density with temperature $T$ and $l_{\rm los}$ is the length along the line of sight.
Practically, dividing the temperature range into bins and considering the vertical line of sight
\revise{($l_{\rm los} = r$)}, 
the DEM and associated emission measure distribution (EMD) are numerically obtained as follows
\begin{align}
    {\rm DEM} (T_i) \Delta T_i \equiv {\rm EMD} (T_i) = n_e^2 \left( T_i \right) \Delta r (T_i), \label{eq:DEM_practical}
\end{align}
where $n_e (T_i)$ is the total number density of an electron that exhibits a temperature in $[T_i-\Delta T_i/2,T_i+\Delta T_i/2]$ and $\Delta r (T_i)$ is the total radial extension of where $T_i-\Delta T_i/2 \le T < T_i+\Delta T_i/2$.
The DEM is calculated in the temperature range of $10^4 {\rm \ K} \le T \le 10^7 {\rm \ K}$ with an equal spacing in the logarithmic scale of $T$.
\revise{The DEM in the range of $T <10^4 {\rm \ K}$ is not calculated because, in the low-temperature range, the atmosphere is not optically thin and the DEM loses its meaning.}
Note that the unit of EMD is ${\rm cm^{-5}}$, while the ``volume'' EMD,
which has often been used in the literature \citep{Gudel_2003_ApJ,Scelsi_2005_AA},
represents the distribution of an emission measure over the whole coronal volume and has a different unit (${\rm cm^{-3}}$).

The high-temperature ($T>10^6 {\rm \ K}$) corona is successfully reproduced in the fiducial case.
Since we are imposing the phenomenological, mean-field formulation of the coronal heating (field braiding/turbulence), the heating tends to be more constant in time and more uniform in space than the actual three-dimensional case in which the heating is intermittent in time and space.
\sout{Meanwhile, the time-averaged heating rate should be similar between our one-dimensional approximation and full three-dimensional simulation, and thus, the time-averaged properties are reliable.}
\revise{Nevertheless, the time-averaged heating rate should be similar between the 1D approximation and the 3D simulation, because the previous 3D simulation of solar wind yielded a similar mean field to the 1D simulation with turbulence phenomenology \citep{Shoda_2019_ApJ}.}

\begin{figure}[t!]
\centering
\includegraphics[width=80mm]{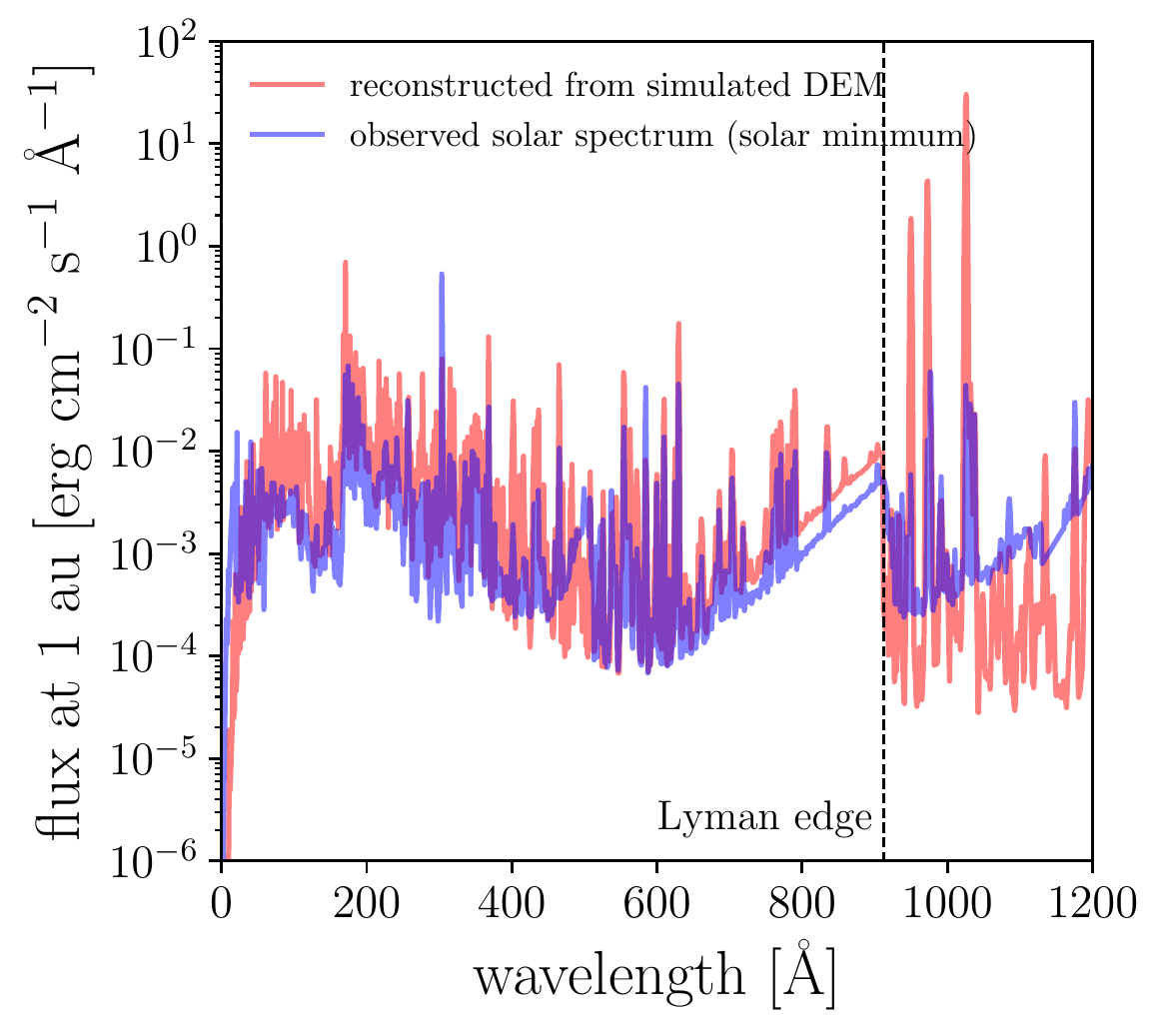}
\vspace{0.5em}
\caption{
Observed (blue) and simulated (red) spectral flux density of the Sun measured at $1 {\rm \ au}$.
The observed spectrum is retrieved in the solar activity minimum.
The two spectra are in a good agreement below the Lyman edge ($\le 912 \ \AA$).
}
\label{fig:spectrum_fiducial}
\vspace{0em}
\end{figure}

Under the assumption that the upper atmosphere ($T \ge 10^4 {\rm \ K}$) is optically thin in the wavelength of interest,
the XUV spectrum is obtained from the EMD using the open-source package ChiantiPy based on CHIANTI database ver 10.0 \citep{Del_Zanna_2021_ApJ}.
In particular, the specific intensity $I_\lambda$ was calculated from the EMD using the ChiantiPy.core.Spectrum module
\revise{with the coronal abundance given by \citet{Schmelz_2012_ApJ}}.
\revise{Although the radiative loss function $\Lambda(T)$ is constructed with photospheric abundance, because the loss function is nearly independent of the FIP effect in the radiation-dominated temperature range $T \le 3 \times 10^5 {\rm \ K}$, the inconsistency in abundance do not violate the simulation results.}
\sout{Given} \revise{Assuming} that the stellar corona is a uniformly bright sphere, the spectral flux density at the heliocentric distance $r$ is deduced by \citep[see, e.g., Section 1.3 of][]{Rybicki_1979_book}
\begin{align}
    F_\lambda = \pi I_\lambda \left( \frac{R_\odot}{r} \right)^2,
\end{align}
which yields the X-ray and EUV luminosities as
\begin{align}
    L_{\rm X} = 4 \pi r^2 \int_{5 {\rm \ \AA}}^{100 {\rm \ \AA}} d\lambda \ F_\lambda, \ \ \ \ L_{\rm EUV} = 4 \pi r^2 \int_{100 {\rm \ \AA}}^{912 {\rm \ \AA}} d\lambda \ F_\lambda.
    \label{eq:Lx_L_euv_definition} 
\end{align}
\revise{
In terms of energy, X-ray photons are in the range of $0.12-2.48 {\rm \ keV}$.
Caution must be exercised when calibrating the X-rays because a subtle difference in the bandpass of the instrument can result in large differences in the derived response function and X-ray luminosity \citep{Zhuleku_2020_AA}.
For the procedure of translation between different instruments, see \citet{Judge_2003_ApJ}.}

To test the capability of our model in the prediction of XUV spectrum, 
we compare in Figure \ref{fig:spectrum_fiducial} the spectral flux density obtained from our simulation (red) and that from the observation in a solar activity minimum (blue).
The observed spectrum is obtained from the coordinated observation in the Whole Heliosphere Interval \citep[WHI, from March 20, 2008 to April 16, 2008][]{Woods_2009_GRL,Chamberlin_2009_GRL}.
Figure~\ref{fig:spectrum_fiducial} shows that, below the Lyman edge ($\le 912 \ \AA$), the simulated spectrum is in a good agreement with the observed spectrum. 
Above the Lyman edge ($\ge 912 \ \AA$), the continuum is underestimated, and the emission lines are overestimated in the simulated spectrum, possibly because the optically thin approximation is inadequate in this wavelength range.
In this study, because the focus is on the spectrum below the Lyman edge, the simulation is validated with respect to spectrum prediction.

\subsection{Loop-length dependence: Density and temperature}

The coronal loop length is a fundamental parameter that affects the coronal density and temperature.
Here, we show the relation between the time-averaged coronal properties and the coronal loop length.
For simplicity, we fix the magnetic filling factor to the fiducial (solar) value: $f_\ast = f_\odot = 0.01$.
Hereinafter, the time-averaged value of $X$ will be denoted by $X_{\rm ave}$.

\begin{figure}[t!]
\centering
\includegraphics[width=80mm]{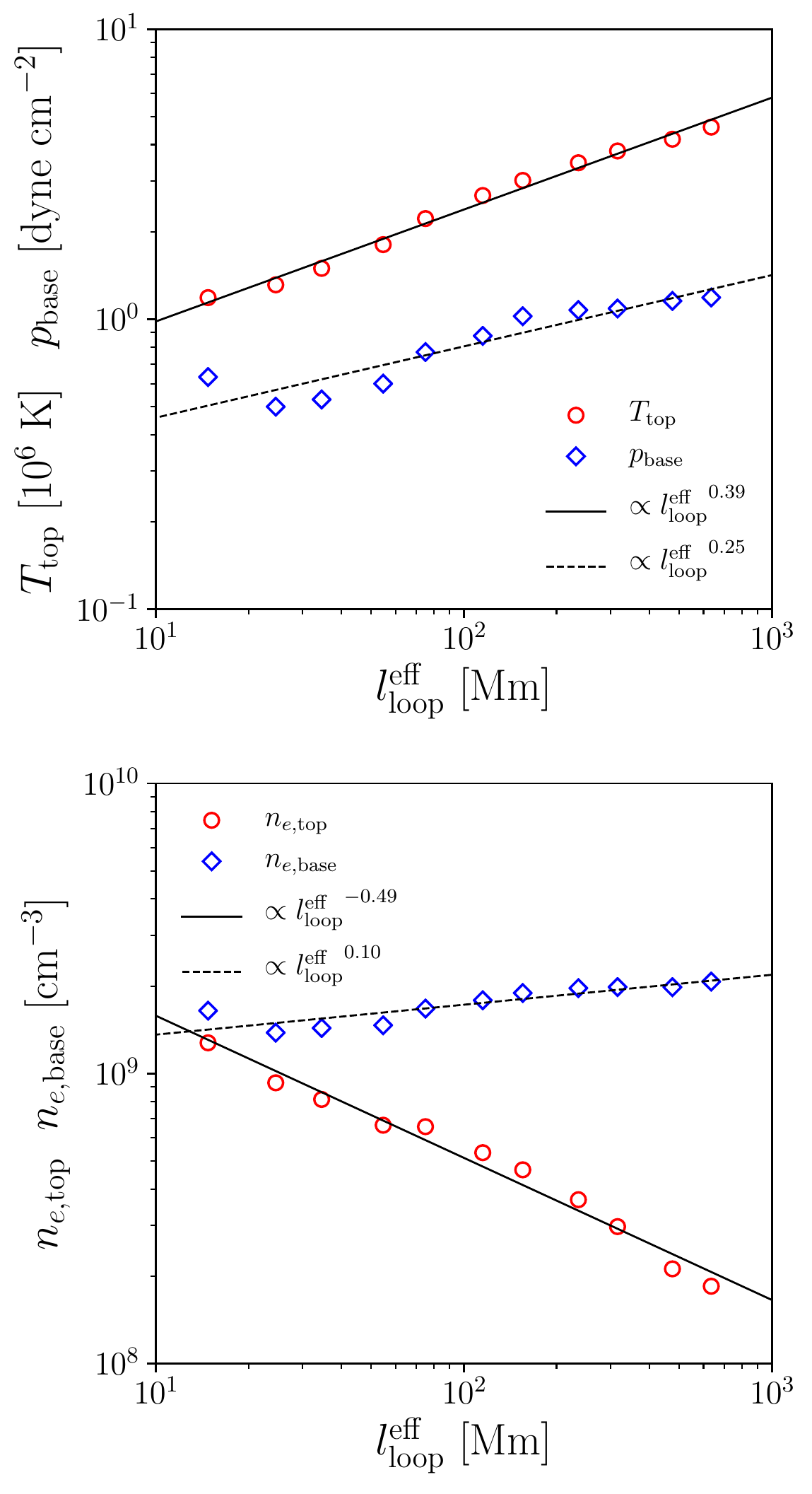}
\vspace{0.5em}
\caption{
Top: relation between the effective half loop length $l_{\rm loop}^{\rm eff}$ (see Eq. (\ref{eq:definition_effective_half_loop_length}) for definition) and the time-averaged loop-top temperature ($T_{\rm top}$, red circles) and coronal-base pressure ($p_{\rm base}$, blue diamonds).
Bottom: relation between the effective half loop length $l_{\rm loop}^{\rm eff}$ and the time-averaged loop-top electron density ($n_{e,{\rm top}}$, red circles) and coronal-base electron density ($n_{e,{\rm base}}$, blue diamonds).
In both panels, lines represent the power-law fittings to the symbols.
}
\label{fig:loop_length_vs_loop_properties_fexp100}
\vspace{0em}
\end{figure}

In the discussion on the behaviour of the coronal properties, the results must be compared with the analytical RTV scaling law \citep{Rosner_1978_ApJ}.
Note that the half-loop length in the RTV scaling law denotes the length from the transition region to the apex of the loop, whereas $l_{\rm loop}$ stands for the length from the stellar surface to the apex of the loop.
For better comparison, instead of $l_{\rm loop}$, we use the effective half-loop length $l_{\rm loop}^{\rm eff}$, which denotes the coronal length, and is defined as
\begin{align}
    l_{\rm loop}^{\rm eff} = \int_{s_{\rm TR}}^{l_{\rm loop}} ds, \label{eq:definition_effective_half_loop_length}
\end{align}
where $s_{\rm TR} \ (<l_{\rm loop})$ is where $T_{\rm ave} = 10^5 {\rm \ K}$.

Another factor that should be considered is the gravitational stratification.
The effective half-loop length often exceeds the coronal pressure scale height.
In such cases, the loop-top pressure in the original RTV scaling law is replaced by the coronal-base pressure \citep{Serio_1981_ApJ}.
Given that the pressure is continuous across the transition region, we define the coronal-base pressure $p_{\rm base}$ as the pressure measured at $T_{\rm ave} = 10^5 {\rm \ K}$.
For the coronal electron density, both the loop-top value $n_{e,{\rm top}}$ and coronal-base value $n_{e,{\rm base}}$ are measured.
In contrast to pressure, the density is discontinuous across the transition region, and therefore the definition of the coronal-base density is not trivial.
Here, we define $n_{e,{\rm base}}$ as the value measured in the coronal base:
\begin{align}
    n_{e,{\rm base}} \equiv \frac{1}{s_{{\rm cor},2}-s_{{\rm cor},1}} \int^{s_{{\rm cor},2}}_{s_{{\rm cor},1}} n_{e,{\rm ave}} \ ds,
\end{align}
where we set $s_{{\rm cor},1} = 10 {\rm \ Mm}$ and $s_{{\rm cor},2} = 15 {\rm \ Mm}$.
The loop-top density and temperature are given by
\begin{align}
    n_{e,{\rm top}} = n_e \left( s=l_{\rm loop} \right), \ \ \ \ 
    T_{\rm top} = T \left( s= l_{\rm loop} \right).
\end{align}

Figure \ref{fig:loop_length_vs_loop_properties_fexp100} shows the $l_{\rm loop}^{\rm eff}$-dependencies of the time-averaged coronal properties (density, temperature and pressure).
The loop-top temperature and coronal-base pressure obey a power--law relation with respect to the effective half-loop length, which is formulated as
\begin{align}
    T_{\rm top} &\propto {l_{\rm loop}^{\rm eff}}^{0.39}, 
    \label{eq:loop_top_temperature_scaling} \\
    p_{\rm base} &\propto {l_{\rm loop}^{\rm eff}}^{0.25}.
    \label{eq:coronal_base_pressure_scaling}
\end{align}
The (generalised) RTV scaling law predicts that the loop-top temperature obeys the following relation
\begin{align}
    T_{\rm top}^{\rm RTV} \propto \left( p_{\rm base} l_{\rm loop}^{\rm eff} \right)^{1/3},
    \label{eq:RTV_scaling_law_temperature}
\end{align}
where we dismiss the exponential correction term as it is negligible \citep{Serio_1981_ApJ}.
A comparison of Eqs. (\ref{eq:loop_top_temperature_scaling}), (\ref{eq:coronal_base_pressure_scaling}), and (\ref{eq:RTV_scaling_law_temperature}) reveals that the simulation results are consistent with the RTV scaling law.

An alternative form of the RTV scaling law predicts a relation among the coronal energy flux $F_{\rm cor}$, loop length $l_{\rm loop}^{\rm eff}$, and loop-top temperature $T_{\rm top}^{\rm RTV}$, and is expressed as
\begin{align}
    T_{\rm top}^{\rm RTV} \propto \left( {F_{\rm cor}} {l_{\rm loop}^{\rm eff}} \right)^{2/7}.
    \label{eq:loop_top_temperature_scaling_RTV}
\end{align}
A comparison of Eqs. (\ref{eq:loop_top_temperature_scaling}) and (\ref{eq:loop_top_temperature_scaling_RTV}) indicates that the energy flux injected into the corona is larger for longer loops.
\sout{,which is directly validated in the following section.}
\revise{In terms of the heating rate per unit volume $Q$, the RTV predictions,
\begin{align}
    T_{\rm top}^{\rm RTV} &\propto Q^{2/7} {l_{\rm loop}^{\rm eff}}^{4/7}, \label{eq:RTV_TQL} \\ 
    p_{\rm base}^{\rm RTV} &\propto Q^{6/7} {l_{\rm loop}^{\rm eff}}^{5/7}, \label{eq:RTV_PQL}
\end{align}
and simulation results from Eqs. (\ref{eq:loop_top_temperature_scaling}) and (\ref{eq:coronal_base_pressure_scaling}) indicate that $Q = F_{\rm cor} / l_{\rm loop}^{\rm eff}$ is a decreasing function of $l_{\rm loop}^{\rm eff}$.
These conclusions shall be directly validated in the following section.}

The bottom panel of Figure \ref{fig:loop_length_vs_loop_properties_fexp100} shows the variations in loop-top and coronal-base electron densities over $l_{\rm loop}^{\rm eff}$.
Given that the RTV scaling law predicts a larger loop-top density at a higher loop-top temperature for a uniform corona, the decrease in the loop-top density is attributed to gravitational stratification.
Note that the coronal-base density exhibits a weaker dependence on $l_{\rm loop}^{\rm eff}$ than the coronal-base pressure,
which is contradictory if the coronal-base temperature is constant in $l_{\rm loop}^{\rm eff}$.
It may be interpreted that the coronal-base temperature increases with $l_{\rm loop}^{\rm eff}$ in response to the increasing $T_{\rm top}$ with $l_{\rm loop}^{\rm eff}$.

\subsection{Loop-length dependence: Energy flux and XUV emission
\label{sec:energetics_loop_length}}

\begin{figure}[t!]
\centering
\includegraphics[width=80mm]{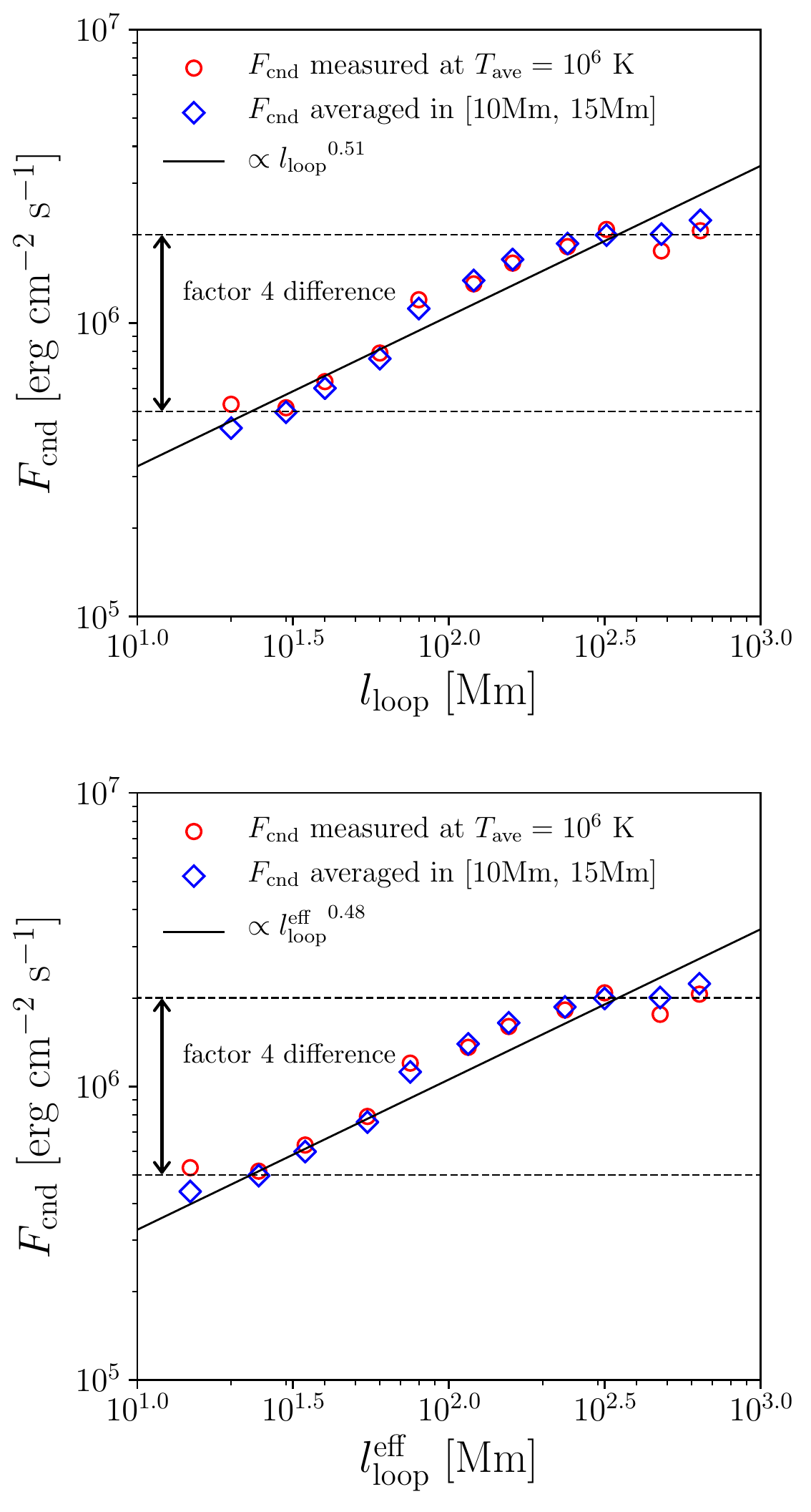}
\vspace{0.5em}
\caption{
Top: half loop length versus coronal conductive flux measured at $T_{\rm ave} = 10^6 {\rm \ K}$ (red circles) 
and averaged in $10 {\rm \ Mm} \le s \le 15 {\rm \ Mm}$ (blue diamonds).
Also shown by the solid line is the power-law fitting to the blue diamonds.
Bottom: same as the top panel but with respect to the effective half loop length $l_{\rm loop}^{\rm eff}$.
}
\label{fig:loop_length_vs_conductive_flux_fexp100}
\vspace{0em}
\end{figure}

For a better interpretation of the behaviour of coronal properties, the variation in the energy flux entering the corona $F_{\rm cor}$ with the effective half-loop length $l_{\rm loop}^{\rm eff}$ must be revealed.

Directly measuring $F_{\rm cor}$ from the simulation data is, however, not a trivial task.
To obtain $F_{\rm cor}$, one needs to measure the energy flux at the base of the corona, which moves in time and is broadened after time averaging.
Because a significant amount of energy is reflected back at the transition region, 
a slight difference in the position of the coronal base yields a significant error or uncertainty in $F_{\rm cor}$.
Therefore, instead of directly measuring $F_{\rm cor}$, 
we measure the backward conductive flux $F_{\rm cnd}$, which should be balanced with $F_{\rm cor}$ by energy conservation.

The top and 
\sout{middle} 
\revise{bottom} 
panels in Figure \ref{fig:loop_length_vs_conductive_flux_fexp100} show the loop-length ($l_{\rm loop}$ and $l_{\rm loop}^{\rm eff}$) dependence of the coronal conductive flux.
The panels display the simulation runs with a fixed magnetic filling factor of $f_\ast = 1.00 \times 10^{-2}$.
The red circles and blue diamonds represent the conductive flux measured at $T_{\rm ave} = 1.0 \times 10^6 {\rm \ K}$ and averaged over $10 {\rm \ Mm} \le s \le 15 {\rm \ Mm}$, respectively.
The black solid lines represent the power-law fittings to the blue diamonds.
The coronal conductive flux increases with the loop length.
However, the trend deviates from a simple power law.
When the loop length is sufficiently small ($l_{\rm loop}^{\rm eff} \lesssim 10^{1.5} {\rm \ Mm}$) or large ($l_{\rm loop}^{\rm eff} \gtrsim 10^{2.5} {\rm \ Mm}$), the conductive flux weakly depends on the loop length.
The energy flux increases around $l_{\rm loop}^{\rm eff} \sim 10^{2.0} {\rm \ Mm}$,
with the minimum and maximum values differing by a factor of $4$.

An approximate power-law fit to the blue diamonds yields the following scaling law 
\begin{align}
    F_{\rm cnd} \propto {l_{\rm loop}^{\rm eff}}^{0.48},
    \label{eq:scaling_loop_length_conductive_flux}
\end{align}
which, in combination with the RTV scaling law, predicts
\begin{align}
    T_{\rm top}^{\rm RTV} \propto \left( F_{\rm cor} l_{\rm loop}^{\rm eff} \right)^{2/7} \propto {l_{\rm loop}^{\rm eff}}^{0.42}.
\end{align}
The simulated dependence of $T_{\rm top}$ on $l_{\rm loop}^{\rm eff}$,
Eq. (\ref{eq:loop_top_temperature_scaling}),
is reproduced by the semi-analytical arguments.
Thus, the results shall be explained semi-analytically once the theoretical behaviour of $F_{\rm cor}$ (or equivalently $F_{\rm cnd}$) has been derived.
In Section \ref{sec:analytical_model}, we propose a simple model to produce $F_{\rm cor}$.

\begin{figure}[t!]
\centering
\includegraphics[width=80mm]{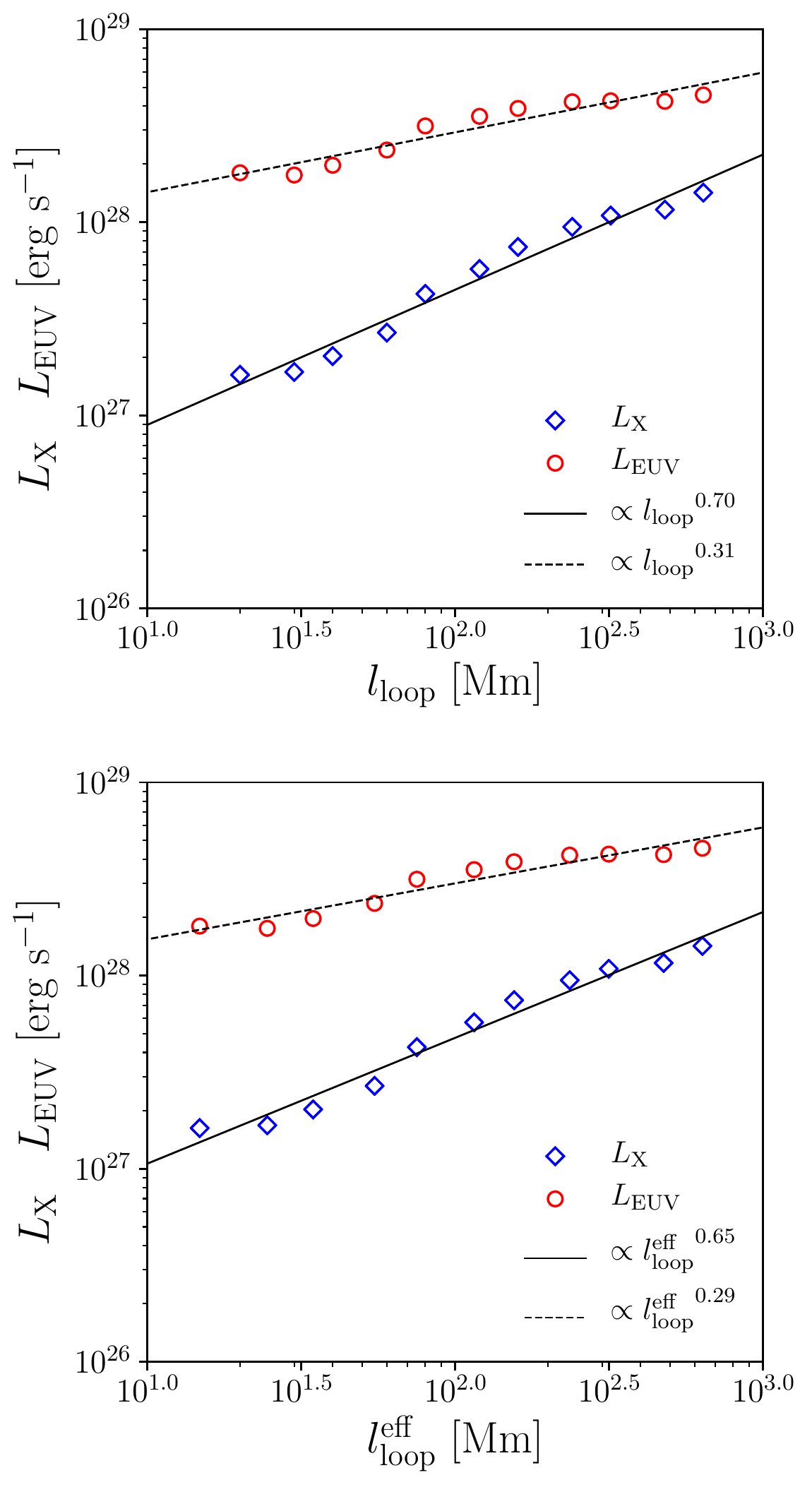}
\vspace{0.5em}
\caption{
(Effective) loop length versus X-ray luminosity $L_{\rm X}$ (red circles) and EUV luminosity $L_{\rm EUV}$ (blue diamonds).
The solid and dashed lines show the power-law fittings to the simulation results.
The top and bottom panels show the relation with respect to the half loop length $l_{\rm loop}$ and the effective hal loop length $l_{\rm loop}^{\rm eff}$ (see Eq. (\ref{eq:definition_effective_half_loop_length})), respectively.
}
\label{fig:loop_length_vs_XUV_fexp100}
\vspace{0em}
\end{figure} 

The enhanced energy injection to the corona produces enhanced XUV emissions.
Figure \ref{fig:loop_length_vs_XUV_fexp100} depicts the loop-length dependence of the predicted XUV luminosity.
For each coronal loop calculation, the XUV spectrum is calculated as in Figure \ref{fig:spectrum_fiducial} and converted to XUV luminosity through Eq. (\ref{eq:Lx_L_euv_definition}).
Both $L_X$ and $L_{\rm EUV}$ exhibit increasing trends as those found in the coronal energy flux $F_{\rm cnd}$. 
In particular, $L_X$ exhibits this trend significantly.
The inferred power--law relations are
\begin{align}
    &L_{\rm EUV} \propto {l_{\rm loop}^{\rm eff}}^{0.29}, \label{eq:Leuv_loop_length_simulation_result} \\
    &L_{\rm X} \propto {l_{\rm loop}^{\rm eff}}^{0.65}. \label{eq:Lx_loop_length_simulation_result}
\end{align}

\begin{figure}[t!]
\centering
\includegraphics[width=80mm]{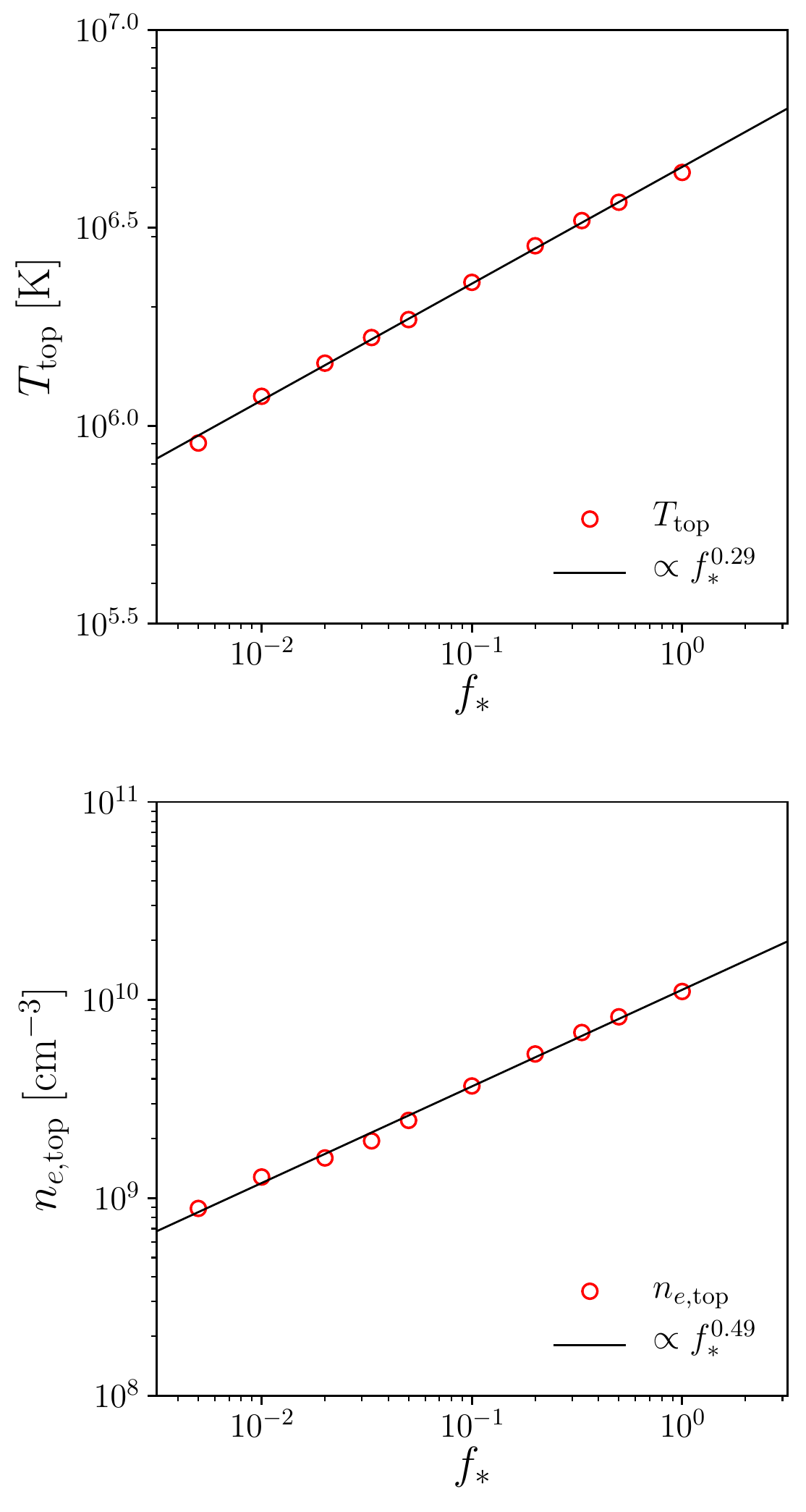}
\vspace{0.5em}
\caption{
Magnetic filling factor on the photosphere $f_\ast$ versus time-averaged loop-top temperature ($T_{\rm top}$, top panel) and loop-top electron density ($n_{e,{\rm top}}$, bottom panel).
The half loop length is fixed to $l_{\rm loop} = 20 {\rm \ Mm}$.
Red circles show the simulation results and the solid lines show the power-law fittings to them.
}
\label{fig:filling_factor_vs_temperature_density_L20}
\vspace{0em}
\end{figure}

The dependence of $L_{\rm X}$ on $l_{\rm loop}^{\rm eff}$ is further explained semi-analytically.
The X-ray luminosity should be proportional to the emission measure of the corona, i.e.,
\begin{align}
    L_X \propto n_{\rm cor}^2 l_{\rm loop}^{\rm eff},
    \label{eq:Xray_luminosity}
\end{align}
where $n_{\rm cor}$ is the typical number density of the coronal loop \revise{that lies between $n_{\rm base}$ and $n_{\rm top}$}.
\sout{Given}
\revise{Assuming}
that $n_{\rm cor}$ follows the prediction of the RTV scaling law,
\begin{align}
    n_{\rm cor} \propto {T_{\rm top}^{\rm RTV}}^2 / l_{\rm loop}^{\rm eff} 
    \propto {F_{\rm cor}}^{4/7} {l_{\rm loop}^{\rm eff}}^{-3/7},
\end{align}
the X-ray luminosity is connected to the coronal energy flux $F_{\rm cor}$ and loop length $l_{\rm loop}^{\rm eff}$ by
\begin{align}
    L_X \propto {F_{\rm cor}}^{8/7} {l_{\rm loop}^{\rm eff}}^{1/7} \propto {l_{\rm loop}^{\rm eff}}^{0.69},
    \label{eq:Xray_luminosity_RTV}
\end{align}
where the simulated scaling law Eq. (\ref{eq:scaling_loop_length_conductive_flux}) is used in the second proportional relation.
The actual scaling relation Eq. (\ref{eq:Lx_loop_length_simulation_result}) is in good agreement with the semi-analytical prediction Eq. (\ref{eq:Xray_luminosity_RTV}), 
validating the RTV scaling law in describing the stellar coronal properties.

Note that EUV luminosity $L_{\rm EUV}$ is weakly dependent on the loop length compared with $L_X$ because a portion of the EUV photons originates from the upper chromosphere and transition region, which are barely affected by the variation in the coronal loop length. 
Within the investigated loop-length range, the ratio $L_{\rm EUV}/L_{\rm X}$ decreases from $\sim 10$ to $\sim 3$ as the effective loop length increases.

\subsection{Filling-factor dependence: density and temperature}

\begin{figure}[t!]
\centering
\includegraphics[width=80mm]{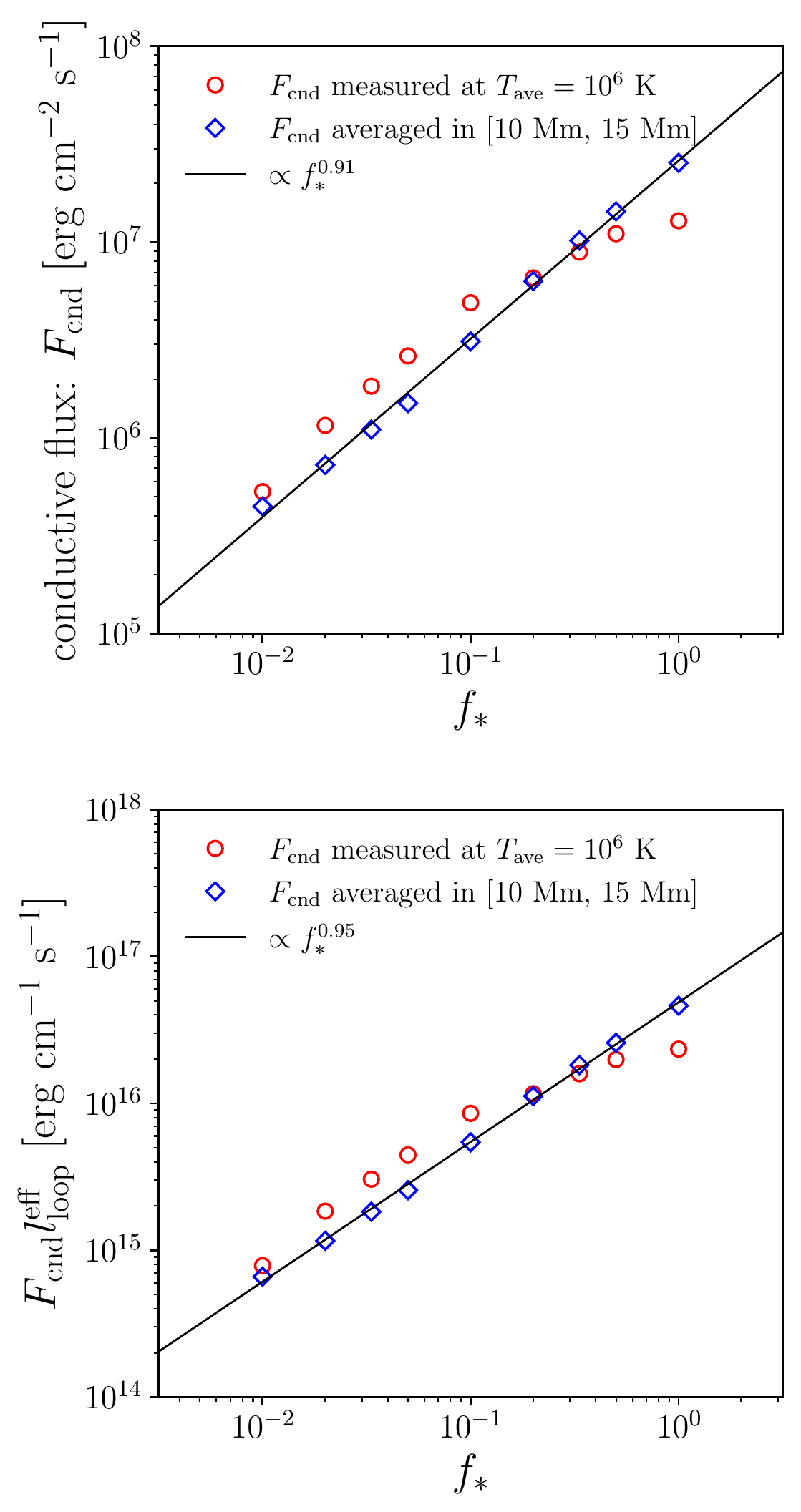}
\vspace{0.5em}
\caption{
Top: magnetic filling factor \revise{on the photosphere} $f_\ast$ versus backward conductive flux measured in the corona.
The half loop length is fixed to $l_{\rm loop} = 20 {\rm \ Mm}$.
The red circles and blue diamonds represent the flux measured at $T_{\rm ave} = 10^6 {\rm \ K}$ and
averaged in $10 {\rm \ Mm} \le s \le 15 {\rm \ Mm}$.
The black solid line is the power-law fitting to the blue diamonds.
Bottom: same as the top panel but the vertical axis denotes
\revise{$F_{\rm cnd} {l_{\rm loop}^{\rm eff}}$}.
}
\label{fig:filling_factor_vs_conductive_flux}
\vspace{0em}
\end{figure}

\begin{figure*}[t!]
\centering
\includegraphics[width=180mm]{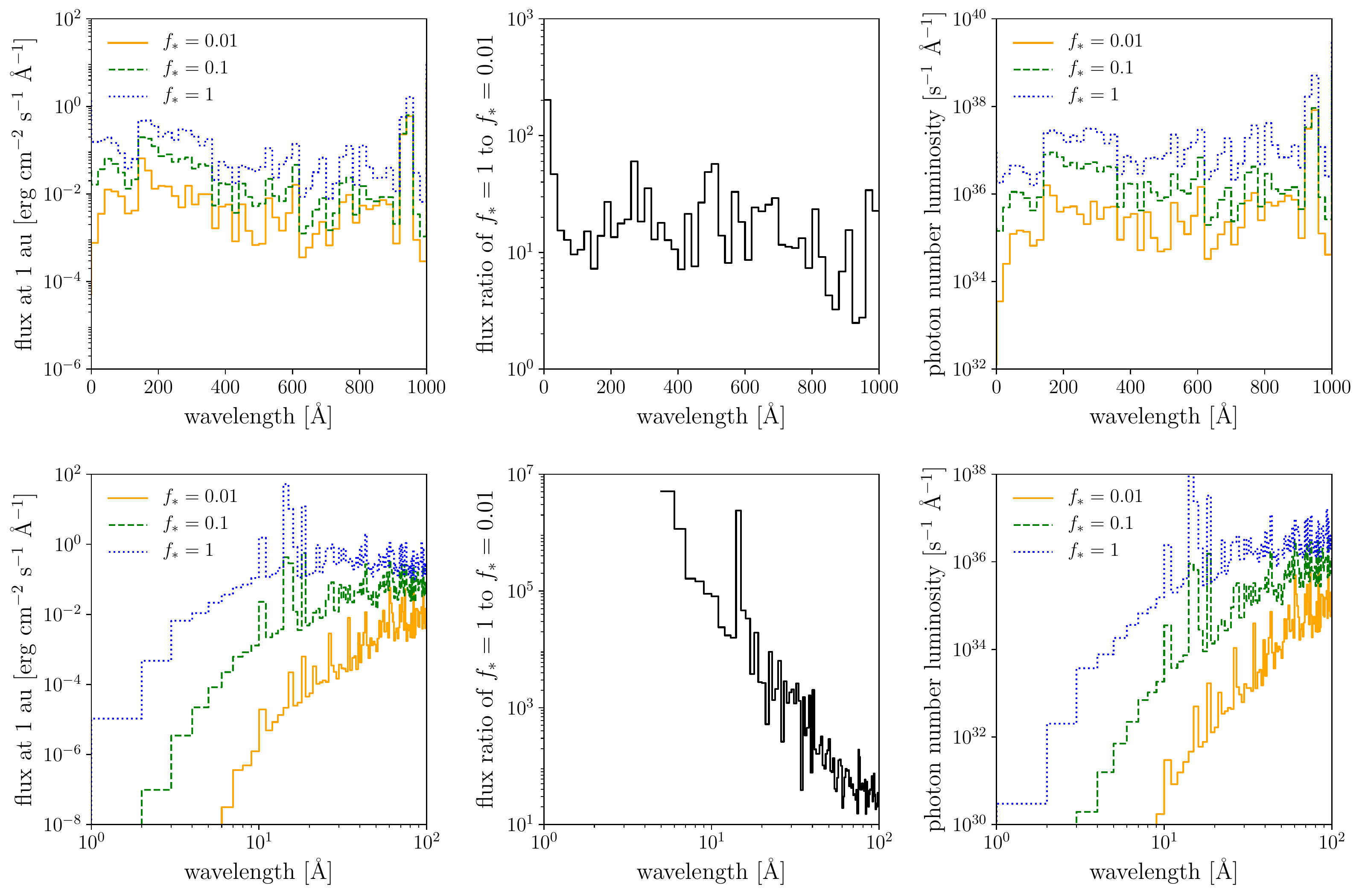}
\vspace{0.5em}
\caption{
Left: spectral energy flux at $1 {\rm \ au}$ for $f_\ast =0.01$ (orange, solid), $f_\ast=0.1$ (green, dashed), and $f_\ast =1$ (blue, dotted).
Center: energy-flux ratio of $f_\ast =1$ to $f_\ast = 0.01$.
Right: photon-number luminosity spectrum for $f_\ast =0.01$ (orange, solid), $f_\ast=0.1$ (green, dashed), and $f_\ast =1$ (blue, dotted).
Bottom panels are the same as top panels but with logarithmic $x$ axis that emphasize the short-wavelength region.
For better visualisation, spectra are averaged over $20 {\rm \ \AA}$-bins in the top panels.
}
\label{fig:filling_factor_vs_XUV_photon_spectrum}
\vspace{0em}
\end{figure*}

The magnetic filling factor of the photosphere appears to have scaled with the Rossby number, or equivalently the magnetic activity level \citep{Saar_1996_proceedings,Saar_2001_proceedings,Reiners_2009_ApJ}.
Hence, the dependence of the coronal properties on the filling factor is worth investigating as a proxy of the dependence of the stellar corona on the activity level.
The simulation results with various magnetic filling factors shall be discussed below, with the half-loop length fixed to $l_{\rm loop} = 20 \ {\rm Mm}$.

Figure \ref{fig:filling_factor_vs_temperature_density_L20} shows the filling-factor dependence of the time-averaged loop-top temperature $T_{\rm top}$ and electron density $n_{e,{\rm top}}$.
The coronal density and temperature increase with an increase in the magnetic filling factor following a power--law relationship
\begin{align}
    &T_{\rm top} \propto f_\ast^{0.29}, \label{eq:scaling_te_ff} \\
    &n_{e,{\rm top}} \propto f_\ast^{0.49}. \label{eq:scaling_ne_ff}
\end{align}
Although not explicitly demonstrated, the effective half-loop length also weakly depends on $f_\ast$ as
\begin{align}
    l_{\rm loop}^{\rm eff} \propto f_\ast^{0.04}. \label{eq:scaling_lloop_ff}
\end{align}
We note that the RTV scaling law semi-analytically predicts a relation of
\begin{align}
    T_{\rm top}^{\rm RTV} \propto \sqrt{n_{e,{\rm top}} l_{\rm loop}^{\rm eff}} \propto f_\ast^{0.27},
\end{align}
which is close to the simulation result Eq. (\ref{eq:scaling_te_ff}).
\revise{Alternatively, in terms of the heating rate per unit volume $Q$, the RTV scaling law coupled with Eq.s (\ref{eq:scaling_te_ff}) and (\ref{eq:scaling_lloop_ff}) yield
\begin{align}
    Q \propto f^{0.935},
\end{align}
while the RTV scaling law Eq.s (\ref{eq:scaling_ne_ff}) and (\ref{eq:scaling_lloop_ff}) yield
\begin{align}
    Q \propto f^{0.868},
\end{align}
The two trends are mostly consistent with each other.
}

The RTV scaling law predicts that the loop-top temperature $T_{\rm top}$ depends on the coronal energy flux $F_{\rm cor}$ and effective half-loop length $l_{\rm loop}^{\rm eff}$ as Eq. (\ref{eq:loop_top_temperature_scaling_RTV}).
A comparison between Eqs. (\ref{eq:loop_top_temperature_scaling_RTV}) and (\ref{eq:scaling_te_ff}) reveals the mostly proportional relationship between $F_{\rm cor}$ and $f_\ast$:
\begin{align}
    F_{\rm cor} {l_{\rm loop}^{\rm eff}} \propto f_\ast, \label{eq:coronal_energy_flux_ff_RTV}
\end{align}
which shall be further investigated in the following section.

\subsection{Filling-factor dependence: energy flux and XUV emission \label{sec:ff_vs_energetics}}

The energetics of the corona with various magnetic filling factors shall be discussed here.
The top panel of Figure \ref{fig:filling_factor_vs_conductive_flux} shows the relation between the filling factor $f_\ast$ and the coronal backward conductive flux measured at $T_{\rm ave} = 10^6 \ {\rm K}$ (red circles) and averaged over $10 {\rm \ Mm} \le s \le 15 {\rm \ Mm}$ (blue diamonds).
The two conductive fluxes exhibit different behaviours, with the red circles appearing to saturate within the large $f_\ast$ limit.
This difference may arise from the different physical meanings of $T_{\rm ave} = 10^6 {\rm \ K}$, where the values of the red circles are measured.
$T_{\rm ave} = 10^6 {\rm \ K}$ is located near the loop top for the small $f_\ast$ cases and near the transition region for the large ones.
Therefore, the blue diamonds may be interpreted as the coronal energy flux.

Given the dependence of the loop-top temperature and RTV scaling law on the filling factor, the relation among the coronal energy flux $F_{\rm cor}$, effective loop length $l_{\rm loop}^{\rm eff}$, and filling factor $f_\ast$ is predicted by Eq. (\ref{eq:coronal_energy_flux_ff_RTV}).
This prediction is confirmed in the bottom panel of Figure \ref{fig:filling_factor_vs_conductive_flux}, which shows the following scaling relation:
\begin{align}
    F_{\rm cor} {l_{\rm loop}^{\rm eff}} \propto f_\ast^{0.95}. \label{eq:energy_flux_filling_factor}
\end{align}
This consistency indicates that the RTV scaling law must be applicable to stellar coronae with varying activity levels.
\sout{The simulation validates the arguments of RTV-based stellar coronal energetics based on the RTV scaling law}
\revise{The simulation results validate the previous studies on stellar coronal energetics based on the RTV scaling law \citep{Shibata_2002_ApJ,Takasao_2020_ApJ}}.

Figure \ref{fig:filling_factor_vs_XUV_photon_spectrum} shows the spectra of XUV emission for different magnetic filling factors.
For all wavelengths below the Lyman edge ($912 \ {\rm \AA}$), the XUV emissions increase as the magnetic filling factor increases.
This is a natural consequence of the increased coronal density shown in Figure \ref{fig:filling_factor_vs_temperature_density_L20}.
In particular, as seen in the central panels of Figure \ref{fig:filling_factor_vs_XUV_photon_spectrum},
the enhancement of the energy flux from $f_\ast=0.01$ to $f_\ast=1$ is centred on $\sim 10$ with a variance of factor $\sim 5$
in the EUV range ($100 {\rm \ \AA} < \lambda \le 912 {\rm \ \AA}$), 
whereas a significantly larger enhancement occurs in the X-ray range ($5 {\rm \ \AA} < \lambda \le 100 {\rm \ \AA}$).
This is because of the sensitivity of the X-ray emissivity to the temperature.
Also shown in the right panels of Figure \ref{fig:filling_factor_vs_XUV_photon_spectrum} are the spectra of the total photon number emitted from the star per unit time (photon number luminosity)
\revise{given by $ 4 \pi r^2 \lambda F_{\lambda}/(hc)$}.
For all $f_\ast$, the spectrum of the photon number luminosity is flatter than the energy spectrum, implying that the photon population with respect to wavelength is approximately uniform in the XUV range.

\begin{figure}[t!]
\centering
\includegraphics[width=80mm]{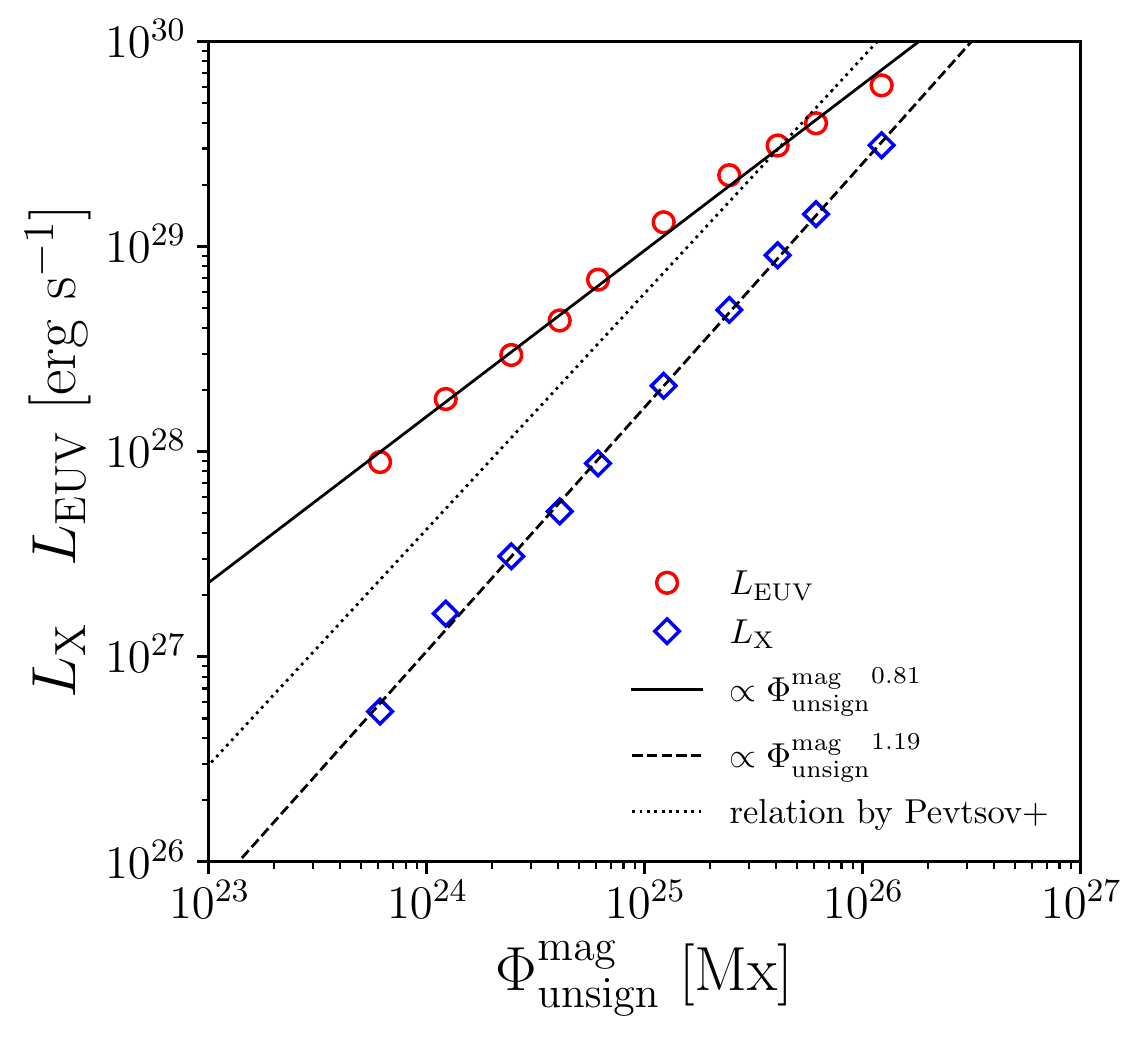}
\vspace{0.5em}
\caption{
Unsigned magnetic flux $\Phi_{\rm unsign}^{\rm \revise{mag}}$ versus X-ray luminosity $L_{\rm X}$ (blue diamonds) and EUV luminosity $L_{\rm EUV}$ (red circles).
Results for the fixed half loop length $l_{\rm loop}=20 {\rm \ Mm}$ are shown.
Solid and dashed lines are the power-law fittings to the simulation results.
Also shown by the dotted line is the observational relation by \citet{Pevtsov_2003_ApJ}.
}
\label{fig:Phi_vs_Lx_Leuv_L20}
\vspace{0em}
\end{figure}

The magnetic-flux dependencies of the X-ray luminosity $L_{\rm X}$ and EUV luminosity $L_{\rm EUV}$ are also investigated.
Figure \ref{fig:Phi_vs_Lx_Leuv_L20} reveals the variation in $L_{\rm X}$ (blue diamonds) and $L_{\rm EUV}$ (red circles) with the unsigned magnetic flux $\Phi_{\rm unsign}^{\rm \revise{mag}}$ defined by
\begin{align}
    {\Phi_{\rm unsign}^{\rm \revise{mag}}} = 4 \pi R_\odot^2 B_{s,\ast} f_\ast.
\end{align}
\revise{Because the photospheric magnetic field is fixed, the filling factor $f_\ast$ is the only variable in the definition of ${\Phi_{\rm unsign}^{\rm \revise{mag}}}$.}
We find that both $L_{\rm X}$ and $L_{\rm EUV}$ follow power laws with respect to ${\Phi_{\rm unsign}^{\rm \revise{mag}}}$ and are expressed as follows
\begin{align}
    &L_{\rm EUV} \propto {\Phi_{\rm unsign}^{\rm \revise{mag}}}^{0.81}, \label{eq:Leuv_scaling} \\
    &L_{\rm X} \propto {\Phi_{\rm unsign}^{\rm \revise{mag}}}^{1.19},
\end{align}
which are represented by the solid and dashed lines in Figure \ref{fig:Phi_vs_Lx_Leuv_L20}, respectively.
A scaling law is observationally discovered between the unsigned magnetic flux and the X-ray luminosity \citep{Pevtsov_2003_ApJ}
\begin{align}
    &L_{\rm X}^{\rm obs} \propto {\Phi_{\rm unsign}^{\rm \revise{mag}}}^{1.15},
\end{align}
which is shown by the dotted line in Figure \ref{fig:Phi_vs_Lx_Leuv_L20}.
Although the magnitudes of the X-ray luminosity from the simulation are smaller than the observed values, 
possibly because the proposed model reproduces the coronal properties in the activity minimum (a quiet, quasi-steady corona without flaring activities), 
the power-law index from the simulation (1.19) is nearly identical to that from the observation (1.15).
The reproduction of the observational ${\Phi_{\rm unsign}^{\rm \revise{mag}}}$--$L_{\rm X}$ relation provides another support to our model.
\revise{The proposed ${\Phi_{\rm unsign}^{\rm \revise{mag}}}$--$L_{\rm EUV}$ relationship, Eq. (\ref{eq:Leuv_scaling}), can be tested by the Sun-as-a-star observations in the EUV range \citep[e.g.][]{Toriumi_2020_ApJ}.}

\begin{figure}[t!]
\centering
\includegraphics[width=80mm]{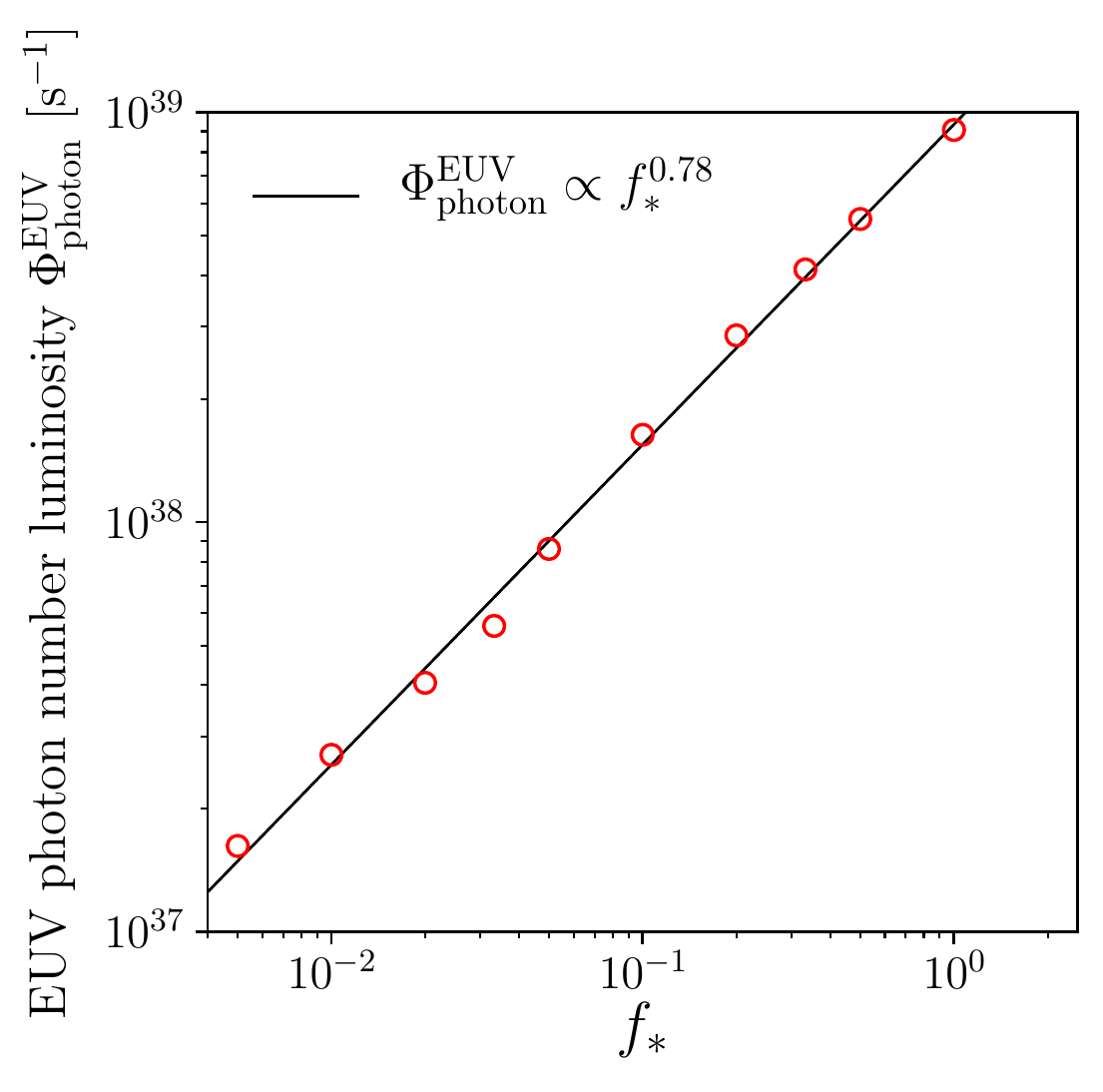}
\vspace{0.5em}
\caption{
Magnetic filling factor \revise{on the photosphere $f_\ast$} versus photon number luminosity in the EUV range.
Each simulation run corresponds to each red circle.
Also shown by the black solid line shows the power-law fitting to the simulation results.
}
\label{fig:filling_factor_vs_photon_luminosity}
\vspace{0em}
\end{figure}

The filling-factor dependence of the EUV photon number luminosity $\Phi_{\rm photon}^{\rm EUV}$ is also investigated.
$\Phi_{\rm photon}^{\rm EUV}$ is defined as the total number of EUV photons emitted from the star,
\revise{i.e.,
\begin{align}
    \Phi_{\rm photon}^{\rm EUV} = 4 \pi r^2 \int_{100 \ {\rm \AA}}^{912 \ {\rm \AA}} d\lambda \ \frac{\lambda F_{\lambda}}{hc},
\end{align}
which acts as 
}
a proxy of the photo-ionised hydrogen number per unit time.
Figure \ref{fig:filling_factor_vs_photon_luminosity} shows the EUV photon number luminosity with varying magnetic filling factors.
As with $L_{\rm EUV}$, $\Phi_{\rm photon}^{\rm EUV}$ also follows a power law with respect to $f_\ast$, which is given by
\begin{align}
    \Phi_{\rm photon}^{\rm EUV} \propto f_\ast^{\revise{0.78}} \propto {\Phi_{\rm unsign}^{\rm \revise{mag}}}^{\revise{0.78}}.
    \label{eq:phi_photon_euv_scaling}
\end{align}
A comparison of Eq.s (\ref{eq:Leuv_scaling}) and (\ref{eq:phi_photon_euv_scaling}), 
reveals that $L_{\rm EUV}$ and $\Phi_{\rm photon}^{\rm EUV}$ follow marginally different power laws with respect to $f_\ast$ (or equivalently $\Phi_{\rm unsign}^{\rm \revise{mag}}$).

\subsection{$L_{\rm X}$--$L_{\rm EUV}$ and $L_{\rm X}$--$\Phi^{\rm EUV}_{\rm photon}$ relations}

One of the objectives of this study is to propose a method to estimate the stellar EUV luminosity $L_{\rm EUV}$ from other observable quantities.
Because the X-ray luminosity $L_{\rm X}$ is measurable and should be correlated with $L_{\rm EUV}$ \citep[see e.g.][]{Sanz-Forcada_2011_AA}, the relationship between $L_{\rm X}$ and $L_{\rm EUV}$ over a wide range of filling factors and loop lengths warrants investigation.

\begin{figure}[t!]
\centering
\includegraphics[width=80mm]{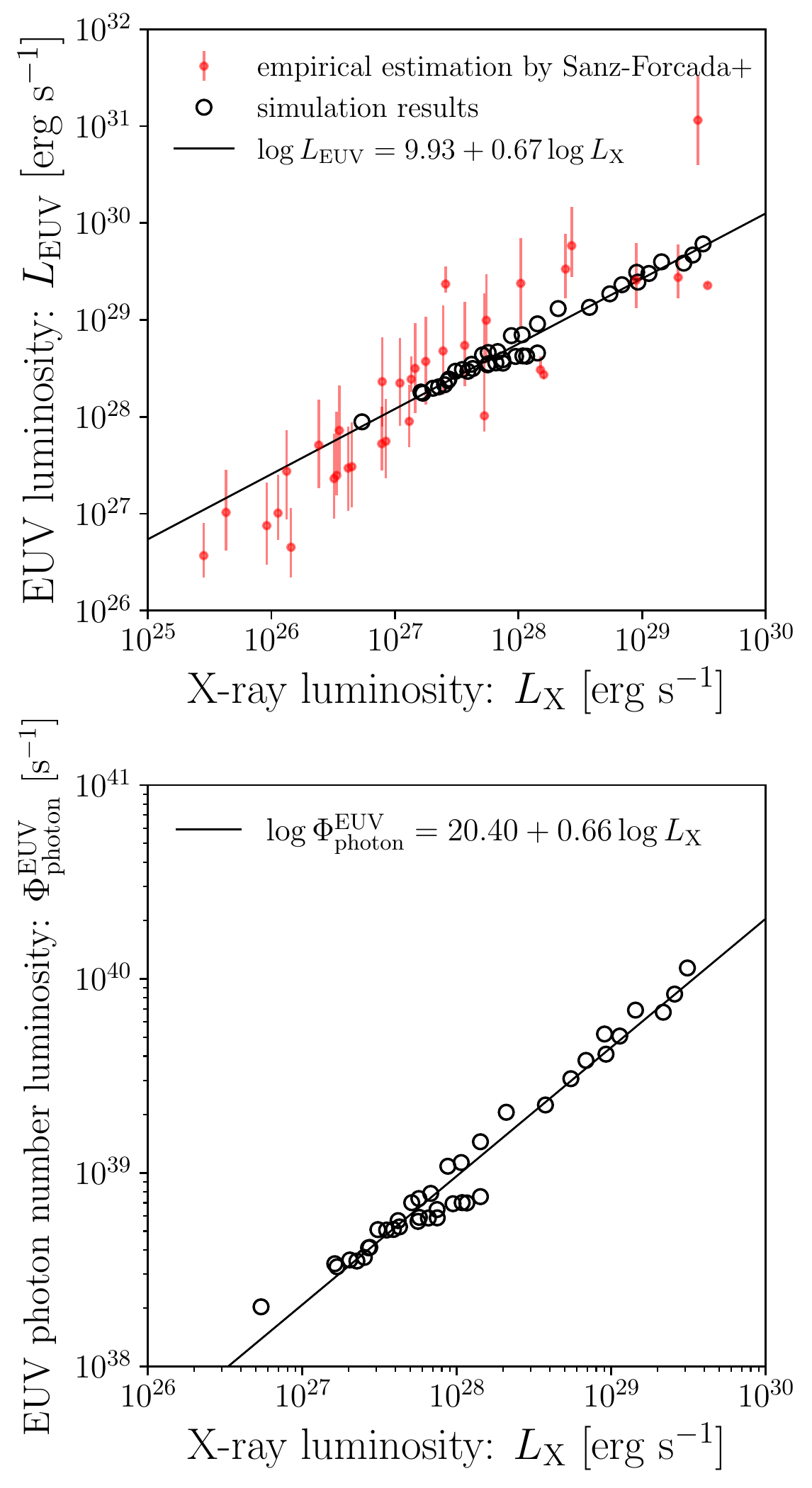}
\vspace{0.5em}
\caption{
Top: correlation between the X-ray luminosity $L_{\rm X}$ and EUV luminosity $L_{\rm EUV}$.
X-ray and EUV are defined in terms of wavelength $\lambda$ as $5 {\rm \ \AA} < \lambda \le 100 {\rm \ \AA}$ and $100 {\rm \ \AA} < \lambda \le 912 {\rm \ \AA}$, respectively.
Each black circle corresponds to the simulation run listed in Table \ref{table:run}, and the black solid line shows the power-law fitting to them.
Also shown by the red points with vertical error bars are the empirical estimation by \citet{Sanz-Forcada_2011_AA}.
Bottom: same as the top panel but for the X-ray luminosity $L_{\rm X}$ and photon number luminosity $\Phi^{\rm EUV}_{\rm photon}$.
}
\label{fig:Lx_vs_Leuv_all_run}
\vspace{0em}
\end{figure}

The top panel of Figure \ref{fig:Lx_vs_Leuv_all_run} highlights the correlation between X-ray luminosity $L_{\rm X}$ and EUV luminosity $L_{\rm EUV}$.
The black circles correspond to the simulation runs listed in Table \ref{table:run}.
Although the simulation runs are spread over wide ranges in the filling factor and loop length, a single power-law fit adequately applies to the $L_{\rm X}$--$L_{\rm EUV}$ relation, which is given by
\begin{align}
    \log L_{\rm EUV} = 9.93 + 0.67 \log L_{\rm X},
    \label{eq:Lx_Leuv_relation}
\end{align}
where $L_{\rm EUV}$ and $L_{\rm X}$ are measured in the unit of ${\rm erg \ s^{-1}}$.
The red points with the vertical error bars also indicate the observational estimation by \citet{Sanz-Forcada_2011_AA}.
The pure theoretical estimation of $L_{\rm EUV}$ in this study conforms with the pure observational estimation of the same by \citet{Sanz-Forcada_2011_AA},
\revise{Although the power-law index of $L_{\rm X}$--$L_{\rm EUV}$ relation is 0.86, according to \citet{Sanz-Forcada_2011_AA}}, the coefficient $0.67$ in Eq. (\ref{eq:Lx_Leuv_relation}) is, in fact, close to the observational value ($0.681$) derived from the Extreme Ultraviolet Explorer (EUVE) observations \citep{Johnstone_2021_AA}.
The fact that the independent methods produces a similar $L_{\rm X}$--$L_{\rm EUV}$ relation validates the proposed model of stellar coronae and their XUV emissions.

The correlation between the X-ray luminosity $L_{\rm X}$ and the EUV photon-number luminosity $\Phi_{\rm photon}^{\rm EUV}$ is also highlighted in the bottom panel of Figure \ref{fig:Lx_vs_Leuv_all_run}.
They, too, follow a power--law relationship given by
\begin{align}
    \log \Phi_{\rm photon}^{\rm EUV} = 20.40 + 0.66 \log L_{\rm X},
    \label{eq:Lx_Phieuv_relation}
\end{align}
where $\Phi_{\rm photon}^{\rm EUV}$ is measured in unit of ${\rm s^{-1}}$.

\section{Analytical arguments on coronal energy flux \label{sec:analytical_model}}

Several scaling relations with respect to the loop length and magnetic filling factor are provided in Section \ref{sec:simulation_result}.
In this section, we analytically describe the coronal energy injection process governing the scaling relations.
Because energy is mainly transported as transverse fluctuations of the velocity and magnetic field, i.e., Alfv\'en waves, the efficiency of Alfv\'en-wave transmission into the corona is key to the analysis.
This section is devoted to reviewing the linear theory of Alfv\'en-wave propagation in a stellar atmosphere and to provide analytical arguments for the origin of the scaling laws discovered through the simulation.

\subsection{Linear theory of Alfv\'en-wave propagation and reflection}

A significant amount of Alfv\'en waves excited at the photosphere is reflected before reaching the solar corona \citep{Cranmer_2005_ApJ,Verdini_2007_ApJ}.
The relationship between the Alfv\'en-wave transmission efficiency and the loop properties in the linear regime remains unclear and shall be discussed in this section.

\begin{figure}[t!]
\centering
\includegraphics[width=90mm]{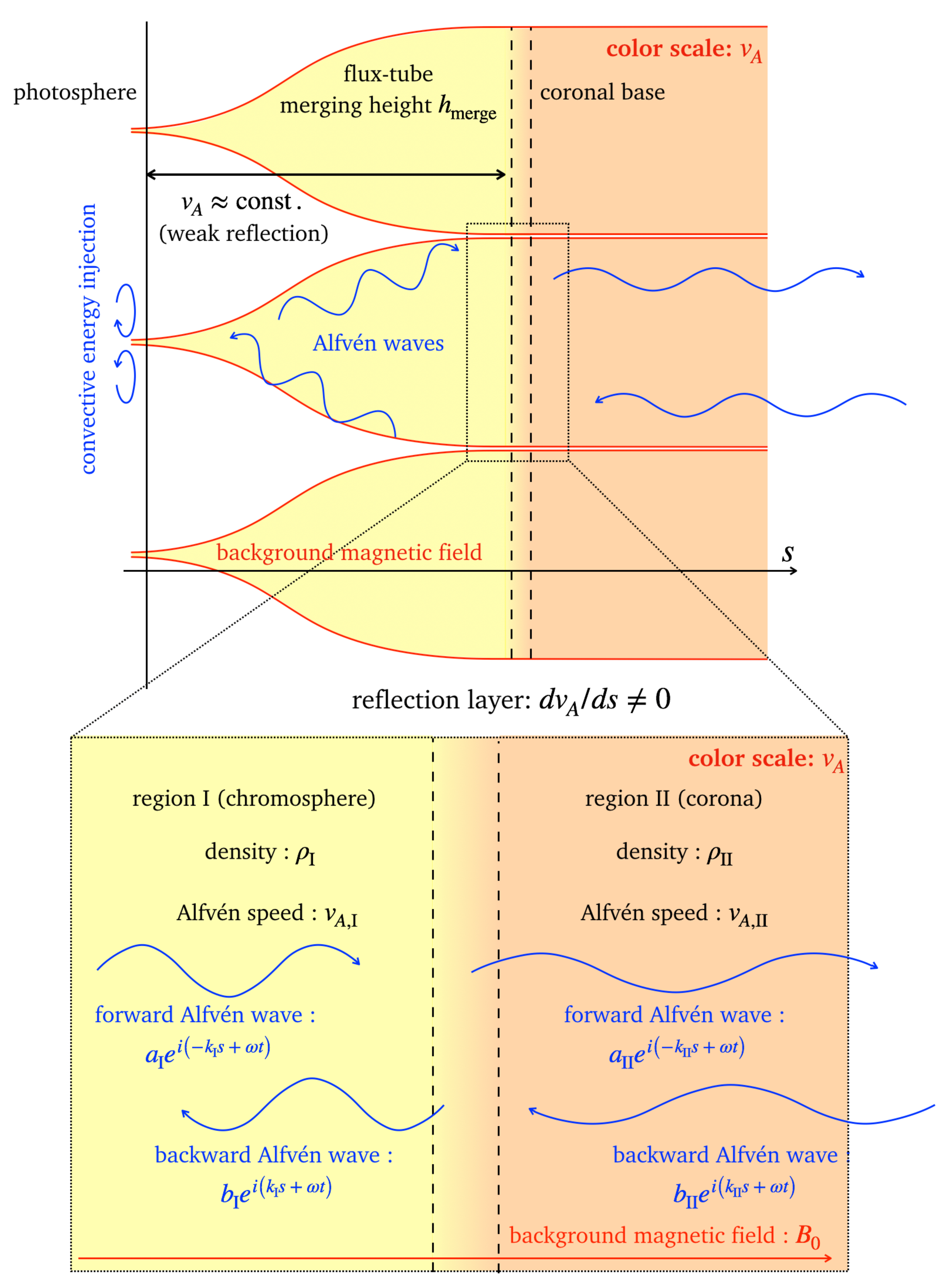}
\vspace{0.5em}
\caption{
A model of Alfv\'en-wave transmission across a free boundary with density jump.
}
\label{fig:Alfven_wave_transmission_schematic}
\vspace{0em}
\end{figure}

Figure \ref{fig:Alfven_wave_transmission_schematic} illustrates the schematic of the concerned system.
The convection--magnetic field interaction excites the Alfv\'en waves propagating upward along the expanding flux tube.
In our simulation setting, the Alfv\'en speed is maintained at a nearly constant value during the tube expansion (Eq.(\ref{eq:filling_factor_formulation})).
Because the Alfv\'en-wave reflection is triggered by a gradient in the Alfv\'en speed \citep{Stein_1971_ApJ,An_1989_ApJ,Vasquez_1990_ApJ},
the Alfv\'en waves reflected in the flux-tube expansion region are feeble.
Therefore, the bulk of the reflection occurs in a layer between the flux-tube merging height (where the flux-tube expansion ends) and the coronal base, which is in between the two vertical dashed lines in Figure \ref{fig:Alfven_wave_transmission_schematic}.

The Alfv\'en waves travelling through the reflection layer is considered in the following analysis.
For simplicity, the thickness of the layer is disregarded.
This approximation is validated by the fact that the reflection-layer thickness ($\sim 10^0 \ {\rm Mm}$) is notably smaller than the typical wavelength of Alfv\'en waves therein ($\sim 10^{1-2} \ {\rm Mm}$).
Then, the problem is simplified as a linear-wave transmission through a discontinuity, as described in the lower part of Figure \ref{fig:Alfven_wave_transmission_schematic}.
Hereinafter, we denote the variables in the region on the left/right of the discontinuity as region I/II, respectively.

The equation describing the Alfv\'en-wave propagation is 
\begin{align}
    \left[ 
    \frac{\partial^2}{\partial t^2} - v_{A,{\rm I/II}}^2 \frac{\partial^2}{\partial s^2} \right] \delta v_{\rm I/II} = 0,
\end{align}
the solution of which is given by
\begin{align}
    \delta v_{\rm I/II}= a_{\rm I/II} e^{i \left( - k_{\rm I/II} s + \omega t \right)} + b_{\rm I/II} e^{i \left( k_{\rm I/II} s + \omega t \right)}.
\end{align}
Note that $a_{\rm I/II}$ ($b_{\rm I/II}$) denotes the amplitudes of the upward (downward) propagating Alfv\'en wave in region I/II, respectively.
Given the frequency $\omega$, the dispersion relation yields the wavenumber $k_{\rm I/II}$ in each region.
\begin{align}
    k_{\rm I/II} = \omega / v_{A,{\rm I/II}}.
\end{align}
The corresponding magnetic-field fluctuation is deduced from the linearised induction equation, that is,
\begin{align}
    \frac{\partial}{\partial t} \delta B_{\rm I/II} = - B_0 \frac{\partial}{\partial s} \delta v_{\rm I/II}.
\end{align}
Because the fluctuations of the velocity and magnetic field are required to be continuous across the boundary, the amplitudes must satisfy the following relations.
\begin{align}
    &a_{\rm I} + b_{\rm I} = a_{\rm II} + b_{\rm II}, \\
    &- k_{\rm I} a_{\rm I} + k_{\rm I} b_{\rm I} = - k_{\rm II} a_{\rm II} + k_{\rm II} b_{\rm II}.
\end{align}
Solving these two equations yield
\begin{align}
    a_{\rm I} =  \frac{k_{\rm I} + k_{\rm II}}{2 k_{\rm I}} a_{\rm II} + \frac{k_{\rm I} - k_{\rm II}}{2 k_{\rm I}} b_{\rm II}.
    \label{eq:alfven_wave_coefficient}
\end{align}

The energy transmission rate $T_{\rm AW}$ is defined as the ratio of the forward-propagating Alfv\'en-wave energy in regions I and II. Namely,
\begin{align}
    T_{\rm AW} = \frac{\rho_{\rm II} v_{A,{\rm II}} \left| a_{\rm II} \right|^2}{\rho_{\rm I} v_{A,{\rm I}} \left| a_{\rm I} \right|^2}.
    \label{eq:alfven_wave_transmission}
\end{align}
We obtain a general form of the energy transmission rate within the limit of $\rho_{\rm II}/\rho_{\rm I} \ll 1$ by combining Eqs. (\ref{eq:alfven_wave_coefficient}) and (\ref{eq:alfven_wave_transmission}).
\begin{align}
    T_{\rm AW} =
    \frac{4 v_{A,{\rm I}} \left| a_{\rm II} \right|^2}{ v_{A,{\rm II}} \left( \left| a_{\rm II} \right|^2 + a_{\rm II} b_{\rm II}^\ast + a_{\rm II}^\ast b_{\rm II} + \left| b_{\rm II} \right|^2 \right)},
\end{align}
where $X^\ast$ denotes the complex conjugate of $X$.
It is interesting to see that the energy transmission rate depends on the values of $a_{\rm II}$ and $b_{\rm II}$, i.e., the wave population in region II (corona).

The transmission rate in the two limiting cases are investigated below.
When $b_{\rm II} = 0$ (no Alfv\'en waves propagating toward the chromosphere), the energy transmission rate is given by \citep{Hollweg_1984_SolPhys}
\begin{align}
    T_{\rm AW} = 4 \frac{v_{A,{\rm I}}}{v_{A,{\rm II}}} = 4 \sqrt{\frac{\rho_{\rm II}}{\rho_{\rm I}}}.
\end{align}
The other limiting case is $b_{\rm II} = a_{\rm II}$ (forward and reflected Alfv\'en waves are equally abundant in the corona), where the transmission rate is given by
\begin{align}
    T_{\rm AW} = \frac{v_{A,{\rm I}}}{v_{A,{\rm II}}} = \sqrt{\frac{\rho_{\rm II}}{\rho_{\rm I}}}.
\end{align}

To summarise, the transmission rate of Alfv\'en wave is given by
\begin{align}
    T_{\rm AW} = c_{\rm ref} \sqrt{\frac{\rho_{\rm II}}{\rho_{\rm I}}},  \ \ \ \ 1 \le c_{\rm ref} \le 4 \label{eq:transmission_rate}
\end{align}
where the value of $c_{\rm ref}$ depends on the population ratio of the forward and backward Alfv\'en waves in region II.

In the numerical model implemented in this study,
the density in the region I should be the mass density at the flux-tube merging height $s = h_{\rm merge}$.
In the flux-tube model employed in this model, $h_{\rm merge}$ is expressed as
\begin{align}
    f_\ast \exp \left( \frac{h_{\rm merge}}{H_{\rm mag}} \right) = 1, \ \ \ \ H_{\rm mag} = 2.5 H_\ast,
\end{align}
where $H_\ast$ is the density scale height at the photosphere (see Eq. (\ref{eq:scale_height_photosphere}) for definition).
Assuming that the density scale height is uniform in $0 \le s \le h_{\rm merge}$, 
we can write the mass density at the merging height $\rho_{\rm merge}$ as
\begin{align}
    \rho_{\rm merge} \approx \rho_\ast \exp \left( - \frac{h_{\rm merge}}{H_\ast} \right) = \rho_\ast f_\ast^{2/5}.
\end{align}
Then, using Eq. (\ref{eq:alfven_wave_transmission}), we obtain the Alfv\'en-wave transmission rate for our model as 
\begin{align}
    T_{\rm AW} \approx c_{\rm ref} \sqrt{\frac{\rho_{\rm cor}}{\rho_{\rm merge}}} = c_{\rm ref} \sqrt{\frac{\rho_{\rm cor}}{\rho_\ast}} f_\ast^{-1/5}, \label{eq:transmission_rate_this_model}
\end{align}
where $\rho_{\rm cor}$ denotes the mass density at the coronal base.
Note that a large $f_\ast$ could lead to a small transmission rate, when $\rho_{\rm cor}$ does not increase faster than $f_\ast^{2/5}$.

\subsection{Loop-length dependence of the coronal energy flux}

In Section \ref{sec:energetics_loop_length},
it is shown that the coronal energy flux increases with $l_{\rm loop}$ but saturates at approximately four times the value of the short-loop limit (see Figure \ref{fig:loop_length_vs_conductive_flux_fexp100}).
This behaviour is naturally explained by the theory of Alfv\'en-wave transmission.

The bottom panel of Figure \ref{fig:loop_length_vs_loop_properties_fexp100} shows that the mass density of the coronal base remains mostly constant with respect to the loop length.
Then, according to Eq. (\ref{eq:transmission_rate_this_model}),
the change in the coronal energy flux is attributed to the change in the value of $c_{\rm ref}$.
Because $1 \le c_{\rm ref} \le 4$, 
the coronal energy flux should saturate at approximately four times the basal value, as observed in Figure \ref{fig:loop_length_vs_conductive_flux_fexp100}.

The enhancement of the coronal energy flux at $l_{\rm loop}^{\rm eff} \approx 10^2 {\rm \ Mm}$ can be explained in terms of the timescales of Alfv\'en-wave propagation and dissipation.
The backward waves in the corona are originated from the other base of the corona. 
The backward waves are subject to the turbulent dissipation during their propagation, which results in a smaller amplitude at the reflection layer than the amplitude of the forward waves.
The population ratio, $\left| \vec{z}^+_\perp \right|^2/\left| \vec{z}^-_\perp \right|^2$, should thus depends on the ratio of the Alfv\'en-crossing time $\tau_A$ to the turbulent dissipation time $\tau_{\rm diss}$, which are respectively defined as
\begin{align}
    \tau_A = \int_{s_{\rm TR}}^{s_{\rm TR} + 2l_{\rm loop}^{\rm eff}} ds/v_A, \ \ \ 
    \tau_{\rm diss} = \frac{1}{2 l_{\rm loop}^{\rm eff}} \int_{s_{\rm TR}}^{s_{\rm TR} + 2l_{\rm loop}^{\rm eff}} \frac{\lambda_\perp}{c_d z_{\perp,{\rm rms}}^+} ds. \label{eq:Alfven_wave_timescale}
\end{align}
Specifically,
\begin{enumerate}
    \item when $\tau_A/\tau_{\rm diss} < 1$,
    the coronal Alfv\'en waves propagating from one coronal base to the other hardly experience any turbulent dissipation.
    Therefore, the population ratio $\left| \vec{z}^+_\perp \right|^2/\left| \vec{z}^-_\perp \right|^2$ should tend to unity in this case. \\[-1.0em]
    \item When $\tau_A/\tau_{\rm diss} > 1$,
    the coronal Alfv\'en waves significantly dissipate during the propagation through the corona.
    Thus, the population ratio $\left| \vec{z}^+_\perp \right|^2/\left| \vec{z}^-_\perp \right|^2$ should be far from unity in this case.
\end{enumerate}

\begin{figure}[t!]
\centering
\includegraphics[width=80mm]{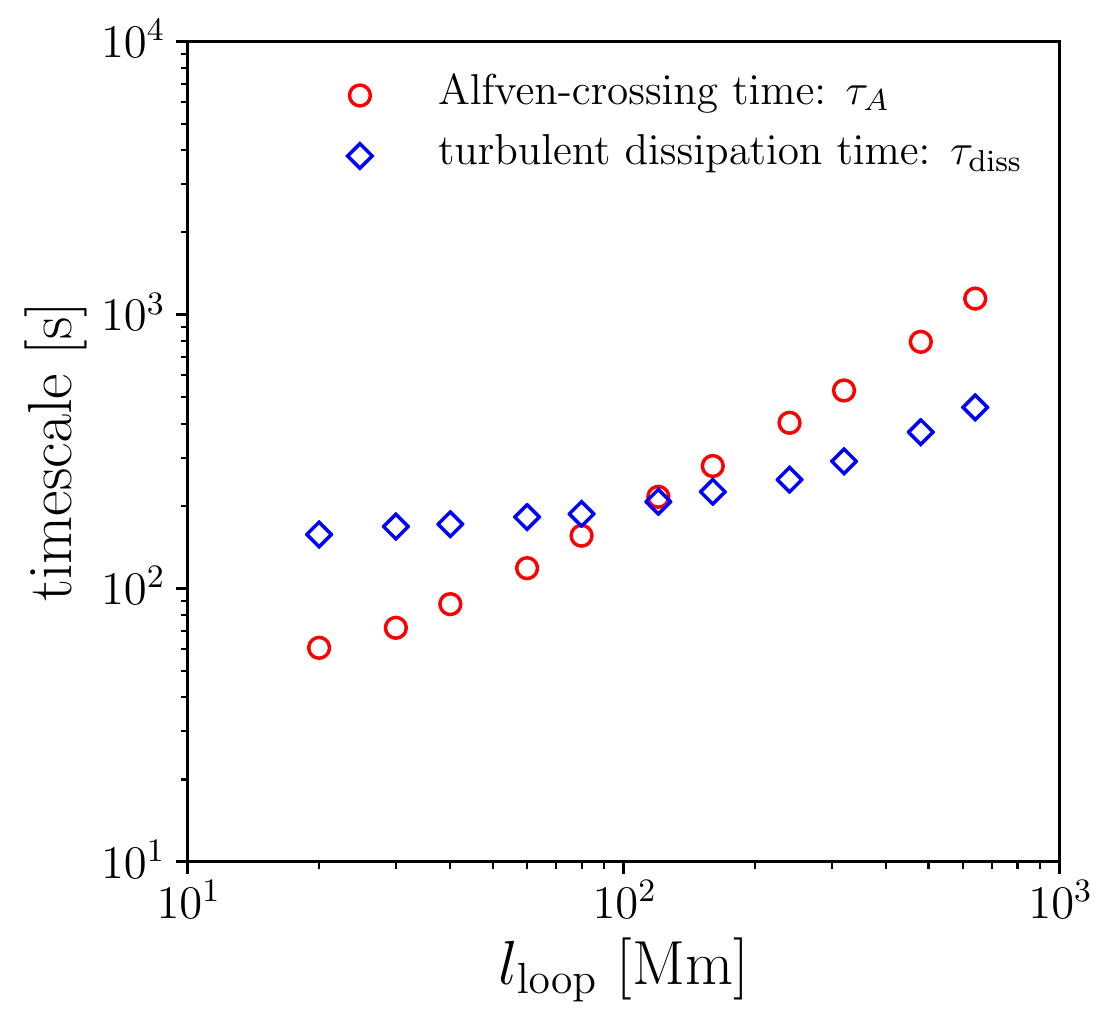}
\vspace{0.5em}
\caption{
Half loop length versus timescale of Alfv\'en-wave crossing ($\tau_A$ red circles) and Alfv\'en-wave turbulent dissipation ($\tau_{\rm diss}$, blue diamonds).
For the definitions of $\tau_A$ and $\tau_{\rm diss}$, see Eq. (\ref{eq:Alfven_wave_timescale}).
}
\label{fig:loop_length_vs_Alfven_timescales}
\vspace{0em}
\end{figure} 

This argument is substantiated by Figure \ref{fig:loop_length_vs_Alfven_timescales}, in which $\tau_A$ (red circles) and $\tau_{\rm diss}$ (blue diamonds) are plotted against the half-loop length $l_{\rm loop}$.
The Alfv\'en crossing time is shorter than the turbulent dissipation time in $l_{\rm loop} < 10^2 {\rm \ Mm}$ and vice versa in $l_{\rm loop} > 10^2 {\rm \ Mm}$.
Therefore, it is likely that, when $l_{\rm loop} < 10^2 {\rm \ Mm}$, the weak turbulent dissipation yields $\left| \vec{z}^+_\perp \right|^2/\left| \vec{z}^-_\perp \right|^2 \sim 1$ and $c_{\rm ref} \approx 1$.
In the opposite limiting case of $l_{\rm loop} > 10^2 {\rm \ Mm}$, the backward Alfv\'en waves significantly decay in the coronal loop, resulting in $\left| \vec{z}^+_\perp \right|^2/\left| \vec{z}^-_\perp \right|^2 \ll 1$ and $c_{\rm ref} \approx 4$.
These arguments underscore the importance of the timescale of turbulent dissipation in a coronal loop.

\subsection{Filling-factor dependence of the coronal energy flux}

In Section \ref{sec:ff_vs_energetics},
the coronal energy flux is found to follow a power law as a function of $f_\ast$.
\begin{align}
    F_{\rm cor} \propto f_\ast^{0.91},
\end{align}
or alternatively
\begin{align}
    F_{\rm cor} l_{\rm loop}^{\rm eff} \propto f_\ast^{0.95}. \label{eq:scaling_fcorl2}
\end{align}
These power-law relations are explained by the combined effect of the RTV scaling law and the energy transmission rate Eq.(\ref{eq:transmission_rate_this_model}).
In the absence of the Alfv\'en-wave dissipation in the chromosphere, 
the energy flux injected into the corona is given by
\begin{align}
    F_{\rm cor} = F_{A,\ast} f_\ast T_{\rm AW} \propto c_{\rm ref} \rho_{\rm cor}^{1/2} f_\ast^{4/5}, 
    \label{eq:coronal_energy_flux_analytical}
\end{align}
where $F_{A,\ast}$ is the upward Alfv\'en-wave energy flux at the footpoint of the flux tube,
which is fixed for all the simulation runs.
According to the RTV scaling law, 
the density and the coronal energy flux are related by
\begin{align}
    \rho_{\rm cor} \propto F_{\rm cor}^{4/7} {l_{\rm loop}^{\rm eff}}^{-3/7}.
    \label{eq:RTV_scaling_law_density}
\end{align}
Eq.s (\ref{eq:coronal_energy_flux_analytical}) and (\ref{eq:RTV_scaling_law_density}) yield
\begin{align}
    F_{\rm cor} l_{\rm loop}^{\rm eff} \propto
    c_{\rm ref} {l_{\rm loop}^{\rm eff}}^{7/10} f_\ast^{28/25} \propto c_{\rm ref} f_\ast^{1.15},
\end{align}
where we use $l_{\rm loop}^{\rm eff} \propto f_\ast^{0.04}$.
As addressed in the previous section, the value of $c_{\rm ref}$ depends on the balance between the Alfv\'en-crossing timescale $\tau_A$ and the turbulent dissipation timescale $\tau_{\rm diss}$ in the coronal loop.
Given that a larger $f_\ast$ yields a smaller $\tau_A/\tau_{\rm diff}$ (as well as a smaller $c_{\rm ref}$),
the predicted dependence of $c_{\rm ref}$ on $f_\ast$ is
\begin{align}
    c_{\rm ref} \propto f_\ast^{-\alpha_{\rm ref}},
    \label{eq:scaling_cref_filling_factor}
\end{align}
where $\alpha_{\rm ref}>0$.
Given the limited range of $c_{\rm ref}$ ($1 \le c_{\rm ref} \le 4$), the power index $\alpha_{\rm ref}$ is small.
Then, the semi-analytical scaling relation,
\begin{align}
    F_{\rm cor} l_{\rm loop}^{\rm eff} \propto f_\ast^{1.15-\alpha_{\rm ref}},
\end{align}
conforms with the obtained scaling relation Eq. (\ref{eq:scaling_fcorl2}).

\section{Discussion \label{sec:discussion}}

\subsection{Inferred ${\rm Ro}$--$L_{\rm X}$ relation}

\begin{figure}[t!]
\centering
\includegraphics[width=80mm]{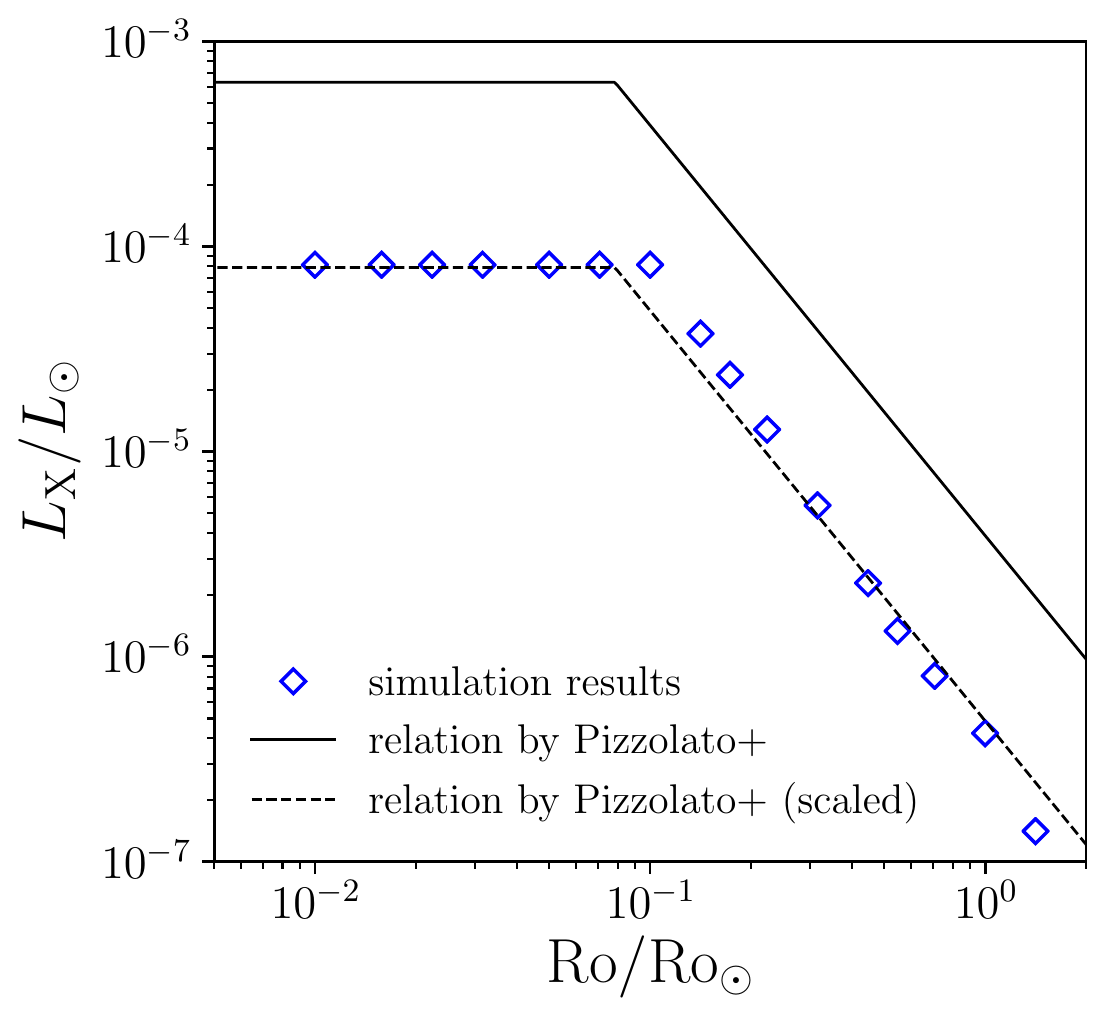}
\vspace{0.5em}
\caption{
Rossby number versus normalized X-ray luminosity (blue diamonds).
The Rossby number and the filling factor are connected by Eq. (\ref{eq:ff_Ro_relation}).
Results for the fixed half loop length $l_{\rm loop}=20 {\rm \ Mm}$ are shown.
Also shown by the solid and dashed lines are the observational relation by \citet{Pizzolato_2003_AA} and its scaled-down version (1/8).
}
\label{fig:Ro_vs_Lx_L20}
\vspace{0em}
\end{figure}

The X-ray behaviour with respect to the stellar Rossby number has been established in the literature \citep{Pizzolato_2003_AA,Wright_2011_ApJ,Wright_2016_Nature,Wright_2018_MNRAS,Magaudda_2020_AA}.
By prescribing a relation between the filling factor $f_\ast$ and the Rossby number ${\rm Ro}$, we compare the model prediction with the observational relation.
The magnetic filling factor follows a power-law relation with respect to the Rossby number \citep{Saar_1996_proceedings,Saar_2001_proceedings,Reiners_2009_ApJ}.
Given that the magnetic activity saturates at ${\rm Ro}/{\rm Ro}_\odot \approx 0.1$ and $f_\odot \approx 0.01$, the power--law relationship is assumed as follows.
\begin{align}
    f_\ast = \min \left[1, \ f_\odot \left( \frac{{\rm Ro}}{{\rm Ro}_\odot} \right)^{-2} \right]. \label{eq:ff_Ro_relation}
\end{align}

Figure \ref{fig:Ro_vs_Lx_L20} shows the relation between ${\rm Ro}/{\rm Ro}_\odot$ given by Eq. (\ref{eq:ff_Ro_relation}) and $L_{\rm X}/L_\odot$.
The blue diamonds represent the simulation results, and the black solid line represents the observational relation observed in Sun-like stars \citep{Pizzolato_2003_AA}.
Although a gap exists between the simulation results and the observational relation, when the observational relation is scaled down by a factor of 8 (represented by the dashed line), the simulation results conform with the observations.
Given that the observational data points have a large scatter (an order of magnitude in $L_{\rm X}/L_\odot$), and that the proposed model corresponds to the activity minimum where the X-ray luminosity is 10--20 times smaller than the activity maximum \citep{Johnstone_2015_AA},
the simulation results adequately corroborate with the observations.

\subsection{Missing physics and limitations}

Although our model self-consistently solves the physical processes of energy injection into the corona, turbulent coronal heating, and thermal responses of the coronal loop, 
several other physical processes have not been considered.
\begin{enumerate}
    \item The impulsive nature of the coronal heating is disregarded.
    The phenomenological turbulent dissipation implemented in this study yields the time-averaged heating rate, whereas the actual coronal heating is highly intermittent in space and time, as indicated by previous simulation studies \citep{Einaudi_1996_ApJ,Cassak_2008_ApJ,Dahlburg_2016_ApJ,Kanella_2018_AA} and observations \citep{Cirtain_2013_Nature,Testa_2014_Science,Ishikawa_2017_NatAs,Antolin_2021_NatAs}.
    In the presence of localised heating, the DEM extends to the high-temperature side, possibly forming the bi-modal structure observed in active stellar coronae \citep{Gudel_1997_ApJ}.
    The accuracy of the DEM obtained in this 1D simulation must be tested in the future.
    \\[-1.0em]
    \item The coronal loop is assumed to be isolated from the ambient plasma. However, previous studies have stated that the interface between the coronal flux tube and its surrounding is a preferential region for heating by phase mixing and resonant absorption \citep{Ionson_1978_ApJ,Heyvaerts_1983_AA,Hollweg_1988_JGR,Sakurai_1991_SolPhys,Hood_2002_RSPSA,Pagano_2019_AA,Van_Damme_2020_AA}.
    In fact, resonant absorption is observed to occur in the solar corona \citep{Antolin_2015_ApJ,Okamoto_2015_ApJ}.
    Although there is no scientific consensus on the ability of phase mixing and/or resonant absorption to feed sufficient thermal energy to the corona, the relation between these processes and XUV emissions must be investigated.
    \\[-1.0em]
    \item Energy injection by the emergence of magnetic field from the stellar interior \citep[magnetic flux emergence,][]{Takasao_2013_PASJ} was dismissed. 
    Recent observations of the solar corona indicate that the coronal loops are preferentially heated near the bottom \citep{Berghmans_2021_arXiv}, possibly by its interaction (magnetic reconnection) with adjacent small-scale loops \citep{Chen_2021_arXiv}.
    Because the (small-scale) coronal loops are formed by the magnetic flux emergence, these observations imply that a fraction of the coronal energy must be supplied by the emergence.
    Indeed, a recent work on the solar wind indicates the considerable role of magnetic flux emergence in the injection of energy to the open-field regions \citep{Wang_2020_ApJ}.
    \\[-1.0em]
    \item Throughout this study, the energy spectrum is reconstructed from EMD (DEM) under the optically thin approximation.
    Although this approximation has been widely used \citep{Sanz-Forcada_2011_AA,Duvvuri_2021_arXiv}, it must be tested in future under a realistic treatment of radiative transfer.
    For example, the Lyman continuum appears to be optically thick, with an extended source region in the upper chromosphere \citep{Avrett_2008_ApJ}.
    Other studies have claimed that a simple optically thin treatment of the coronal lines may produce incorrect results \citep{Schrijver_1994_AA}.
    To solve this issue, the frequency-dependent radiative transfer equation must be solved in the future.
    \\[-1.0em]
    \item In converting the single-loop properties retrieved from the simulation results to a global quantity such as $L_{\rm X}$, we assume that the stellar corona has a uniform brightness.
    The brightness of the actual stellar coronae, however, may be highly structured, as seen in the solar corona.
    Because emissions from a single active region depends on its size, the size distribution of active regions over the entire stellar surface affects the measured distribution of the total coronal emissions from the star \citep{Takasao_2020_ApJ}.
    The effect of coronal structuring must be considered in the future by, for instance, the global modelling of stellar coronae \citep{van_der_Holst_2014_ApJ,Alvarado-Gomez_2016_AA,Oran_2017_ApJ,Airapetian_2021_arXiv}.
\end{enumerate}

\subsection{Extension to other types of stars}

In this study, we focused on the main-sequence Sun-like stars that exhibit solar mass, solar radius, solar luminosity, and solar metallicity.
Therefore, the application of Eq. (\ref{eq:Lx_Leuv_relation}) or Eq. (\ref{eq:Lx_Phieuv_relation}) to other classes of stars would require additional considerations.
\begin{enumerate}
    \item M-type stars are of particular interest in the context of exoplanet because they occupy a large fraction of the local stellar population \citep{Bochanski_2010_AJ,Winters_2019_AJ} and often host terrestrial planets within the habitable zones \citep{Bonfils_2013_AA,Dressing_2015_ApJ}.
    The applicability of the proposed theoretical predictions to lower-mass stars (especially, M dwarfs) must be investigated in the future.
    Given the lower metallicity of old M-type stars, the dependence of XUV emissions on metallicity must also be studied \citep{Washinoue_2019_ApJ}.
    According to Figure 14 in \citet{Johnstone_2021_AA}, a single power law between $L_{\rm X}$ and $L_{\rm EUV}$ could be found in low-mass stars with various spectral types, and thus Eqs. (\ref{eq:Lx_Leuv_relation}) and (\ref{eq:Lx_Phieuv_relation}) may be directly applicable to lower-mass stars.
    \\[-1.0em]
    \item The XUV emissions of pre-main-sequence stars significantly alter the evolution of proto-planetary discs and planet formation \citep{Gorti_2009_ApJ,Nakatani_2018_ApJ,Wang_2019_ApJ}.
    In the case of pre-main-sequence star accretion (classical T-Tauri stars), two effects of accretion on XUV emissions must be considered.
    First, the XUV emissions from the accretion shocks could be comparable to or even greater than those from the corona \citep{Gunther_2007_AA}.
    Second, because the accretion to the stellar surface may excite velocity disturbances on the photosphere, the Alfv\'en-wave energy injected into the corona is likely to be enhanced \citep{Cranmer_2008_ApJ,Cranmer_2009_ApJ}.
    A model that considers these accretion effects is required to theoretically predict the XUV emissions from pre-main-sequence stars.
\end{enumerate}

\section{Conclusion \label{sec:conclusion}}

In this study, we theoretically investigate the coronal properties of the Sun and Sun-like stars.
The simulation domain extends from the solar surface to the corona, with energy transport from the surface and dissipation in the corona being solved in a self-consistent manner.
The thermal structuring of the stellar atmosphere is also reproduced by implementing thermal conduction and radiative cooling.
The behaviours of the coronal properties with respect to the loop length and magnetic filling factor are explained by a combination of the behaviour of the coronal energy flux and the RTV scaling law.
With the parameter survey, we reproduce the nearly linear relation between the unsigned magnetic flux and the X-ray luminosity.
Scaling relations between the EUV properties (luminosity and photon number luminosity) and the X-ray luminosity are discovered, which will be useful in constraining the stellar EUV parameters from the observable quantity $L_{\rm X}$.

The authors thank Drs. Haruhisa Iijima, Kosuke Namekata, and Riouhei Nakatani for valuable comments.
Numerical computations were carried out on the Cray XC50 at the Center for Computational Astrophysics (CfCA), National Astronomical Observatory of Japan.
M.S. is supported by a Grant-in-Aid for Japan Society for the Promotion of Science (JSPS) Fellows and by the NINS program for cross-disciplinary study (grant Nos. 01321802 and 01311904) on Turbulence, Transport, and Heating Dynamics in Laboratory and Solar/ Astrophysical Plasmas: “SoLaBo-X.” 
S.T. is supported by JSPS KAKENHI grant Nos. JP18K13579 and JP21H04487.
This work made use of matplotlib, a Python library for publication quality graphics \citep{Hunter_2007_CSE}, and NumPy \citep{van_der_Walt_2011_CSE}.

\bibliographystyle{aa}

\begin{thebibliography}{185}
\expandafter\ifx\csname natexlab\endcsname\relax\def\natexlab#1{#1}\fi

\bibitem[{{Airapetian} {et~al.}(2017){Airapetian}, {Glocer}, {Khazanov},
  {Loyd}, {France}, {Sojka}, {Danchi}, \& {Liemohn}}]{Airapetian_2017_ApJ}
{Airapetian}, V.~S., {Glocer}, A., {Khazanov}, G.~V., {et~al.} 2017, \apjl,
  836, L3

\bibitem[{{Airapetian} {et~al.}(2021){Airapetian}, {Jin}, {Lueftinger}, {Boro
  Saikia}, {Kochukhov}, {Guedel}, {Van Der Holst}, \&
  {Manchester}}]{Airapetian_2021_arXiv}
{Airapetian}, V.~S., {Jin}, M., {Lueftinger}, T., {et~al.} 2021, arXiv
  e-prints, arXiv:2106.01284

\bibitem[{{Alfv{\'e}n}(1947)}]{Alfven_1947_MNRAS}
{Alfv{\'e}n}, H. 1947, \mnras, 107, 211

\bibitem[{{Alvarado-G{\'o}mez} {et~al.}(2016){Alvarado-G{\'o}mez}, {Hussain},
  {Cohen}, {Drake}, {Garraffo}, {Grunhut}, \&
  {Gombosi}}]{Alvarado-Gomez_2016_AA}
{Alvarado-G{\'o}mez}, J.~D., {Hussain}, G.~A.~J., {Cohen}, O., {et~al.} 2016,
  \aap, 588, A28

\bibitem[{{An} {et~al.}(1989){An}, {Musielak}, {Moore}, \&
  {Suess}}]{An_1989_ApJ}
{An}, C.~H., {Musielak}, Z.~E., {Moore}, R.~L., \& {Suess}, S.~T. 1989, \apj,
  345, 597

\bibitem[{{Antiochos} {et~al.}(1999){Antiochos}, {MacNeice}, {Spicer}, \&
  {Klimchuk}}]{Antiochos_1999_ApJ}
{Antiochos}, S.~K., {MacNeice}, P.~J., {Spicer}, D.~S., \& {Klimchuk}, J.~A.
  1999, \apj, 512, 985

\bibitem[{{Antiochos} \& {Sturrock}(1978)}]{Antiochos_1978_ApJ}
{Antiochos}, S.~K. \& {Sturrock}, P.~A. 1978, \apj, 220, 1137

\bibitem[{{Antolin} {et~al.}(2015){Antolin}, {Okamoto}, {De Pontieu},
  {Uitenbroek}, {Van Doorsselaere}, \& {Yokoyama}}]{Antolin_2015_ApJ}
{Antolin}, P., {Okamoto}, T.~J., {De Pontieu}, B., {et~al.} 2015, \apj, 809, 72

\bibitem[{{Antolin} {et~al.}(2021){Antolin}, {Pagano}, {Testa}, {Petralia}, \&
  {Reale}}]{Antolin_2021_NatAs}
{Antolin}, P., {Pagano}, P., {Testa}, P., {Petralia}, A., \& {Reale}, F. 2021,
  Nature Astronomy, 5, 54

\bibitem[{{Antolin} \& {Shibata}(2010)}]{Antolin_2010_ApJ}
{Antolin}, P. \& {Shibata}, K. 2010, \apj, 712, 494

\bibitem[{{Aschwanden} \& {Parnell}(2002)}]{Aschwanden_2002_ApJ}
{Aschwanden}, M.~J. \& {Parnell}, C.~E. 2002, \apj, 572, 1048

\bibitem[{{Avrett} \& {Loeser}(2008)}]{Avrett_2008_ApJ}
{Avrett}, E.~H. \& {Loeser}, R. 2008, \apjs, 175, 229

\bibitem[{{Barnes}(2003)}]{Barnes_2003_ApJ}
{Barnes}, S.~A. 2003, \apj, 586, 464

\bibitem[{{Berghmans} {et~al.}(2021){Berghmans}, {Auch{\`e}re}, {Long},
  {Soubri{\'e}}, {Zhukov}, {Mierla}, {Sch{\"u}hle}, {Antolin}, {Parenti},
  {Harra}, {Podladchikova}, {Aznar Cuadrado}, {Buchlin}, {Dolla}, {Verbeeck},
  {Gissot}, {Teriaca}, {Haberreiter}, {Katsiyannis}, {Rodriguez}, {Kraaikamp},
  {Smith}, {Stegen}, {Rochus}, {Halain}, {Jacques}, {Thompson}, \&
  {Inhester}}]{Berghmans_2021_arXiv}
{Berghmans}, D., {Auch{\`e}re}, F., {Long}, D.~M., {et~al.} 2021, arXiv
  e-prints, arXiv:2104.03382

\bibitem[{{Bochanski} {et~al.}(2010){Bochanski}, {Hawley}, {Covey}, {West},
  {Reid}, {Golimowski}, \& {Ivezi{\'c}}}]{Bochanski_2010_AJ}
{Bochanski}, J.~J., {Hawley}, S.~L., {Covey}, K.~R., {et~al.} 2010, \aj, 139,
  2679

\bibitem[{{Boerner} {et~al.}(2012){Boerner}, {Edwards}, {Lemen}, {Rausch},
  {Schrijver}, {Shine}, {Shing}, {Stern}, {Tarbell}, {Title}, {Wolfson},
  {Soufli}, {Spiller}, {Gullikson}, {McKenzie}, {Windt}, {Golub}, {Podgorski},
  {Testa}, \& {Weber}}]{Boerner_2012_SolPhys}
{Boerner}, P., {Edwards}, C., {Lemen}, J., {et~al.} 2012, \solphys, 275, 41

\bibitem[{{Bonfils} {et~al.}(2013){Bonfils}, {Delfosse}, {Udry}, {Forveille},
  {Mayor}, {Perrier}, {Bouchy}, {Gillon}, {Lovis}, {Pepe}, {Queloz}, {Santos},
  {S{\'e}gransan}, \& {Bertaux}}]{Bonfils_2013_AA}
{Bonfils}, X., {Delfosse}, X., {Udry}, S., {et~al.} 2013, \aap, 549, A109

\bibitem[{{Bradshaw} \& {Cargill}(2013)}]{Bradshaw_2013_ApJ}
{Bradshaw}, S.~J. \& {Cargill}, P.~J. 2013, \apj, 770, 12

\bibitem[{{Cassak} {et~al.}(2008){Cassak}, {Mullan}, \&
  {Shay}}]{Cassak_2008_ApJ}
{Cassak}, P.~A., {Mullan}, D.~J., \& {Shay}, M.~A. 2008, \apjl, 676, L69

\bibitem[{{Chamberlin} {et~al.}(2009){Chamberlin}, {Woods}, {Crotser},
  {Eparvier}, {Hock}, \& {Woodraska}}]{Chamberlin_2009_GRL}
{Chamberlin}, P.~C., {Woods}, T.~N., {Crotser}, D.~A., {et~al.} 2009, \grl, 36,
  L05102

\bibitem[{{Chandran} \& {Perez}(2019)}]{Chandran_2019_JPP}
{Chandran}, B. D.~G. \& {Perez}, J.~C. 2019, Journal of Plasma Physics, 85,
  905850409

\bibitem[{{Chen} {et~al.}(2021){Chen}, {Przybylski}, {Peter}, {Tian},
  {Auch{\`e}re}, \& {Berghmans}}]{Chen_2021_arXiv}
{Chen}, Y., {Przybylski}, D., {Peter}, H., {et~al.} 2021, arXiv e-prints,
  arXiv:2104.10940

\bibitem[{{Chitta} {et~al.}(2012){Chitta}, {van Ballegooijen}, {Rouppe van der
  Voort}, {DeLuca}, \& {Kariyappa}}]{Chitta_2012_ApJ}
{Chitta}, L.~P., {van Ballegooijen}, A.~A., {Rouppe van der Voort}, L.,
  {DeLuca}, E.~E., \& {Kariyappa}, R. 2012, \apj, 752, 48

\bibitem[{{Cho} \& {Lazarian}(2003)}]{Cho_2003_MNRAS}
{Cho}, J. \& {Lazarian}, A. 2003, \mnras, 345, 325

\bibitem[{{Cho} \& {Vishniac}(2000)}]{Cho_2000_ApJ}
{Cho}, J. \& {Vishniac}, E.~T. 2000, \apj, 539, 273

\bibitem[{{Cirtain} {et~al.}(2013){Cirtain}, {Golub}, {Winebarger}, {de
  Pontieu}, {Kobayashi}, {Moore}, {Walsh}, {Korreck}, {Weber}, {McCauley},
  {Title}, {Kuzin}, \& {Deforest}}]{Cirtain_2013_Nature}
{Cirtain}, J.~W., {Golub}, L., {Winebarger}, A.~R., {et~al.} 2013, \nat, 493,
  501

\bibitem[{{Claire} {et~al.}(2012){Claire}, {Sheets}, {Cohen}, {Ribas},
  {Meadows}, \& {Catling}}]{Claire_2012_ApJ}
{Claire}, M.~W., {Sheets}, J., {Cohen}, M., {et~al.} 2012, \apj, 757, 95

\bibitem[{{Cranmer}(2008)}]{Cranmer_2008_ApJ}
{Cranmer}, S.~R. 2008, \apj, 689, 316

\bibitem[{{Cranmer}(2009)}]{Cranmer_2009_ApJ}
{Cranmer}, S.~R. 2009, \apj, 706, 824

\bibitem[{{Cranmer}(2017)}]{Cranmer_2017_ApJ}
{Cranmer}, S.~R. 2017, \apj, 840, 114

\bibitem[{{Cranmer} \& {van Ballegooijen}(2005)}]{Cranmer_2005_ApJ}
{Cranmer}, S.~R. \& {van Ballegooijen}, A.~A. 2005, \apjs, 156, 265

\bibitem[{{Dahlburg} {et~al.}(2016){Dahlburg}, {Einaudi}, {Taylor},
  {Ugarte-Urra}, {Warren}, {Rappazzo}, \& {Velli}}]{Dahlburg_2016_ApJ}
{Dahlburg}, R.~B., {Einaudi}, G., {Taylor}, B.~D., {et~al.} 2016, \apj, 817, 47

\bibitem[{{De Pontieu} {et~al.}(2007){De Pontieu}, {McIntosh}, {Carlsson},
  {Hansteen}, {Tarbell}, {Schrijver}, {Title}, {Shine}, {Tsuneta}, {Katsukawa},
  {Ichimoto}, {Suematsu}, {Shimizu}, \& {Nagata}}]{De_Pontieu_2007_Science}
{De Pontieu}, B., {McIntosh}, S.~W., {Carlsson}, M., {et~al.} 2007, Science,
  318, 1574

\bibitem[{{Del Zanna} {et~al.}(2021){Del Zanna}, {Dere}, {Young}, \&
  {Landi}}]{Del_Zanna_2021_ApJ}
{Del Zanna}, G., {Dere}, K.~P., {Young}, P.~R., \& {Landi}, E. 2021, \apj, 909,
  38

\bibitem[{{Dere} {et~al.}(1997){Dere}, {Landi}, {Mason}, {Monsignori Fossi}, \&
  {Young}}]{Dere_1997_AA}
{Dere}, K.~P., {Landi}, E., {Mason}, H.~E., {Monsignori Fossi}, B.~C., \&
  {Young}, P.~R. 1997, \aaps, 125, 149

\bibitem[{{Diamond-Lowe} {et~al.}(2021){Diamond-Lowe}, {Youngblood},
  {Charbonneau}, {King}, {Teal}, {Bastelberger}, {Corrales}, \&
  {Kempton}}]{Diamond-Lowe_2021_arXiv}
{Diamond-Lowe}, H., {Youngblood}, A., {Charbonneau}, D., {et~al.} 2021, arXiv
  e-prints, arXiv:2104.10522

\bibitem[{{Dmitruk} {et~al.}(2002){Dmitruk}, {Matthaeus}, {Milano}, {Oughton},
  {Zank}, \& {Mullan}}]{Dmitruk_2002_ApJ}
{Dmitruk}, P., {Matthaeus}, W.~H., {Milano}, L.~J., {et~al.} 2002, \apj, 575,
  571

\bibitem[{{Dressing} \& {Charbonneau}(2015)}]{Dressing_2015_ApJ}
{Dressing}, C.~D. \& {Charbonneau}, D. 2015, \apj, 807, 45

\bibitem[{{Duvvuri} {et~al.}(2021){Duvvuri}, {Pineda}, {Berta-Thompson},
  {Brown}, {France}, {Kowalski}, {Redfield}, {Tilipman}, {Vieytes}, {Wilson},
  {Youngblood}, {Froning}, {Linsky}, {Loyd}, {Mauas}, {Miguel}, {Newton},
  {Rugheimer}, \& {Schneider}}]{Duvvuri_2021_arXiv}
{Duvvuri}, G.~M., {Pineda}, J.~S., {Berta-Thompson}, Z.~K., {et~al.} 2021,
  arXiv e-prints, arXiv:2102.08493

\bibitem[{{Edl{\'e}n}(1943)}]{Edlen_1943}
{Edl{\'e}n}, B. 1943, \zap, 22, 30

\bibitem[{{Ehrenreich} {et~al.}(2015){Ehrenreich}, {Bourrier}, {Wheatley},
  {Lecavelier des Etangs}, {H{\'e}brard}, {Udry}, {Bonfils}, {Delfosse},
  {D{\'e}sert}, {Sing}, \& {Vidal-Madjar}}]{Ehrenreich_2015_Nature}
{Ehrenreich}, D., {Bourrier}, V., {Wheatley}, P.~J., {et~al.} 2015, \nat, 522,
  459

\bibitem[{{Einaudi} {et~al.}(1996){Einaudi}, {Velli}, {Politano}, \&
  {Pouquet}}]{Einaudi_1996_ApJ}
{Einaudi}, G., {Velli}, M., {Politano}, H., \& {Pouquet}, A. 1996, \apjl, 457,
  L113

\bibitem[{{Felipe} {et~al.}(2018){Felipe}, {Kuckein}, \&
  {Thaler}}]{Felipe_2018_AA}
{Felipe}, T., {Kuckein}, C., \& {Thaler}, I. 2018, \aap, 617, A39

\bibitem[{{Fontenla} {et~al.}(1990){Fontenla}, {Avrett}, \&
  {Loeser}}]{Fontenla_1990_ApJ}
{Fontenla}, J.~M., {Avrett}, E.~H., \& {Loeser}, R. 1990, \apj, 355, 700

\bibitem[{{France} {et~al.}(2018){France}, {Arulanantham}, {Fossati}, {Lanza},
  {Loyd}, {Redfield}, \& {Schneider}}]{France_2018_ApJ}
{France}, K., {Arulanantham}, N., {Fossati}, L., {et~al.} 2018, \apjs, 239, 16

\bibitem[{{Galsgaard} \& {Nordlund}(1996)}]{Galsgaard_1996_JGR}
{Galsgaard}, K. \& {Nordlund}, {\r{A}}. 1996, \jgr, 101, 13445

\bibitem[{{Goodman} \& {Judge}(2012)}]{Goodman_2012_ApJ}
{Goodman}, M.~L. \& {Judge}, P.~G. 2012, \apj, 751, 75

\bibitem[{{Gorti} \& {Hollenbach}(2009)}]{Gorti_2009_ApJ}
{Gorti}, U. \& {Hollenbach}, D. 2009, \apj, 690, 1539

\bibitem[{{Gottlieb} {et~al.}(2001){Gottlieb}, {Shu}, \&
  {Tadmor}}]{Gottlieb_2001_SIAMR}
{Gottlieb}, S., {Shu}, C.-W., \& {Tadmor}, E. 2001, SIAM Review, 43, 89

\bibitem[{{G{\"u}del} {et~al.}(2003){G{\"u}del}, {Audard}, {Kashyap}, {Drake},
  \& {Guinan}}]{Gudel_2003_ApJ}
{G{\"u}del}, M., {Audard}, M., {Kashyap}, V.~L., {Drake}, J.~J., \& {Guinan},
  E.~F. 2003, \apj, 582, 423

\bibitem[{{G{\"u}del} {et~al.}(1997){G{\"u}del}, {Guinan}, \&
  {Skinner}}]{Gudel_1997_ApJ}
{G{\"u}del}, M., {Guinan}, E.~F., \& {Skinner}, S.~L. 1997, \apj, 483, 947

\bibitem[{{Gudiksen} \& {Nordlund}(2005)}]{Gudiksen_2005_ApJ}
{Gudiksen}, B.~V. \& {Nordlund}, {\r{A}}. 2005, \apj, 618, 1020

\bibitem[{{Guinan} {et~al.}(2016){Guinan}, {Engle}, \&
  {Durbin}}]{Guinan_2016_ApJ}
{Guinan}, E.~F., {Engle}, S.~G., \& {Durbin}, A. 2016, \apj, 821, 81

\bibitem[{{G{\"u}nther} {et~al.}(2007){G{\"u}nther}, {Schmitt}, {Robrade}, \&
  {Liefke}}]{Gunther_2007_AA}
{G{\"u}nther}, H.~M., {Schmitt}, J.~H.~M.~M., {Robrade}, J., \& {Liefke}, C.
  2007, \aap, 466, 1111

\bibitem[{{Hansteen} {et~al.}(2015){Hansteen}, {Guerreiro}, {De Pontieu}, \&
  {Carlsson}}]{Hansteen_2015_ApJ}
{Hansteen}, V., {Guerreiro}, N., {De Pontieu}, B., \& {Carlsson}, M. 2015,
  \apj, 811, 106

\bibitem[{{Heyvaerts} \& {Priest}(1983)}]{Heyvaerts_1983_AA}
{Heyvaerts}, J. \& {Priest}, E.~R. 1983, \aap, 117, 220

\bibitem[{{Hollweg}(1984)}]{Hollweg_1984_SolPhys}
{Hollweg}, J.~V. 1984, \solphys, 91, 269

\bibitem[{{Hollweg} {et~al.}(1982){Hollweg}, {Jackson}, \&
  {Galloway}}]{Hollweg_1982_SolPhys}
{Hollweg}, J.~V., {Jackson}, S., \& {Galloway}, D. 1982, \solphys, 75, 35

\bibitem[{{Hollweg} \& {Yang}(1988)}]{Hollweg_1988_JGR}
{Hollweg}, J.~V. \& {Yang}, G. 1988, \jgr, 93, 5423

\bibitem[{{Hood} {et~al.}(2002){Hood}, {Brooks}, \& {Wright}}]{Hood_2002_RSPSA}
{Hood}, A.~W., {Brooks}, S.~J., \& {Wright}, A.~N. 2002, Proceedings of the
  Royal Society of London Series A, 458, 2307

\bibitem[{{Hossain} {et~al.}(1995){Hossain}, {Gray}, {Pontius}, {Matthaeus}, \&
  {Oughton}}]{Hossain_1995_PhysFluids}
{Hossain}, M., {Gray}, P.~C., {Pontius}, Duane~H., J., {Matthaeus}, W.~H., \&
  {Oughton}, S. 1995, Physics of Fluids, 7, 2886

\bibitem[{{Hunter}(2007)}]{Hunter_2007_CSE}
{Hunter}, J.~D. 2007, Computing in Science and Engineering, 9, 90

\bibitem[{{Iijima}(2016)}]{Iijima_2016_PhD}
{Iijima}, H. 2016, PhD thesis, Department of Earth and Planetary Science,
  School of Science, The University of Tokyo, Japan

\bibitem[{{Iijima} \& {Imada}(2021)}]{Iijima_2021_arXiv}
{Iijima}, H. \& {Imada}, S. 2021, arXiv e-prints, arXiv:2106.00864

\bibitem[{{Ionson}(1978)}]{Ionson_1978_ApJ}
{Ionson}, J.~A. 1978, \apj, 226, 650

\bibitem[{{Irwin} \& {Bouvier}(2009)}]{Irwin_2009_proceedings}
{Irwin}, J. \& {Bouvier}, J. 2009, in The Ages of Stars, ed. E.~E. {Mamajek},
  D.~R. {Soderblom}, \& R.~F.~G. {Wyse}, Vol. 258, 363--374

\bibitem[{{Ishikawa} {et~al.}(2021){Ishikawa}, {Trujillo Bueno}, {del Pino
  Aleman}, {Okamoto}, {McKenzie}, {Auchere}, {Kano}, {Song}, {Yoshida},
  {Rachmeler}, {Kobayashi}, {Hara}, {Kubo}, {Narukage}, {Sakao}, {Shimizu},
  {Suematsu}, {Bethge}, {De Pontieu}, {Sainz Dalda}, {Vigil}, {Winebarger},
  {Alsina Ballester}, {Belluzzi}, {Stepan}, {Asensio Ramos}, {Carlsson}, \&
  {Leenaarts}}]{Ishikawa_2021_arXiv}
{Ishikawa}, R., {Trujillo Bueno}, J., {del Pino Aleman}, T., {et~al.} 2021,
  arXiv e-prints, arXiv:2103.01583

\bibitem[{{Ishikawa} {et~al.}(2017){Ishikawa}, {Glesener}, {Krucker},
  {Christe}, {Buitrago-Casas}, {Narukage}, \&
  {Vievering}}]{Ishikawa_2017_NatAs}
{Ishikawa}, S.-n., {Glesener}, L., {Krucker}, S., {et~al.} 2017, Nature
  Astronomy, 1, 771

\bibitem[{{Jess} {et~al.}(2016){Jess}, {Reznikova}, {Ryans}, {Christian},
  {Keys}, {Mathioudakis}, {Mackay}, {Krishna Prasad}, {Banerjee}, {Grant},
  {Yau}, \& {Diamond}}]{Jess_2016_NatPhys}
{Jess}, D.~B., {Reznikova}, V.~E., {Ryans}, R. S.~I., {et~al.} 2016, Nature
  Physics, 12, 179

\bibitem[{{Johnston} \& {Bradshaw}(2019)}]{Johnston_2019_ApJ}
{Johnston}, C.~D. \& {Bradshaw}, S.~J. 2019, \apjl, 873, L22

\bibitem[{{Johnston} {et~al.}(2017){Johnston}, {Hood}, {Cargill}, \& {De
  Moortel}}]{Johnston_2017_AA}
{Johnston}, C.~D., {Hood}, A.~W., {Cargill}, P.~J., \& {De Moortel}, I. 2017,
  \aap, 597, A81

\bibitem[{{Johnston} {et~al.}(2021){Johnston}, {Hood}, {De Moortel}, {Pagano},
  \& {Howson}}]{Johnston_2021_arXiv}
{Johnston}, C.~D., {Hood}, A.~W., {De Moortel}, I., {Pagano}, P., \& {Howson},
  T.~A. 2021, arXiv e-prints, arXiv:2106.03989

\bibitem[{{Johnstone} {et~al.}(2021){Johnstone}, {Bartel}, \&
  {G{\"u}del}}]{Johnstone_2021_AA}
{Johnstone}, C.~P., {Bartel}, M., \& {G{\"u}del}, M. 2021, \aap, 649, A96

\bibitem[{{Johnstone} \& {G{\"u}del}(2015)}]{Johnstone_2015_AA}
{Johnstone}, C.~P. \& {G{\"u}del}, M. 2015, \aap, 578, A129

\bibitem[{{Judge} {et~al.}(2003){Judge}, {Solomon}, \&
  {Ayres}}]{Judge_2003_ApJ}
{Judge}, P.~G., {Solomon}, S.~C., \& {Ayres}, T.~R. 2003, \apj, 593, 534

\bibitem[{{Kanella} \& {Gudiksen}(2018)}]{Kanella_2018_AA}
{Kanella}, C. \& {Gudiksen}, B.~V. 2018, \aap, 617, A50

\bibitem[{{Katsukawa} \& {Tsuneta}(2005)}]{Katsukawa_2005_ApJ}
{Katsukawa}, Y. \& {Tsuneta}, S. 2005, \apj, 621, 498

\bibitem[{{Kawaler}(1988)}]{Kawaler_1988_ApJ}
{Kawaler}, S.~D. 1988, \apj, 333, 236

\bibitem[{{Keller} {et~al.}(2004){Keller}, {Sch{\"u}ssler}, {V{\"o}gler}, \&
  {Zakharov}}]{Keller_2004_ApJ}
{Keller}, C.~U., {Sch{\"u}ssler}, M., {V{\"o}gler}, A., \& {Zakharov}, V. 2004,
  \apjl, 607, L59

\bibitem[{{Klimchuk}(2006)}]{Klimchuk_2006_SolPhys}
{Klimchuk}, J.~A. 2006, \solphys, 234, 41

\bibitem[{{Klimchuk} {et~al.}(1992){Klimchuk}, {Lemen}, {Feldman}, {Tsuneta},
  \& {Uchida}}]{Klimchuk_1992_PASJ}
{Klimchuk}, J.~A., {Lemen}, J.~R., {Feldman}, U., {Tsuneta}, S., \& {Uchida},
  Y. 1992, \pasj, 44, L181

\bibitem[{{Klimchuk} {et~al.}(2008){Klimchuk}, {Patsourakos}, \&
  {Cargill}}]{Klimchuk_2008_ApJ}
{Klimchuk}, J.~A., {Patsourakos}, S., \& {Cargill}, P.~J. 2008, \apj, 682, 1351

\bibitem[{{Kochukhov} {et~al.}(2020){Kochukhov}, {Hackman}, {Lehtinen}, \&
  {Wehrhahn}}]{Kochukhov_2020_AA}
{Kochukhov}, O., {Hackman}, T., {Lehtinen}, J.~J., \& {Wehrhahn}, A. 2020,
  \aap, 635, A142

\bibitem[{{Kraft}(1967)}]{Kraft_1967_ApJ}
{Kraft}, R.~P. 1967, \apj, 150, 551

\bibitem[{{Kudoh} \& {Shibata}(1999)}]{Kudoh_1999_ApJ}
{Kudoh}, T. \& {Shibata}, K. 1999, \apj, 514, 493

\bibitem[{{Lanzafame}(1995)}]{Lanzafame_1995_AA}
{Lanzafame}, A.~C. 1995, \aap, 302, 839

\bibitem[{{Lecavelier des Etangs} {et~al.}(2012){Lecavelier des Etangs},
  {Bourrier}, {Wheatley}, {Dupuy}, {Ehrenreich}, {Vidal-Madjar}, {H{\'e}brard},
  {Ballester}, {D{\'e}sert}, {Ferlet}, \&
  {Sing}}]{Lecavelier_des_Etangs_2012_AA}
{Lecavelier des Etangs}, A., {Bourrier}, V., {Wheatley}, P.~J., {et~al.} 2012,
  \aap, 543, L4

\bibitem[{{Lemen} {et~al.}(2012){Lemen}, {Title}, {Akin}, {Boerner}, {Chou},
  {Drake}, {Duncan}, {Edwards}, {Friedlaender}, {Heyman}, {Hurlburt}, {Katz},
  {Kushner}, {Levay}, {Lindgren}, {Mathur}, {McFeaters}, {Mitchell}, {Rehse},
  {Schrijver}, {Springer}, {Stern}, {Tarbell}, {Wuelser}, {Wolfson}, {Yanari},
  {Bookbinder}, {Cheimets}, {Caldwell}, {Deluca}, {Gates}, {Golub}, {Park},
  {Podgorski}, {Bush}, {Scherrer}, {Gummin}, {Smith}, {Auker}, {Jerram},
  {Pool}, {Soufli}, {Windt}, {Beardsley}, {Clapp}, {Lang}, \&
  {Waltham}}]{Lemen_2012_SolPhys}
{Lemen}, J.~R., {Title}, A.~M., {Akin}, D.~J., {et~al.} 2012, \solphys, 275, 17

\bibitem[{{Linsky} {et~al.}(2014){Linsky}, {Fontenla}, \&
  {France}}]{Linsky_2014_ApJ}
{Linsky}, J.~L., {Fontenla}, J., \& {France}, K. 2014, \apj, 780, 61

\bibitem[{{Magaudda} {et~al.}(2020){Magaudda}, {Stelzer}, {Covey}, {Raetz},
  {Matt}, \& {Scholz}}]{Magaudda_2020_AA}
{Magaudda}, E., {Stelzer}, B., {Covey}, K.~R., {et~al.} 2020, \aap, 638, A20

\bibitem[{{Malanushenko} {et~al.}(2021){Malanushenko}, {Cheung}, {DeForest},
  {Klimchuk}, \& {Rempel}}]{Malanushenko_2021_arXiv}
{Malanushenko}, A., {Cheung}, M.~C.~M., {DeForest}, C.~E., {Klimchuk}, J.~A.,
  \& {Rempel}, M. 2021, arXiv e-prints, arXiv:2106.14877

\bibitem[{{Matsumoto} \& {Shibata}(2010)}]{Matsumoto_2010_ApJ}
{Matsumoto}, T. \& {Shibata}, K. 2010, \apj, 710, 1857

\bibitem[{{Matt} {et~al.}(2015){Matt}, {Brun}, {Baraffe}, {Bouvier}, \&
  {Chabrier}}]{Matt_2015_ApJ}
{Matt}, S.~P., {Brun}, A.~S., {Baraffe}, I., {Bouvier}, J., \& {Chabrier}, G.
  2015, \apjl, 799, L23

\bibitem[{{Matthaeus} {et~al.}(1999){Matthaeus}, {Zank}, {Oughton}, {Mullan},
  \& {Dmitruk}}]{Matthaeus_1999_ApJ}
{Matthaeus}, W.~H., {Zank}, G.~P., {Oughton}, S., {Mullan}, D.~J., \&
  {Dmitruk}, P. 1999, \apjl, 523, L93

\bibitem[{{McIntosh} {et~al.}(2011){McIntosh}, {de Pontieu}, {Carlsson},
  {Hansteen}, {Boerner}, \& {Goossens}}]{McIntosh_2011_Nature}
{McIntosh}, S.~W., {de Pontieu}, B., {Carlsson}, M., {et~al.} 2011, \nat, 475,
  477

\bibitem[{{Meyer} {et~al.}(2012){Meyer}, {Balsara}, \&
  {Aslam}}]{Meyer_2012_MNRAS}
{Meyer}, C.~D., {Balsara}, D.~S., \& {Aslam}, T.~D. 2012, \mnras, 422, 2102

\bibitem[{{Meyer} {et~al.}(2014){Meyer}, {Balsara}, \&
  {Aslam}}]{Meyer_2014_JCP}
{Meyer}, C.~D., {Balsara}, D.~S., \& {Aslam}, T.~D. 2014, Journal of
  Computational Physics, 257, 594

\bibitem[{{Miyoshi} \& {Kusano}(2005)}]{Miyoshi_2005_JCP}
{Miyoshi}, T. \& {Kusano}, K. 2005, Journal of Computational Physics, 208, 315

\bibitem[{{Moriyasu} {et~al.}(2004){Moriyasu}, {Kudoh}, {Yokoyama}, \&
  {Shibata}}]{Moriyasu_2004_ApJ}
{Moriyasu}, S., {Kudoh}, T., {Yokoyama}, T., \& {Shibata}, K. 2004, \apjl, 601,
  L107

\bibitem[{{Nakariakov} \& {Ofman}(2001)}]{Nakariakov_2001_AA}
{Nakariakov}, V.~M. \& {Ofman}, L. 2001, \aap, 372, L53

\bibitem[{{Nakatani} {et~al.}(2018){Nakatani}, {Hosokawa}, {Yoshida}, {Nomura},
  \& {Kuiper}}]{Nakatani_2018_ApJ}
{Nakatani}, R., {Hosokawa}, T., {Yoshida}, N., {Nomura}, H., \& {Kuiper}, R.
  2018, \apj, 865, 75

\bibitem[{{Okamoto} {et~al.}(2015){Okamoto}, {Antolin}, {De Pontieu},
  {Uitenbroek}, {Van Doorsselaere}, \& {Yokoyama}}]{Okamoto_2015_ApJ}
{Okamoto}, T.~J., {Antolin}, P., {De Pontieu}, B., {et~al.} 2015, \apj, 809, 71

\bibitem[{{Oran} {et~al.}(2017){Oran}, {Landi}, {van der Holst}, {Sokolov}, \&
  {Gombosi}}]{Oran_2017_ApJ}
{Oran}, R., {Landi}, E., {van der Holst}, B., {Sokolov}, I.~V., \& {Gombosi},
  T.~I. 2017, \apj, 845, 98

\bibitem[{{Osterbrock}(1961)}]{Osterbrock_1961_ApJ}
{Osterbrock}, D.~E. 1961, \apj, 134, 347

\bibitem[{{Owen} \& {Wu}(2013)}]{Owen_2013_ApJ}
{Owen}, J.~E. \& {Wu}, Y. 2013, \apj, 775, 105

\bibitem[{{Pagano} \& {De Moortel}(2019)}]{Pagano_2019_AA}
{Pagano}, P. \& {De Moortel}, I. 2019, \aap, 623, A37

\bibitem[{{Pallavicini} {et~al.}(1981){Pallavicini}, {Golub}, {Rosner},
  {Vaiana}, {Ayres}, \& {Linsky}}]{Pallavicini_1981_ApJ}
{Pallavicini}, R., {Golub}, L., {Rosner}, R., {et~al.} 1981, \apj, 248, 279

\bibitem[{{Parker}(1972)}]{Parker_1972_ApJ}
{Parker}, E.~N. 1972, \apj, 174, 499

\bibitem[{{Parker}(1983)}]{Parker_1983_ApJ}
{Parker}, E.~N. 1983, \apj, 264, 642

\bibitem[{{Parker}(1988)}]{Parker_1988_ApJ}
{Parker}, E.~N. 1988, \apj, 330, 474

\bibitem[{{Peres} {et~al.}(1982){Peres}, {Serio}, {Vaiana}, \&
  {Rosner}}]{Peres_1982_ApJ}
{Peres}, G., {Serio}, S., {Vaiana}, G.~S., \& {Rosner}, R. 1982, \apj, 252, 791

\bibitem[{{Pesnell} {et~al.}(2012){Pesnell}, {Thompson}, \&
  {Chamberlin}}]{Pesnell_2012_SolPhys}
{Pesnell}, W.~D., {Thompson}, B.~J., \& {Chamberlin}, P.~C. 2012, \solphys,
  275, 3

\bibitem[{{Pevtsov} {et~al.}(2003){Pevtsov}, {Fisher}, {Acton}, {Longcope},
  {Johns-Krull}, {Kankelborg}, \& {Metcalf}}]{Pevtsov_2003_ApJ}
{Pevtsov}, A.~A., {Fisher}, G.~H., {Acton}, L.~W., {et~al.} 2003, \apj, 598,
  1387

\bibitem[{{Pizzolato} {et~al.}(2003){Pizzolato}, {Maggio}, {Micela},
  {Sciortino}, \& {Ventura}}]{Pizzolato_2003_AA}
{Pizzolato}, N., {Maggio}, A., {Micela}, G., {Sciortino}, S., \& {Ventura}, P.
  2003, \aap, 397, 147

\bibitem[{{Priest}(2014)}]{Priest_2014}
{Priest}, E. 2014, {Magnetohydrodynamics of the Sun}

\bibitem[{{Rappazzo} {et~al.}(2007){Rappazzo}, {Velli}, {Einaudi}, \&
  {Dahlburg}}]{Rappazzo_2007_ApJ}
{Rappazzo}, A.~F., {Velli}, M., {Einaudi}, G., \& {Dahlburg}, R.~B. 2007,
  \apjl, 657, L47

\bibitem[{{Rappazzo} {et~al.}(2008){Rappazzo}, {Velli}, {Einaudi}, \&
  {Dahlburg}}]{Rappazzo_2008_ApJ}
{Rappazzo}, A.~F., {Velli}, M., {Einaudi}, G., \& {Dahlburg}, R.~B. 2008, \apj,
  677, 1348

\bibitem[{{Reale} {et~al.}(2004){Reale}, {G{\"u}del}, {Peres}, \&
  {Audard}}]{Reale_2004_AA}
{Reale}, F., {G{\"u}del}, M., {Peres}, G., \& {Audard}, M. 2004, \aap, 416, 733

\bibitem[{{Reale} \& {Micela}(1998)}]{Reale_1998_AA}
{Reale}, F. \& {Micela}, G. 1998, \aap, 334, 1028

\bibitem[{{Reiners} {et~al.}(2009){Reiners}, {Basri}, \&
  {Browning}}]{Reiners_2009_ApJ}
{Reiners}, A., {Basri}, G., \& {Browning}, M. 2009, \apj, 692, 538

\bibitem[{{Rempel}(2017)}]{Rempel_2017_ApJ}
{Rempel}, M. 2017, \apj, 834, 10

\bibitem[{{Ribas} {et~al.}(2005){Ribas}, {Guinan}, {G{\"u}del}, \&
  {Audard}}]{Ribas_2005_ApJ}
{Ribas}, I., {Guinan}, E.~F., {G{\"u}del}, M., \& {Audard}, M. 2005, \apj, 622,
  680

\bibitem[{{Rosner} {et~al.}(1978){Rosner}, {Tucker}, \&
  {Vaiana}}]{Rosner_1978_ApJ}
{Rosner}, R., {Tucker}, W.~H., \& {Vaiana}, G.~S. 1978, \apj, 220, 643

\bibitem[{{Rumph} {et~al.}(1994){Rumph}, {Bowyer}, \& {Vennes}}]{Rumph_1994_AJ}
{Rumph}, T., {Bowyer}, S., \& {Vennes}, S. 1994, \aj, 107, 2108

\bibitem[{{Rybicki} \& {Lightman}(1979)}]{Rybicki_1979_book}
{Rybicki}, G.~B. \& {Lightman}, A.~P. 1979, {Radiative processes in
  astrophysics}

\bibitem[{{Saar}(1996)}]{Saar_1996_proceedings}
{Saar}, S.~H. 1996, in Stellar Surface Structure, ed. K.~G. {Strassmeier} \&
  J.~L. {Linsky}, Vol. 176, 237

\bibitem[{{Saar}(2001)}]{Saar_2001_proceedings}
{Saar}, S.~H. 2001, in Astronomical Society of the Pacific Conference Series,
  Vol. 223, 11th Cambridge Workshop on Cool Stars, Stellar Systems and the Sun,
  ed. R.~J. {Garcia Lopez}, R.~{Rebolo}, \& M.~R. {Zapaterio Osorio}, 292

\bibitem[{{Sakurai} {et~al.}(1991){Sakurai}, {Goossens}, \&
  {Hollweg}}]{Sakurai_1991_SolPhys}
{Sakurai}, T., {Goossens}, M., \& {Hollweg}, J.~V. 1991, \solphys, 133, 227

\bibitem[{{Sanz-Forcada} {et~al.}(2011){Sanz-Forcada}, {Micela}, {Ribas},
  {Pollock}, {Eiroa}, {Velasco}, {Solano}, \&
  {Garc{\'\i}a-{\'A}lvarez}}]{Sanz-Forcada_2011_AA}
{Sanz-Forcada}, J., {Micela}, G., {Ribas}, I., {et~al.} 2011, \aap, 532, A6

\bibitem[{{Scelsi} {et~al.}(2005){Scelsi}, {Maggio}, {Peres}, \&
  {Pallavicini}}]{Scelsi_2005_AA}
{Scelsi}, L., {Maggio}, A., {Peres}, G., \& {Pallavicini}, R. 2005, \aap, 432,
  671

\bibitem[{{Schmelz} {et~al.}(2012){Schmelz}, {Reames}, {von Steiger}, \&
  {Basu}}]{Schmelz_2012_ApJ}
{Schmelz}, J.~T., {Reames}, D.~V., {von Steiger}, R., \& {Basu}, S. 2012, \apj,
  755, 33

\bibitem[{{Schrijver} {et~al.}(1994){Schrijver}, {van den Oord}, \&
  {Mewe}}]{Schrijver_1994_AA}
{Schrijver}, C.~J., {van den Oord}, G.~H.~J., \& {Mewe}, R. 1994, \aap, 289,
  L23

\bibitem[{{See} {et~al.}(2019){See}, {Matt}, {Folsom}, {Boro Saikia}, {Donati},
  {Fares}, {Finley}, {H{\'e}brard}, {Jardine}, {Jeffers}, {Lehmann}, {Marsden},
  {Mengel}, {Morin}, {Petit}, {Vidotto}, {Waite}, \& {BCool
  Collaboration}}]{See_2019_ApJ}
{See}, V., {Matt}, S.~P., {Folsom}, C.~P., {et~al.} 2019, \apj, 876, 118

\bibitem[{{Serio} {et~al.}(1981){Serio}, {Peres}, {Vaiana}, {Golub}, \&
  {Rosner}}]{Serio_1981_ApJ}
{Serio}, S., {Peres}, G., {Vaiana}, G.~S., {Golub}, L., \& {Rosner}, R. 1981,
  \apj, 243, 288

\bibitem[{{Shebalin} {et~al.}(1983){Shebalin}, {Matthaeus}, \&
  {Montgomery}}]{Shebalin_1983_JPP}
{Shebalin}, J.~V., {Matthaeus}, W.~H., \& {Montgomery}, D. 1983, Journal of
  Plasma Physics, 29, 525

\bibitem[{{Shibata} \& {Yokoyama}(2002)}]{Shibata_2002_ApJ}
{Shibata}, K. \& {Yokoyama}, T. 2002, \apj, 577, 422

\bibitem[{{Shimizu}(1995)}]{Shimizu_1995_PASJ}
{Shimizu}, T. 1995, \pasj, 47, 251

\bibitem[{{Shoda} {et~al.}(2019){Shoda}, {Suzuki}, {Asgari-Targhi}, \&
  {Yokoyama}}]{Shoda_2019_ApJ}
{Shoda}, M., {Suzuki}, T.~K., {Asgari-Targhi}, M., \& {Yokoyama}, T. 2019,
  \apjl, 880, L2

\bibitem[{{Shoda} {et~al.}(2020){Shoda}, {Suzuki}, {Matt}, {Cranmer},
  {Vidotto}, {Strugarek}, {See}, {R{\'e}ville}, {Finley}, \&
  {Brun}}]{Shoda_2020_ApJ}
{Shoda}, M., {Suzuki}, T.~K., {Matt}, S.~P., {et~al.} 2020, \apj, 896, 123

\bibitem[{{Shoda} {et~al.}(2018){Shoda}, {Yokoyama}, \&
  {Suzuki}}]{Shoda_2018_ApJ_a_self-consistent_model}
{Shoda}, M., {Yokoyama}, T., \& {Suzuki}, T.~K. 2018, \apj, 853, 190

\bibitem[{{Shu} \& {Osher}(1988)}]{Shu_1988_JCP}
{Shu}, C.-W. \& {Osher}, S. 1988, Journal of Computational Physics, 77, 439

\bibitem[{{Skumanich}(1972)}]{Skumanich_1972_ApJ}
{Skumanich}, A. 1972, \apj, 171, 565

\bibitem[{{Spitzer} \& {H{\"a}rm}(1953)}]{Spitzer_1953_PhysRev}
{Spitzer}, L. \& {H{\"a}rm}, R. 1953, Physical Review, 89, 977

\bibitem[{{Spruit} \& {Zweibel}(1979)}]{Spruit_1979_SolPhys}
{Spruit}, H.~C. \& {Zweibel}, E.~G. 1979, \solphys, 62, 15

\bibitem[{{Sreejith} {et~al.}(2020){Sreejith}, {Fossati}, {Youngblood},
  {France}, \& {Ambily}}]{Sreejith_2020_AA}
{Sreejith}, A.~G., {Fossati}, L., {Youngblood}, A., {France}, K., \& {Ambily},
  S. 2020, \aap, 644, A67

\bibitem[{{Srivastava} {et~al.}(2017){Srivastava}, {Shetye}, {Murawski},
  {Doyle}, {Stangalini}, {Scullion}, {Ray}, {W{\'o}jcik}, \&
  {Dwivedi}}]{Srivastava_2017_ScientificReports}
{Srivastava}, A.~K., {Shetye}, J., {Murawski}, K., {et~al.} 2017, Scientific
  Reports, 7, 43147

\bibitem[{{Stein}(1971)}]{Stein_1971_ApJ}
{Stein}, R.~F. 1971, \apjs, 22, 419

\bibitem[{{Steiner} {et~al.}(1998){Steiner}, {Grossmann-Doerth}, {Kn{\"o}lker},
  \& {Sch{\"u}ssler}}]{Steiner_1998_ApJ}
{Steiner}, O., {Grossmann-Doerth}, U., {Kn{\"o}lker}, M., \& {Sch{\"u}ssler},
  M. 1998, \apj, 495, 468

\bibitem[{{Sturrock} \& {Uchida}(1981)}]{Sturrock_1981_ApJ}
{Sturrock}, P.~A. \& {Uchida}, Y. 1981, \apj, 246, 331

\bibitem[{{Suresh} \& {Huynh}(1997)}]{Suresh_1997_JCP}
{Suresh}, A. \& {Huynh}, H.~T. 1997, Journal of Computational Physics, 136, 83

\bibitem[{{Suzuki} \& {Inutsuka}(2005)}]{Suzuki_2005_ApJ}
{Suzuki}, T.~K. \& {Inutsuka}, S.-i. 2005, \apjl, 632, L49

\bibitem[{{Takasao} {et~al.}(2013){Takasao}, {Isobe}, \&
  {Shibata}}]{Takasao_2013_PASJ}
{Takasao}, S., {Isobe}, H., \& {Shibata}, K. 2013, \pasj, 65, 62

\bibitem[{{Takasao} {et~al.}(2020){Takasao}, {Mitsuishi}, {Shimura}, {Yoshida},
  {Kunitomo}, {Tanaka}, \& {Ishihara}}]{Takasao_2020_ApJ}
{Takasao}, S., {Mitsuishi}, I., {Shimura}, T., {et~al.} 2020, \apj, 901, 70

\bibitem[{{Telleschi} {et~al.}(2005){Telleschi}, {G{\"u}del}, {Briggs},
  {Audard}, {Ness}, \& {Skinner}}]{Telleschi_2005_ApJ}
{Telleschi}, A., {G{\"u}del}, M., {Briggs}, K., {et~al.} 2005, \apj, 622, 653

\bibitem[{{Testa} {et~al.}(2014){Testa}, {De Pontieu}, {Allred}, {Carlsson},
  {Reale}, {Daw}, {Hansteen}, {Martinez-Sykora}, {Liu}, {DeLuca}, {Golub},
  {McKillop}, {Reeves}, {Saar}, {Tian}, {Lemen}, {Title}, {Boerner},
  {Hurlburt}, {Tarbell}, {Wuelser}, {Kleint}, {Kankelborg}, \&
  {Jaeggli}}]{Testa_2014_Science}
{Testa}, P., {De Pontieu}, B., {Allred}, J., {et~al.} 2014, Science, 346,
  1255724

\bibitem[{{Toriumi} {et~al.}(2020){Toriumi}, {Airapetian}, {Hudson},
  {Schrijver}, {Cheung}, \& {DeRosa}}]{Toriumi_2020_ApJ}
{Toriumi}, S., {Airapetian}, V.~S., {Hudson}, H.~S., {et~al.} 2020, \apj, 902,
  36

\bibitem[{{Tsuneta} {et~al.}(2008){Tsuneta}, {Ichimoto}, {Katsukawa}, {Lites},
  {Matsuzaki}, {Nagata}, {Orozco Su{\'a}rez}, {Shimizu}, {Shimojo}, {Shine},
  {Suematsu}, {Suzuki}, {Tarbell}, \& {Title}}]{Tsuneta_2008_ApJ}
{Tsuneta}, S., {Ichimoto}, K., {Katsukawa}, Y., {et~al.} 2008, \apj, 688, 1374

\bibitem[{{Tu} {et~al.}(2015){Tu}, {Johnstone}, {G{\"u}del}, \&
  {Lammer}}]{Tu_2015_AA}
{Tu}, L., {Johnstone}, C.~P., {G{\"u}del}, M., \& {Lammer}, H. 2015, \aap, 577,
  L3

\bibitem[{{van Ballegooijen}(1986)}]{van_Ballegooijen_1986_ApJ}
{van Ballegooijen}, A.~A. 1986, \apj, 311, 1001

\bibitem[{{van Ballegooijen} \&
  {Asgari-Targhi}(2017)}]{van_Ballegooijen_2017_ApJ}
{van Ballegooijen}, A.~A. \& {Asgari-Targhi}, M. 2017, \apj, 835, 10

\bibitem[{{van Ballegooijen} {et~al.}(2011){van Ballegooijen}, {Asgari-Targhi},
  {Cranmer}, \& {DeLuca}}]{van_Ballegooijen_2011_ApJ}
{van Ballegooijen}, A.~A., {Asgari-Targhi}, M., {Cranmer}, S.~R., \& {DeLuca},
  E.~E. 2011, \apj, 736, 3

\bibitem[{{Van Damme} {et~al.}(2020){Van Damme}, {De Moortel}, {Pagano}, \&
  {Johnston}}]{Van_Damme_2020_AA}
{Van Damme}, H.~J., {De Moortel}, I., {Pagano}, P., \& {Johnston}, C.~D. 2020,
  \aap, 635, A174

\bibitem[{{van der Holst} {et~al.}(2014){van der Holst}, {Sokolov}, {Meng},
  {Jin}, {Manchester}, {T{\'o}th}, \& {Gombosi}}]{van_der_Holst_2014_ApJ}
{van der Holst}, B., {Sokolov}, I.~V., {Meng}, X., {et~al.} 2014, \apj, 782, 81

\bibitem[{{van der Walt} {et~al.}(2011){van der Walt}, {Colbert}, \&
  {Varoquaux}}]{van_der_Walt_2011_CSE}
{van der Walt}, S., {Colbert}, S.~C., \& {Varoquaux}, G. 2011, Computing in
  Science and Engineering, 13, 22

\bibitem[{{Van Kooten} \& {Cranmer}(2017)}]{Van_Kooten_2017_ApJ}
{Van Kooten}, S.~J. \& {Cranmer}, S.~R. 2017, \apj, 850, 64

\bibitem[{{van Leer}(1979)}]{van_Leer_1979_JCP}
{van Leer}, B. 1979, Journal of Computational Physics, 32, 101

\bibitem[{{Vasquez}(1990)}]{Vasquez_1990_ApJ}
{Vasquez}, B.~J. 1990, \apj, 356, 693

\bibitem[{{Verdini} {et~al.}(2019){Verdini}, {Grappin}, \&
  {Montagud-Camps}}]{Verdini_2019_SolPhys}
{Verdini}, A., {Grappin}, R., \& {Montagud-Camps}, V. 2019, \solphys, 294, 65

\bibitem[{{Verdini} {et~al.}(2012){Verdini}, {Grappin}, \&
  {Velli}}]{Verdini_2012_AA}
{Verdini}, A., {Grappin}, R., \& {Velli}, M. 2012, \aap, 538, A70

\bibitem[{{Verdini} \& {Velli}(2007)}]{Verdini_2007_ApJ}
{Verdini}, A. \& {Velli}, M. 2007, \apj, 662, 669

\bibitem[{{Verwichte} {et~al.}(2004){Verwichte}, {Nakariakov}, {Ofman}, \&
  {Deluca}}]{Verwichte_2004_SolPhys}
{Verwichte}, E., {Nakariakov}, V.~M., {Ofman}, L., \& {Deluca}, E.~E. 2004,
  \solphys, 223, 77

\bibitem[{{Vidal-Madjar} {et~al.}(2003){Vidal-Madjar}, {Lecavelier des Etangs},
  {D{\'e}sert}, {Ballester}, {Ferlet}, {H{\'e}brard}, \&
  {Mayor}}]{Vidal-Madjar_2003_Nature}
{Vidal-Madjar}, A., {Lecavelier des Etangs}, A., {D{\'e}sert}, J.~M., {et~al.}
  2003, \nat, 422, 143

\bibitem[{{Vidotto} {et~al.}(2014){Vidotto}, {Gregory}, {Jardine}, {Donati},
  {Petit}, {Morin}, {Folsom}, {Bouvier}, {Cameron}, {Hussain}, {Marsden},
  {Waite}, {Fares}, {Jeffers}, \& {do Nascimento}}]{Vidotto_2014_MNRAS}
{Vidotto}, A.~A., {Gregory}, S.~G., {Jardine}, M., {et~al.} 2014, \mnras, 441,
  2361

\bibitem[{{Wang} {et~al.}(2019){Wang}, {Bai}, \& {Goodman}}]{Wang_2019_ApJ}
{Wang}, L., {Bai}, X.-N., \& {Goodman}, J. 2019, \apj, 874, 90

\bibitem[{{Wang}(2020)}]{Wang_2020_ApJ}
{Wang}, Y.~M. 2020, \apj, 904, 199

\bibitem[{{Washinoue} \& {Suzuki}(2019)}]{Washinoue_2019_ApJ}
{Washinoue}, H. \& {Suzuki}, T.~K. 2019, \apj, 885, 164

\bibitem[{{Weber} \& {Davis}(1967)}]{Weber_1967_ApJ}
{Weber}, E.~J. \& {Davis}, Leverett, J. 1967, \apj, 148, 217

\bibitem[{{Winters} {et~al.}(2019){Winters}, {Henry}, {Jao}, {Subasavage},
  {Chatelain}, {Slatten}, {Riedel}, {Silverstein}, \&
  {Payne}}]{Winters_2019_AJ}
{Winters}, J.~G., {Henry}, T.~J., {Jao}, W.-C., {et~al.} 2019, \aj, 157, 216

\bibitem[{{Withbroe} \& {Noyes}(1977)}]{Withbroe_1977_ARAA}
{Withbroe}, G.~L. \& {Noyes}, R.~W. 1977, \araa, 15, 363

\bibitem[{{Woods} {et~al.}(2009){Woods}, {Chamberlin}, {Harder}, {Hock},
  {Snow}, {Eparvier}, {Fontenla}, {McClintock}, \& {Richard}}]{Woods_2009_GRL}
{Woods}, T.~N., {Chamberlin}, P.~C., {Harder}, J.~W., {et~al.} 2009, \grl, 36,
  L01101

\bibitem[{{Wright} \& {Drake}(2016)}]{Wright_2016_Nature}
{Wright}, N.~J. \& {Drake}, J.~J. 2016, \nat, 535, 526

\bibitem[{{Wright} {et~al.}(2011){Wright}, {Drake}, {Mamajek}, \&
  {Henry}}]{Wright_2011_ApJ}
{Wright}, N.~J., {Drake}, J.~J., {Mamajek}, E.~E., \& {Henry}, G.~W. 2011,
  \apj, 743, 48

\bibitem[{{Wright} {et~al.}(2018){Wright}, {Newton}, {Williams}, {Drake}, \&
  {Yadav}}]{Wright_2018_MNRAS}
{Wright}, N.~J., {Newton}, E.~R., {Williams}, P. K.~G., {Drake}, J.~J., \&
  {Yadav}, R.~K. 2018, \mnras, 479, 2351

\bibitem[{{Youngblood} {et~al.}(2017){Youngblood}, {France}, {Loyd}, {Brown},
  {Mason}, {Schneider}, {Tilley}, {Berta-Thompson}, {Buccino}, {Froning},
  {Hawley}, {Linsky}, {Mauas}, {Redfield}, {Kowalski}, {Miguel}, {Newton},
  {Rugheimer}, {Segura}, {Roberge}, \& {Vieytes}}]{Youngblood_2017_ApJ}
{Youngblood}, A., {France}, K., {Loyd}, R.~O.~P., {et~al.} 2017, \apj, 843, 31

\bibitem[{{Zhuleku} {et~al.}(2020){Zhuleku}, {Warnecke}, \&
  {Peter}}]{Zhuleku_2020_AA}
{Zhuleku}, J., {Warnecke}, J., \& {Peter}, H. 2020, \aap, 640, A119

\end{thebibliography}

\begin{appendix}

\section{\revise{Derivation of basic equations} \label{app:basic_equation}}
In this Appendix, the basic equations are derived from the conventional magnetohydrodynamic equations \citep[e.g.][]{Priest_2014} given by 
\begin{align}
    &\frac{\partial \rho}{\partial t} + \nabla \cdot \left( \rho \vec{v} \right) = 0, \label{eq:MHD_mass_conservation} \\
    &\frac{\partial \vec{v}}{\partial t} + \left( \vec{v} \cdot \nabla \right) \vec{v} = - \frac{1}{\rho} \nabla p + \frac{1}{4 \pi \rho} \left( \nabla \times \vec{B} \right) \times \vec{B}  + \vec{g}, \label{eq:MHD_eom} \\
    &\frac{\partial \vec{B}}{\partial t} + \nabla \times \left( \vec{v} \times \vec{B} \right) = 0, \label{eq:MHD_induction} \\
    &\frac{\partial e}{\partial t} + \nabla \cdot \left[ \left(e + p \right) \vec{v} - \frac{1}{4\pi} \left(\vec{v} \times \vec{B} \right) \times \vec{B} \right] = \rho \vec{v} \cdot \vec{g} + Q_{\rm cnd} + Q_{\rm rad}. \label{eq:MHD_energy}
\end{align}
From Eq. (\ref{eq:nabla_operations}) and (\ref{eq:cross_section}), for any vector field $\vec{X}$, its divergence and rotation are given by
\begin{align}
    \nabla \cdot \vec{X} &= \frac{1}{r^2f} \frac{\partial}{\partial s} \left( X_s r^2 f \right), \label{eq:divergence_appendix} \\
    \nabla \times \vec{X} &= - \frac{1}{r \sqrt{f}} \frac{\partial}{\partial s} \left( X_y r \sqrt{f} \right) \vec{e}_x + \frac{1}{r \sqrt{f}} \frac{\partial}{\partial s} \left( X_x r \sqrt{f} \right) \vec{e}_y. \label{eq:rotation_appendix}
\end{align}
Under the assumption that gravity works only in the field-aligned direction, the gravitational force is simplified as
\begin{align}
    \vec{g} = \left( \vec{g} \cdot \vec{e}_s \right) \vec{e}_s = g_s \vec{e}_s,
\end{align}
where $\vec{e}_s$ is the unit vector in the $s$ direction.
Combining Eq.s (\ref{eq:MHD_mass_conservation}), (\ref{eq:MHD_energy}) and (\ref{eq:divergence_appendix}), the mass and energy conservation laws are written as
\begin{align}
    \frac{\partial}{\partial t} \rho &+ \frac{1}{r^2 f} \frac{\partial}{\partial s} \left( \rho v_s r^2 f \right) = 0, \label{eq:basic_mass_conservation} \\
    \frac{\partial}{\partial t} e &+ \frac{1}{r^2 f} \frac{\partial}{\partial s} \left[ \left[\left(e+p+\frac{B_x^2 + B_y^2}{8\pi} \right) v_s - \frac{B_s}{4 \pi} \left(v_x B_x + v_y B_y \right) \right] r^2 f \right]\nonumber \\
    &= \rho v_s g_s + Q_{\rm cnd} + Q_{\rm rad},
    \label{eq:basic_energy_conservation}
\end{align}
where we use
\begin{align}
    & \left[ \left(e + p \right) \vec{v} - \frac{1}{4\pi} \left(\vec{v} \times \vec{B} \right) \times \vec{B} \right] \cdot \vec{e}_s \nonumber \\
    & = \left(e+p+\frac{B_x^2 + B_y^2}{8\pi} \right) v_s - \frac{B_s}{4 \pi} \left(v_x B_x + v_y B_y \right).
\end{align}

The advection term and Lorentz force in Eq. (\ref{eq:MHD_eom}) are written as
\begin{align}
    \left( \vec{v} \cdot \nabla \right) \vec{v} &=
    \left( \nabla \times \vec{v} \right) \times \vec{v} + \frac{1}{2} \nabla \left( v^2 \right), \nonumber \\
    & = \left[ v_s \frac{\partial}{\partial s} v_s - \frac{1}{2} \left( v_x^2 + v_y^2 \right) \frac{d}{ds} \ln \left( r^2 f \right) \right] \vec{e}_s \nonumber \\
    & + \frac{v_s}{r \sqrt{f}} \frac{\partial}{\partial s} \left( r \sqrt{f} v_x \right) \vec{e}_x + \frac{v_s}{r \sqrt{f}} \frac{\partial}{\partial s} \left( r \sqrt{f} v_y \right) \vec{e}_y, 
\end{align}
\begin{align}
    \left( \nabla \times \vec{B} \right) \times \vec{B} &= \left[ - \frac{B_x}{r \sqrt{f}} \frac{\partial}{\partial s} \left( r \sqrt{f} B_x \right) - \frac{B_y}{r \sqrt{f}} \frac{\partial}{\partial s} \left( r \sqrt{f} B_y \right)  \right] \vec{e}_s \nonumber \\
    &+ \frac{B_s}{r \sqrt{f}} \frac{\partial}{\partial s} \left( r \sqrt{f} B_x \right) \vec{e}_x + \frac{B_s}{r \sqrt{f}} \frac{\partial}{\partial s} \left( r \sqrt{f} B_y \right) \vec{e}_y.
\end{align}
The $s$-component of Eq. (\ref{eq:MHD_eom}) is given by
\begin{align}
    & \frac{\partial}{\partial t} v_s + v_s \frac{\partial}{\partial s} v_s - \frac{1}{2} \left( v_x^2 + v_y^2 \right) \frac{d}{ds} \ln \left( r^2 f \right) + \frac{1}{\rho} \frac{\partial}{\partial s}p \nonumber \\
    & + \frac{1}{4 \pi \rho} \left[ \frac{B_x}{r \sqrt{f}} \frac{\partial}{\partial s} \left( r \sqrt{f} B_x \right) + \frac{B_y}{r \sqrt{f}} \frac{\partial}{\partial s} \left( r \sqrt{f} B_y \right) \right] = g_s,
\end{align}
which is, after some algebra, written in terms of conservation law as
\begin{align}
    \frac{\partial}{\partial t} \left( \rho v_s \right) &+ \frac{1}{r^2 f} \frac{\partial}{\partial s} \left[ \left( \rho v_s^2 + p + \frac{B_x^2 + B_y^2}{8 \pi} \right) r^2 f \right] \nonumber \\
    &= \left[ p + \frac{1}{2} \rho \left( v_x^2 + v_y^2 \right) \right] \frac{d}{ds} \ln \left( r^2 f \right) + \rho g_s, \label{eq:basic_eom_scomponent}
\end{align}
The transverse components ($\vec{v}_\perp = v_x \vec{e}_x + v_y \vec{e}_y$) of Eq. (\ref{eq:MHD_eom}) are
\begin{align}
    \frac{\partial}{\partial t} \vec{v}_\perp + \frac{v_s}{r\sqrt{f}} \frac{\partial}{\partial s} \left( r \sqrt{f} \vec{v}_\perp \right) - \frac{B_s}{4 \pi \rho r \sqrt{f}} \frac{\partial}{\partial s} \left( r \sqrt{f} \vec{B}_\perp \right) = 0,
\end{align}
which is written in terms of conservation law as
\begin{align}
    \frac{\partial}{\partial t} \left( \rho \vec{v}_\perp \right) &+ \frac{1}{r^2 f} \frac{\partial}{\partial s} \left[ \left(\rho v_s \vec{v}_\perp - \frac{B_s \vec{B}_\perp}{4 \pi} \right) r^2 f \right] \nonumber \\
    &= -\frac{1}{2} \left( \rho v_s \vec{v}_\perp - \frac{B_s \vec{B}_\perp}{4 \pi} \right) \frac{d}{ds} \ln \left( r^2 f \right).
\end{align}
The actual basic equation is obtained by adding a phenomenological turbulence term $\rho \vec{D}_\perp^v$ on the right-hand side, i.e., 
\begin{align}
    \frac{\partial}{\partial t} \left( \rho \vec{v}_\perp \right) &+ \frac{1}{r^2 f} \frac{\partial}{\partial s} \left[ \left(\rho v_s \vec{v}_\perp - \frac{B_s \vec{B}_\perp}{4 \pi} \right) r^2 f \right] \nonumber \\
    &= -\frac{1}{2} \left( \rho v_s \vec{v}_\perp - \frac{B_s \vec{B}_\perp}{4 \pi} \right) \frac{d}{ds} \ln \left( r^2 f \right) + \rho \vec{D}_\perp^v.
    \label{eq:basic_eom_transversecomponent}
\end{align}
The rotation of the electromotive force in Eq. (\ref{eq:MHD_induction}) is given by
\begin{align}
    \nabla \times \left( \vec{v} \times \vec{B} \right) = &-\frac{1}{r \sqrt{f}} \frac{\partial}{\partial s} \left[ \left( v_s B_x - v_x B_s \right) r \sqrt{f} \right] \vec{e}_x \nonumber \\
    &- \frac{1}{r \sqrt{f}} \frac{\partial}{\partial s} \left[ \left( v_s B_y - v_y B_s \right) r \sqrt{f} \right] \vec{e}_y.
\end{align}
Then, the transverse components ($\vec{B}_\perp = B_x \vec{e}_x + B_y \vec{e}_y$) of the induction equation are
\begin{align}
    \frac{\partial}{\partial t} \vec{B}_\perp + \frac{1}{r\sqrt{f}} \frac{\partial}{\partial s} \left[ \left( v_s \vec{B}_\perp - \vec{v}_\perp B_s \right) r \sqrt{f} \right] = 0,
\end{align}
which is rewritten in terms of conservation law as
\begin{align}
    \frac{\partial}{\partial t} \vec{B}_\perp &+ \frac{1}{r^2 f} \frac{\partial}{\partial s} \left[ \left( v_s \vec{B}_\perp - \vec{v}_\perp B_s \right) r^2 f \right] \nonumber \\ 
    &= \frac{1}{2} \left( v_s \vec{B}_\perp - \vec{v}_\perp B_s \right) \frac{d}{ds} \ln \left( r^2 f \right).
\end{align}
The basic equation is obtained after adding the phenomenological turbulence term:
\begin{align}
    \frac{\partial}{\partial t} \vec{B}_\perp &+ \frac{1}{r^2 f} \frac{\partial}{\partial s} \left[ \left( v_s \vec{B}_\perp - \vec{v}_\perp B_s \right) r^2 f \right] \nonumber \\ 
    &= \frac{1}{2} \left( v_s \vec{B}_\perp - \vec{v}_\perp B_s \right) \frac{d}{ds} \ln \left( r^2 f \right) + \sqrt{4 \pi \rho} \vec{D}_\perp^b.
    \label{eq:basic_induction_transversecomponent}
\end{align}
Eq.s (\ref{eq:basic_mass_conservation}), (\ref{eq:basic_energy_conservation}), (\ref{eq:basic_eom_scomponent}), (\ref{eq:basic_eom_transversecomponent}) and (\ref{eq:basic_induction_transversecomponent}) are the basic equations used in this study.

\section{Transition-region problem and dependence on numerical methods \label{app:transition_region_problem}}

\begin{table}[t!]
\centering
  \begin{tabular}{l l l }
    \begin{tabular}{l} \hspace{-1em} Riemann solver \end{tabular} 
    & \begin{tabular}{c} $\Delta s_{\rm min}$ \end{tabular}
    & \begin{tabular}{c} filling factor $f_\ast$ \end{tabular}
    \rule[-5.5pt]{0pt}{20pt} \\ \hline \hline
    HLL
    & \begin{tabular}{c} \hspace{0em} 5 km \end{tabular}
    & [1, 0.33, 0.1, 0.05]
    \rule[-5.5pt]{0pt}{20pt} \\
    HLLD
    & \begin{tabular}{c} \hspace{0em} 5 km \end{tabular}
    & [1, 0.33, 0.1, 0.05]
    \rule[-5.5pt]{0pt}{20pt} \\
    HLLD
    & \begin{tabular}{c} \hspace{0em} 2 km \end{tabular}
    & [1, 0.33, 0.1, 0.05]
    \rule[-5.5pt]{0pt}{20pt} \\ \hline
  \end{tabular}
  \vspace{1.0em}
  \caption{
  List of the simulation runs executed to study the transition-region problem.
  Note that $\Delta s_{\rm min}$ corresponds to the spatial resolution of the transition region.
  }
  \vspace{0em}
  \label{table:run_TRproblem}
\end{table}

\begin{figure}[t!]
\centering
\includegraphics[width=85mm]{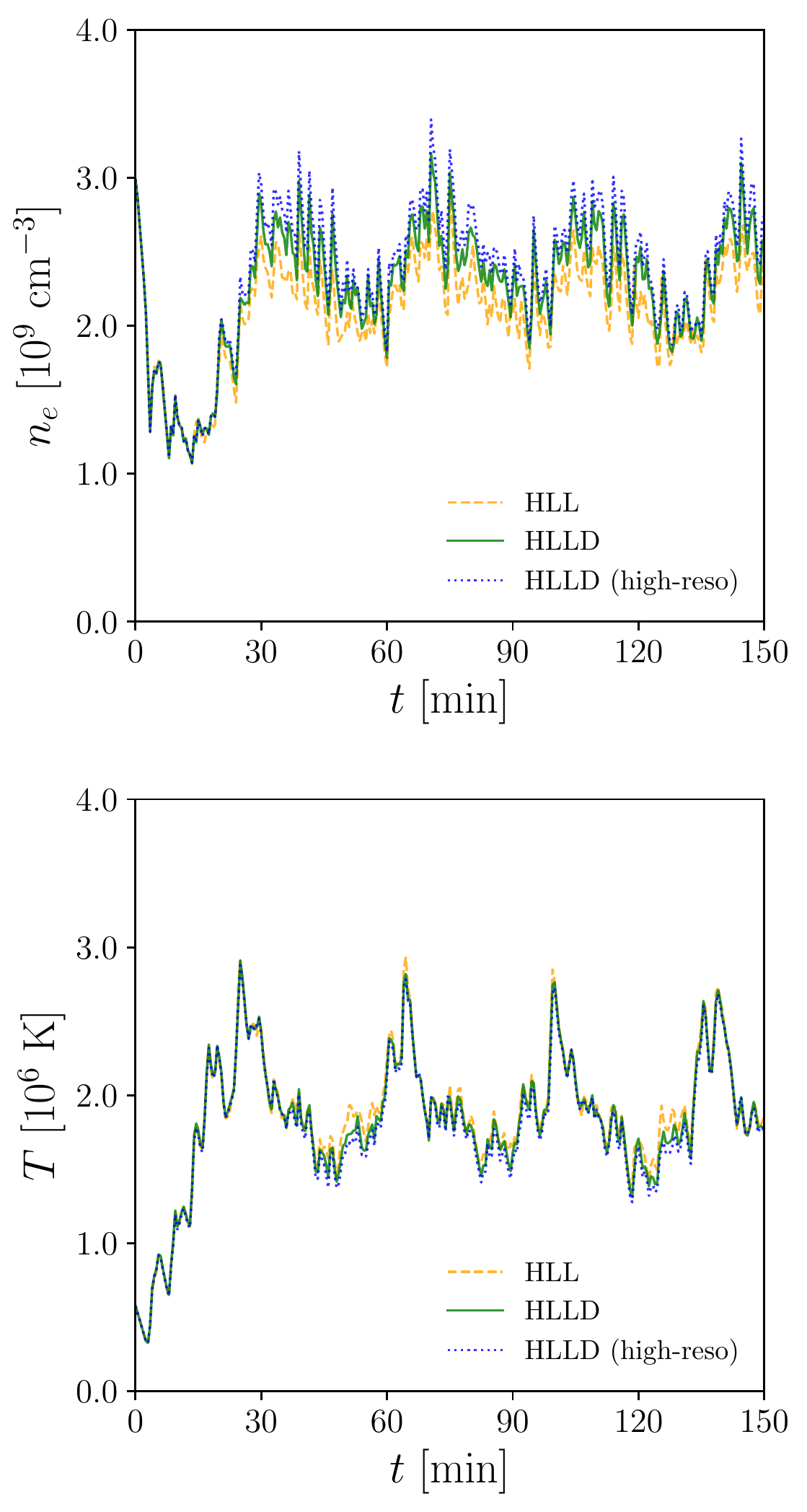}
\vspace{0.5em}
\caption{
Time evolution of the loop-top electron density $n_{e,{\rm top}}$ (top panel) and temperature $T_{\rm top}$ (bottom panel) for fixed filling factor $f_\ast = 0.05$ and half loop length $l_{\rm loop} = 20 {\rm  \ Mm}$.
Three lines show the results with $\Delta s_{\rm min} = 5 {\rm \ km}$ and HLL solver (red), $\Delta s_{\rm min} = 5 {\rm \ km}$ and HLLD solver (green), and $\Delta s_{\rm min} = 2 {\rm \ km}$ and HLLD solver (blue), respectively. 
}
\label{fig:loop_top_evolution_scheme_dependence_fexp20}
\vspace{0em}
\end{figure}

A crucial difficulty in the simulation of coronal heating is that the transition region
needs to be resolved with extremely fine grids, otherwise the coronal density is significantly underestimated \citep{Bradshaw_2013_ApJ}.
This complex problem shall be referred to as the ``transition-region problem''.
Here, we investigate the relation between the coronal density and grid size in the transition region and the numerical scheme.

The numerical settings of the test calculations are listed in Table \ref{table:run_TRproblem}.
Two types of the approximated Riemann solver are used in this test: Harten-van Leer-Lax (HLL) and HLL-Discontinuity (HLLD) schemes.
In exchange for its simplicity, the HLL scheme has the disadvantage that the contact discontinuities (entropy modes) suffer from significant numerical diffusion.
Meanwhile, by explicitly considering the substructures inside the Riemann fan, 
the contact (and rotational) discontinuities are better resolved in the HLLD scheme.
Because the transition region is a type of contact discontinuity, its numerical diffusion in the HLL and HLLD schemes should behave differently even with the same grid spacing. 
Accordingly, the transition-region problem should also change.
Therefore, dependence on the choice of the Riemann solver, as well as the spatial resolution, is also investigated.

\begin{figure}[t!]
\centering
\includegraphics[width=85mm]{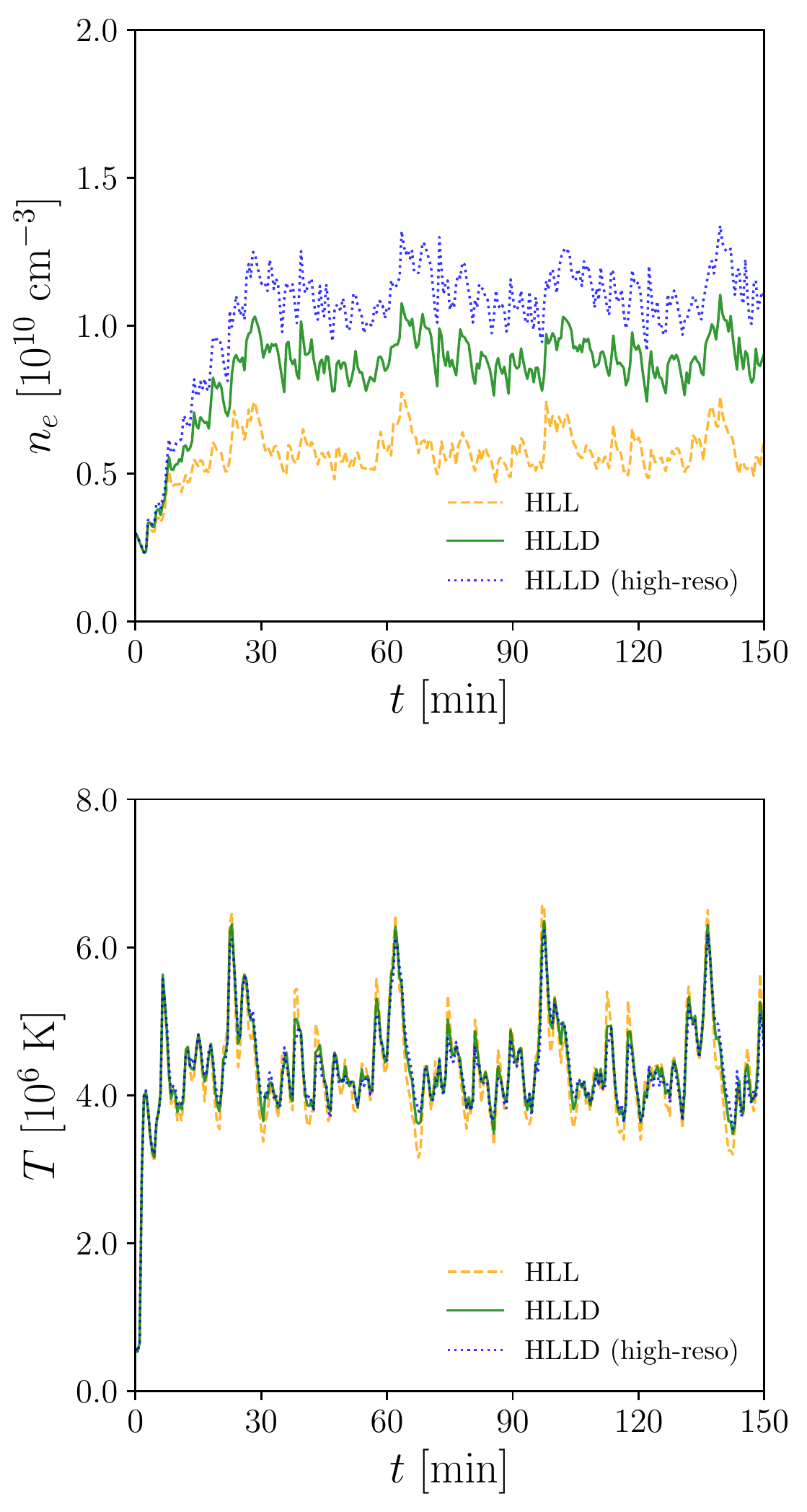}
\vspace{0.5em}
\caption{
Same as Figure \ref{fig:loop_top_evolution_scheme_dependence_fexp20} but for $f_\ast = 1$ and $l_{\rm loop} = 20 {\rm  \ Mm}$.
}
\label{fig:loop_top_evolution_scheme_dependence_fexp1}
\vspace{0em}
\end{figure}

Figure \ref{fig:loop_top_evolution_scheme_dependence_fexp20} shows the time evolution of the loop-top electron density ($n_{e,{\rm top}}$, top panel) and loop-top temperature ($T_{\rm top}$, bottom panel) for a fixed magnetic filling factor at the surface ($f_\ast = 0.05$).
The orange-dashed, green-solid, and blue-dotted lines correspond to the results of the HLL ($\Delta s_{\rm min}=5 {\rm \ km}$), HLLD ($\Delta s_{\rm min}=5 {\rm \ km}$), and HLLD ($\Delta s_{\rm min}=2 {\rm \ km}$) schemes, respectively.
As indicated in the top panel of Figure \ref{fig:loop_top_evolution_scheme_dependence_fexp20}, the three lines exhibit subtle differences in behaviour.
A low resolution with a large numerical diffusion at the transition region yields a small loop-top density, as reported in \citet{Bradshaw_2013_ApJ}.
Meanwhile, the loop-top temperature is nearly independent of the resolution, which is also consistent with \citet{Bradshaw_2013_ApJ}.
Although \citet{Bradshaw_2013_ApJ} state that the transition region must be resolved within 1 km to ensure the numerical convergence of coronal density, the test in this study proves the numerical convergence for $\Delta s_{\rm min} = 5 {\rm \ km}$, when $f_\ast=0.05$.
This gap can be attributed to the different numerical settings.
\citet{Bradshaw_2013_ApJ} consider the thermal response to strong impulsive heating, whereas this study considers quasi-steady uniform coronal heating.

Figure \ref{fig:loop_top_evolution_scheme_dependence_fexp1} is the same as Figure \ref{fig:loop_top_evolution_scheme_dependence_fexp20} but for $f_\ast=1$.
The difference in the loop-top density is more significant than $f_\ast=0.05$.
The HLL and HLLD schemes produce largely different results despite the same grid size at the transition region.
The reason is that, in the HLL method, the numerical diffusivity of the entropy mode is roughly proportional to the speed of the fastest wave. 
At the transition region, the fastest wave speed is the fast magneto-acoustic velocity that increases with the field strength.
Thus, with HLL, $f_\ast =1$ yields a higher wave speed and larger numerical diffusion at the transition region.
Such an enhanced numerical diffusion of the entropy mode is suppressed in the HLLD method because the contact discontinuity inside the Riemann fan is resolved.
Thus, the difference between the HLL and HLLD schemes is more remarkable for $f_\ast=1$, as shown in Figure \ref{fig:loop_top_evolution_scheme_dependence_fexp1}.
The weak dependence of the loop-top temperature on the numerical scheme is also confirmed.

\begin{figure}[t!]
\centering
\includegraphics[width=85mm]{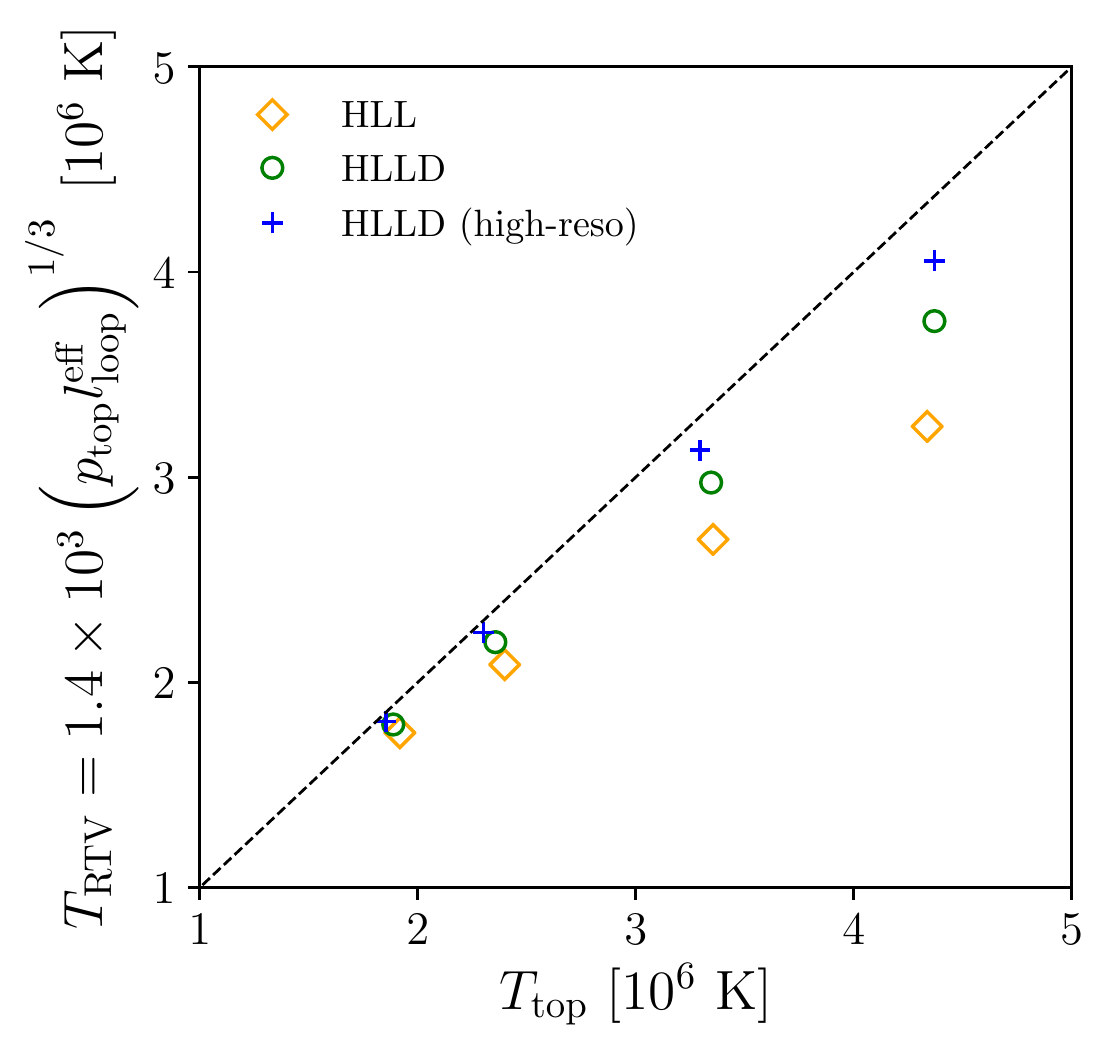}
\vspace{0.5em}
\caption{
This figure compares the time-averaged loop-top temperature $T_{\rm top}$ and the loop-top temperature predicted by the RTV scaling law $T_{\rm RTV}$.
}
\label{fig:RTV_comparison_scheme_dependence}
\vspace{0em}
\end{figure}

To test the numerical convergence of the physical variables in the coronal loop on the actual values,
the simulation result is compared with the results of the RTV scaling law.
Figure \ref{fig:RTV_comparison_scheme_dependence} displays the loop-top temperatures obtained in the simulation and those predicted by the RTV scaling law,
i.e.,
\begin{align}
    T_{\rm RTV} = 1.4 \times 10^{3} {\rm \ K} \ \left( p_{\rm top} l_{\rm loop}^{\rm eff} \right)^{1/3},
\end{align}
where $p_{\rm top}$ is the loop-top pressure and $l_{\rm loop}^{\rm eff}$ is the effective half-loop length,
both of which are measured in the cgs unit.
In Figure \ref{fig:RTV_comparison_scheme_dependence}, as the resolution of the transition region increases, the numerical value monotonically approaches the RTV-predicted value.
Note that the deviation of the results of the simulation from those of the RTV prediction is not necessarily evidence of inaccuracy, although, as a general trend, a higher resolution of the transition region should lead to a value closer to the RTV prediction. Deviations from the RTV value in a low-resolution run is prominent at higher coronal temperatures.
\revise{However, it remains to be seen if the large discrepancy from the RTV value at high coronal temperatures is due to the large numerical diffusion at the transition region or the increase in the required resolution.
The coronal-temperature dependence of the transition-region problem should be investigated in the future.}

\revise{Finally, we note that there are physical mechanisms that can produce the broadening of the transition region, including ambipolar diffusion \citep{Fontenla_1990_ApJ,Lanzafame_1995_AA}.
The transition-region problem should also be revisited considering the additional diffusive processes in future.}

\section{XUV spectrum from different atmospheric layers}

\begin{figure}[t!]
\centering
\includegraphics[width=85mm]{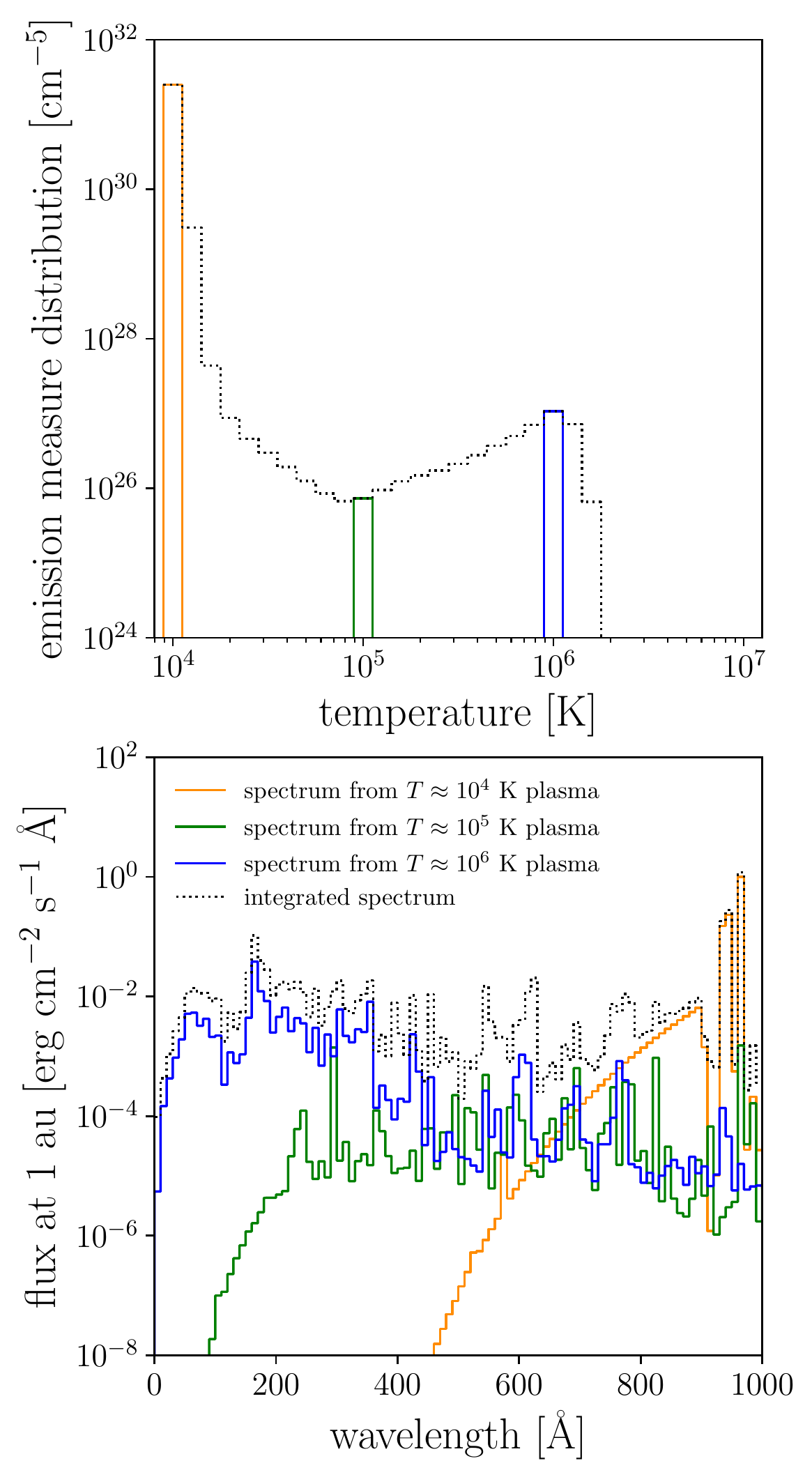}
\vspace{0.5em}
\caption{
This figure shows the emission measure in the limited range of temperature and corresponding spectral energy flux at 1 au.
For better visualisation, the spectral energy fluxes in the bottom panel are averaged over $10 \ \AA$.
Orange, green and blue lines correspond to the emission measures and spectral energy fluxes of $T \approx 10^4 {\rm \ K}$ (chromosphere), $T \approx 10^5 {\rm \ K}$ (transition region) and $T \approx 10^6 {\rm \ K}$ (corona), respectively.
Black dotted lines show the full emission measure distribution and corresponding full spectral energy flux.
}
\label{fig:spectrum_region_divided}
\vspace{0em}
\end{figure}

The XUV photons of a specific wavelength (band) do not originate from the plasma with a specific temperature.
Indeed, narrow-band filters in the Atmospheric Imaging Assembly of the Solar Dynamics Observatory have broad response functions in terms of temperature \citep{Boerner_2012_SolPhys}.
Although there is no direct correlation between temperature and wavelength, it is still useful to understand the energy spectrum emanating from plasma with a specific temperature.

Figure \ref{fig:spectrum_region_divided} shows the correlation between the emission measure distribution in a given temperature bin (top panel) and the corresponding spectral energy flux normalised at 1 au (bottom panel) for the fiducial case ($f_\ast=0.01$, $l_{\rm loop} = 20 {\rm \ Mm}$).
For better visualisation, the spectral energy fluxes averaged over 10 $\AA$ are shown in the bottom panel.
The black dotted line in the top panel represents the total emission measure distribution, while the black dotted line represents the corresponding spectral energy flux in the bottom panel.
We consider three temperature bins centred on $T=10^4 {\rm \ K}$, $T=10^5 {\rm \ K}$, and $T=10^6 {\rm \ K}$ corresponding to the upper chromosphere, transition region, and corona, respectively.
The corresponding spectral energy fluxes are highlighted with the same colours in the bottom panel, which showcase the following three properties.
\begin{enumerate}
    \item Emissions from the upper chromosphere ($T \approx 10^4 {\rm \ K}$) dominate the Lyman continuum but negligible in $\le 600 \ \AA$.
    \\[-1.0em]
    \item Emissions from the transition region ($T \approx 10^5 {\rm \ K}$) are nearly uniform in the middle of the EUV wavelength ($200 \ \AA \le \lambda \le 800 \ \AA$) but negligible in the X-ray range.
    \\[-1.0em]
    \item Emissions from the corona ($T \approx 10^6 {\rm \ K}$) are the dominant sources of stellar X-rays.
    The contribution to the EUV emission is also non-negligible, especially in the shorter wavelength.
\end{enumerate}

As a general trend, the low-energy photons tend to originate from the low-temperature plasma and vice versa.
Note that a non-negligible fraction of EUV photons originates from the chromosphere and the transition region, and therefore including them in the model is essential in predicting the XUV spectrum.

\end{appendix}

\end{document}